\documentclass{cernrep} 
\usepackage{texnames}
\usepackage{bbm}
\usepackage[T1]{fontenc}
\usepackage[bookmarks, colorlinks=true, linktoc=page, pdftex, linkcolor=red, citecolor=red, urlcolor=red]{hyperref}
\usepackage{csquotes}
\begin{document}
\title{Practical Statistics for Particle Physicists}

\author{Luca Lista}   
\institute{Istituto Nazionale di Fisica Nucleare, Sezione di Napoli, Italy}
\maketitle

\begin{abstract}
  These three lectures provide an introduction to the main concepts of statistical data analysis useful for
  precision measurements and searches for new signals in High Energy Physics.
  The frequentist and Bayesian approaches to probability theory are introduced and,
  for both approaches, inference methods are presented. Hypothesis tests
  will be discussed, then  significance and upper limit evaluation will be presented
  with an overview of the modern
  and most advanced techniques adopted for data analysis at the Large Hadron Collider.
\end{abstract}

 
\tableofcontents
\newpage

\section{Introduction}
The main goal of an experimental particle physicist is to make precision measurements
and possibly discover new natural phenomena.
The starting ingredients to this task are particle collisions that
are recorded in form of data delivered by detectors. Data provide
measurements of the position of particle trajectories
or energy releases in the detector, time of particles arrival, etc.
Usually, a large number of collision events are collected by an experiment and
each of such events may containing large amounts of data.
Collision event data are all different from each other due to
the intrinsic randomness of physics process. In Quantum Mechanics
the probability (density) is proportional to the square of the process amplitude
($P \propto |{\cal A}|^2$). Detectors also introduces some degree of randomness
in the data due to fluctuation of the response, like resolution effects, efficiency, etc.
Theory provides prediction of the distributions of measured quantities in data.
Those predictions depend on theory parameters, such as particles masses, particle couplings,
cross section of observed processes, etc.

Given our data sample, we want to either measure the parameters
that appear in the theory (e.g.: determine the top-quark mass
to be: $m_{\mathrm{t}}= 173.44 \pm 0.49$~GeV~\cite{cms_top})
or answer questions about the nature of data. For instance,
as outcome of the search for the Higgs boson at the Large Hadron Collider,
the presence of the boson predicted by Peter Higgs and Fran\c{c}ois Englert
was confirmed providing a quantitative measurement of how strong this evidence was.
Modern experiments search for Dark Matter and they found no convincing evidence so far.
Such searches can provide a range of parameters for theory models
that predict Dark-Matter particle candidates that are allowed or excluded
by the present experimental observation.

In order to achieve the methods that allow to perform the aforementioned measurements
or searches for new signals, first of all a short introduction to probability theory
will be given, in order to master the tools that describe the intrinsic randomness of our data.
Then, methods will be introduced that allow to use probability theory on our data samples in order
to address quantitatively our physics questions.

\section{Probability theory}
\label{sec:probability}
Probability can be defined in different ways, and the applicability of each definition depends
on the kind of claim whose probability we are considering.
One subjective approach expresses the {\it degree of belief/credibility} of a claim,
which may vary from subject to subject.
For repeatable experiments whose outcome is uncertain, probability may be a measure of {\it how frequently} the claim is true.
Repeatable experiments are a subset of the cases where the subjective approach may be applied.

\begin{figure}[htbp]
\centering\includegraphics[width=.495\linewidth]{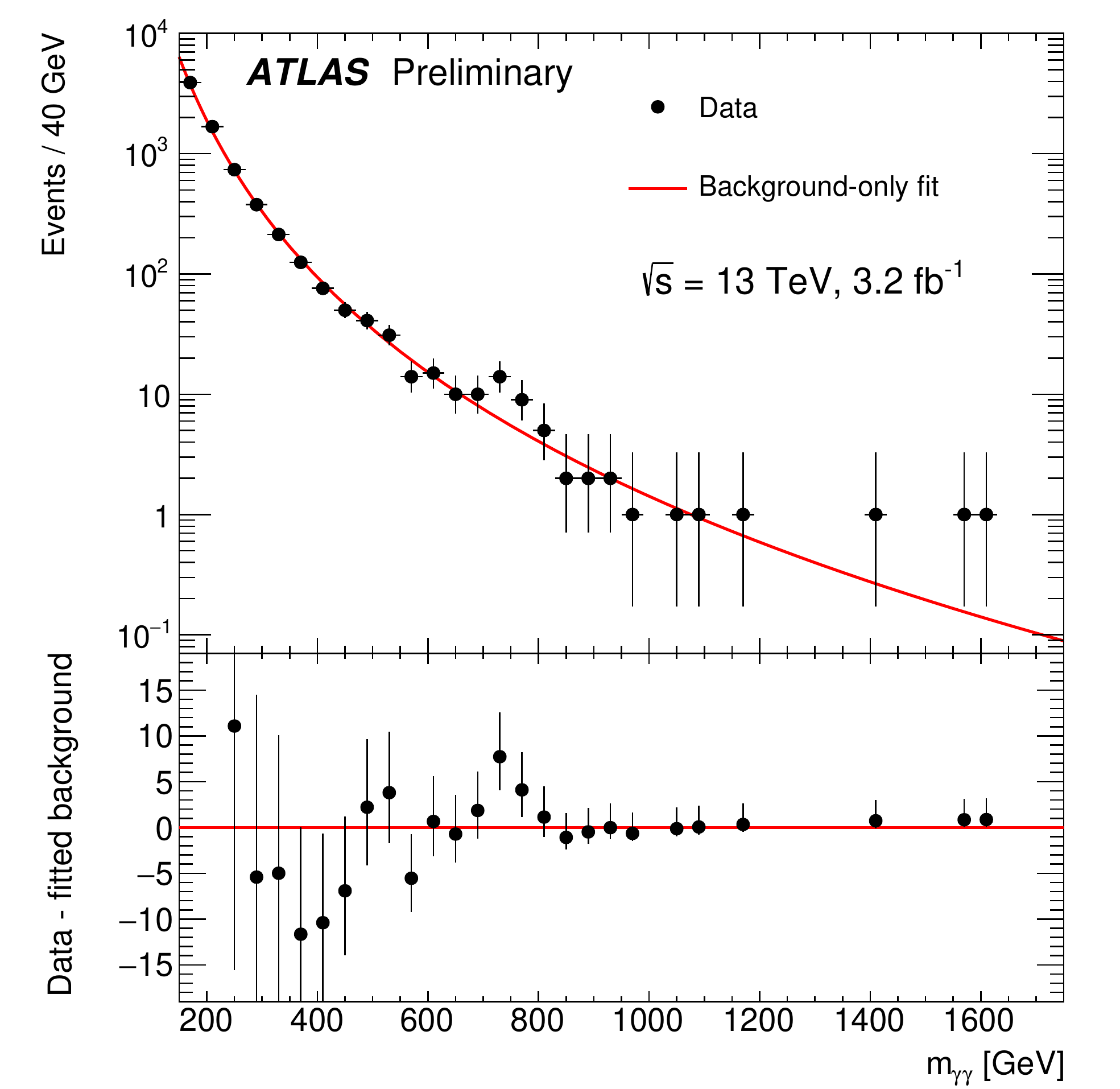}
\caption{Invariant mass distribution of diphoton events selected by ATLAS.
  Figure from Ref.~\cite{ATLASBump}, where details about the analysis are described.}
\label{fig:ATLASBump}
\end{figure}

Examples of probability that can be determined by means of repeatable experiments are the followings:
\begin{itemize}
\item {\it What is the probability to extract an ace in a deck of cards?} \\We can shuffle the deck and extract again the card.
\item {\it What is the probability to win a lottery?} \\Though a specific lottery extraction can't be repeated, we can imagine
  to repeat the extraction process using the same device and extraction procedure.
\item {\it What is the probability that a pion is incorrectly identified as a muon in detector capable of particle identification?}\\
  Most of the experiments have ways to obtain control samples where it's known that only pions are present (e.g.: test
  beams, or specific decay channels, etc.).
  One can count how many pions in a control sample are misidentified as muon.
\item {\it What is the probability that a fluctuation in the background can produce a peak in the $\gamma\gamma$ spectrum with a 
  magnitude at least equal to what has been observed by ATLAS} (Fig.~\ref{fig:ATLASBump}, Ref.~\cite{ATLASBump}){\it?}\\
  At least in principle, the experiment can be repeated with the same running conditions.
  Anyway, this question is different with respect to another possible question: {\it what is the probability that the peak
    is due to a background fluctuation instead of a new signal?} This second question refers to a non-repeatable case.
\end{itemize}
Examples of claims related to non-repeatable situations, instead, are the following:
\begin{itemize}
\item {\it What is the probability that tomorrow it will rain in Geneva?}\\
  The event is related to a specific date in the future. This specific event cannot be repeated.
\item {\it What is the probability that your favorite team will win next championship?}\\
  Though every year there is a championship, a specific one can't be repeated.
\item {\it What is the probability that dinosaurs went extinct because of an asteroid?}\\
  This question is related to an event occurred in the past, but we don't know
  exactly what happened at that time.
\item {\it What is the probability that Dark Matter is made of particles heavier than 1 TeV?}\\
  This question is related to an unknown property of our Universe.
\item {\it What is the probability that climate changes are mainly due to human intervention?}\\
  This question is related to a present event whose cause is unknown.
\end{itemize}
The first examples in the above list are related to events in the future, where it's rather natural
to think in term of degree of belief about some prediction: we can wait and see
if the prediction was true or not. But similar claims can also be related to past
events, or, more in general, to cases where we just don't know whether the claims is true or not.

\subsection{Classical probability}

\begin{figure}[htbp]
\centering\includegraphics[width=.495\linewidth]{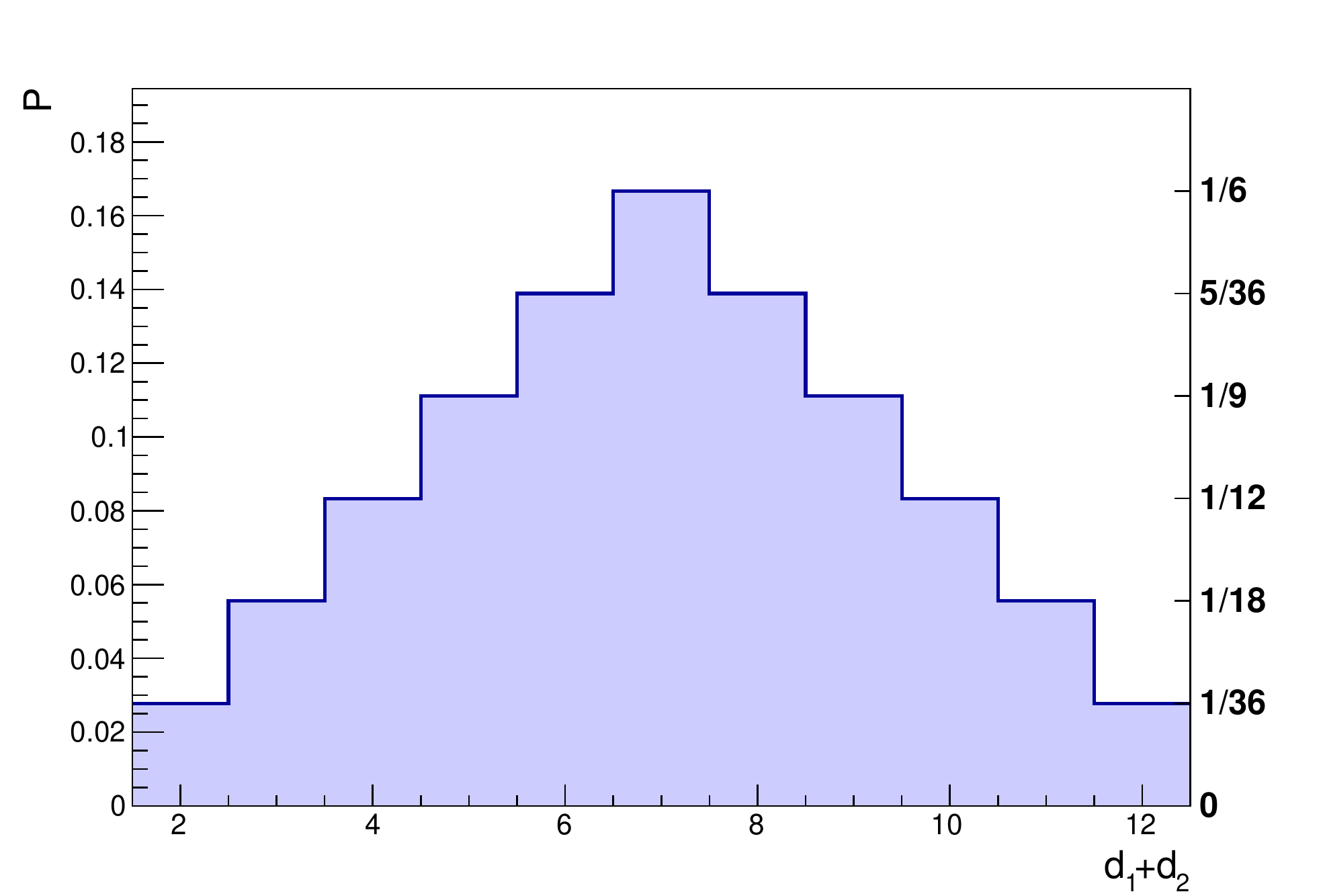}
\caption{Probability distribution of the sum of two dices computed by counting
  all possible elementary outcomes.}
\label{fig:twoDices}
\end{figure}

The simplest way to introduce probability is to consider
the symmetry properties of a random device. Example could be
a tossed coin (outcome may be {\it head} or {\it tail}) or a rolled dice
(outcome may be a number from 1 to 6 for a cubic dice, but dices
also exist with different shapes). According to the original definition
due to Laplace~\cite{Laplace}, we can be ``equally undecided''
about an event outcome due to symmetry properties, and we can assign an equal {\it probability}
to each of the outcomes. If we define an {\it event} from a statement about
the possible outcome of one random extraction (e.g.: {\it a dice roll gives
  an odd number}), the {\it probability} $P$ of that event (i.e.: that the statement is true)
can be defined as:
\begin{equation}
  P = \frac{\text{Number of favorable cases}}{\text{Total number of cases}}\,.
\end{equation}
Probabilities related to composite cases can be computed using {\it combinatorial analysis} by
reducing the composite event of interest into elementary equiprobable events.
The set of all possible elementary events is called {\it sample space} (see also Sec.~\ref{sec:kolmogorov}).
For instance, in Fig.~\ref{fig:twoDices}, the probability to obtain
a given sum of two dices is reported. 
The computation can be done by simply counting the number of elementary cases.
For instance 2 can be obtained only as $1+1$, while 3 can be obtained as
$1+2$ or $2+1$, 4 as $1+3$, $2+2$ or $3+1$, etc.

Textual statements about an event can be translated using set algebra considering
that and/or/not correspond to intersection/union/complement in set algebra.
For instance, the event \textit{``sum of two dices is even \textbf{and} greater than four''}
corresponds to the intersection of two sets:
\begin{equation}
  \{(d_1,d_2): \mathrm{mod}(d_1+d_2,2) = 0\} \cap \{(d_1,d_2): d_1+d_2 > 4\}\,.
\end{equation}

\subsubsection{``Events'' in statistics and in physics}
It's worth at this point to remarking the different meaning that
the word {\it event} usually assumes in statistics and in physics.
In statistics an event is a subset of the sample space.
E.g.: ``the sum of two dices is $\ge 5$''.
In particle physics usually an event is the result of a collision, as recorded by our experiment.
In several concrete cases, an event in statistics may
correspond to many possible collision events.
E.g.: ``$p_{\mathrm{T}}(\gamma) > 40~\mathrm{GeV}$'' may be an event in statistics,
but it may correspond to many events from a data sample that have at least one photon with
transverse momentum greater than 40~GeV.

\subsection{Frequentist probability}

The definition of {\it frequentis} probability
relates probability to the fraction of times an event occurrs, in the limit of very large number
($N\rightarrow\infty$) of repeated trials:
\begin{equation}
  P = \lim_{N\rightarrow\infty}
  \frac{\text{Number of favorable cases}
  }{N=\text{Number of trials}}\,.
  \label{eq:freqProb}
\end{equation}
This definition is exactly realizable only with an infinite number of trials,
which conceptually may be unpleasant. Anyway, physicists may consider this
definition pragmatically acceptable as approximately realizable in a
large, but not infinite, number of cases.
The definition in Eq.~(\ref{eq:freqProb}) is clearly only applicable to
repeatable experiments.

\subsection{Subjective (Bayesian) probability}

Subjective probability expresses one's {\it degree of belief} that a claim is true.
A probability equal to 1 expresses certainty that the claim is true, 0 expresses certainty that the claim
is false. Intermediate values form 0 to 1 quantify how strong the degree of belief
that the claims is true is.
This definition is applicable to all unknown events/claims,
not only repeatable experiments, as it is the case for the frequentist approach.
Each individual may have a different opinion/prejudice about one
claim, so this definition is necessarily {\it subjective}.
Anyway, quantitative rules exist about how
subjective probability should be modified
after learning about some observation/evidence.
Those rules descend from the Bayes theorem (see Sec.~\ref{sec:BayesTh}),
and this gives the name of {\it Bayesian} probability to subjective probability.
Starting from a {\it prior probability}, following some observation,
the probability can be modified into a  {\it posterior probability}.
The more information an individual receives, the more Bayesian probability
is insensitive on prior probability, with the exception of pathological cases
of prior probability. An example of such a case is a prior certainty that a claim is true ({\it dogma}) that
is then falsified by the observation.

\subsection{Komogorov axiomatic approach}
\label{sec:kolmogorov}
An axiomatic definition of probability is due to Kolmogorov~\cite{Kolmogorov},
which  can be applied both to frequentist and Bayesian probabilities.
The axioms assume that $\Omega$ is a sample space, $F$ is an event 
space made of subsets of $\Omega$ ($F\subseteq 2^\Omega$),
and $P$ is a {\it probability measure} that obeys the following three conditions:
\begin{enumerate}
\item $P(E)\ge 0\,,\,\,\, \forall E\in F$
\item $P(\Omega) =1$\,\,\, (normalization condition)
\item $\forall (E_1,\cdots,E_n) \in F^n: E_i\cup E_j = 0\,,\,\,\, P\left(\bigcup\limits_{i=1}^nE_i \right)= \sum\limits_{i=1}^nP(E_i)$
\end{enumerate}
The last condition states that the probability of the union of a set of disjoint events
is equal to the sum of their individual probabilities.

\subsection{Probability distributions}
\begin{figure}[htbp]
\centering\includegraphics[width=.495\linewidth]{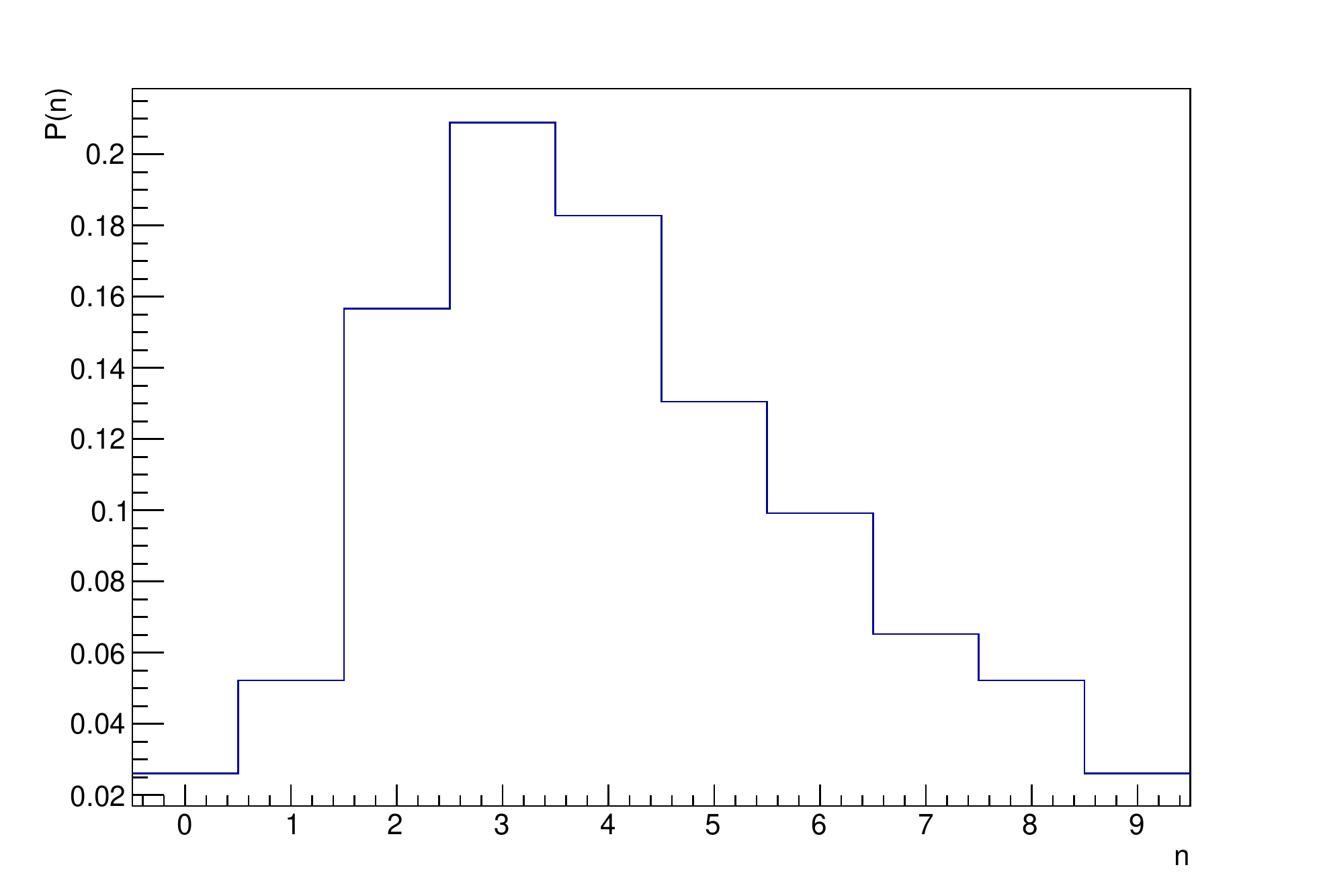}
\caption{Example of probability distribution of a discrete random variable $n$.}
\label{fig:pdfDiscr}
\end{figure}

Given a discrete {\it random variable} $n$, a probability can be assigned to each individual possible
value of $n$:
\begin{equation}
  P(n) = P(\{n\})\,.
\end{equation}
Figure~\ref{fig:pdfDiscr} shows an example of discrete probability distribution.
In case of a continuous variable, the probability 
assigned to an individual value may be zero
(e.g.: $P(\{x\}) = 0$), and a {\it probability density function} (PDF) better quantifies the 
probability content of an interval with finite measure: 
\begin{equation}
  \frac{\mathrm{d}P(x)}{\mathrm{d}x} = f(x)\,,
\end{equation}
and:
\begin{equation}
  P([x_1,x_2]) = \int_{x_1}^{x_2}f(x)\,\mathrm{d}x\,.
  \label{eq:probInt}
\end{equation}
\begin{figure}[htbp]
\centering\includegraphics[width=.495\linewidth]{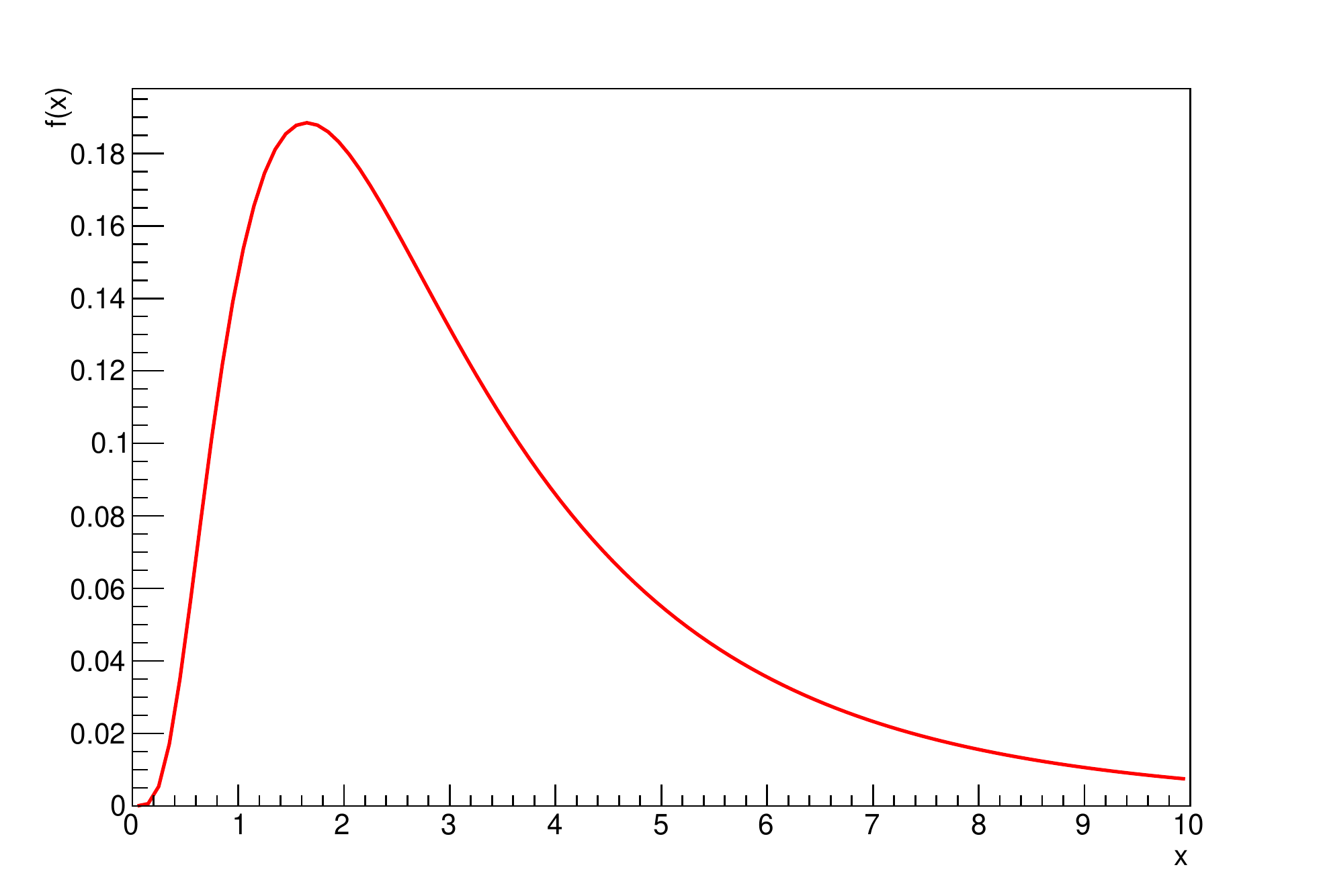}
\caption{Example of probability distribution of a continuous random variable $x$.}
\label{fig:pdfCont}
\end{figure}
Figure~\ref{fig:pdfCont} shows an example of such a continuous distribution.
Discrete and continuous distributions can be
combined using Dirac's delta functions.
For instance, the following PDF: 
\begin{equation}
  \frac{\mathrm{d}P(x)}{\mathrm{d}x} = \frac{1}{2}\delta(x) +\frac{1}{2} f(x)
\end{equation}
corresponds to a 50\% probability to have $x=0$ ($P(\{0\}) = 0.5$)
and $50\%$ probability to have a value $x\ne 0$ distributed according to $f(x)$. 

The {\it cumulative distribution} of a PDF $f$ is defined as:
\begin{equation}
  F(x) = \int_{-\infty}^x f(x)\,\mathrm{d}x\,.
  \label{eq:cumulative}
\end{equation}

\subsection{PDFs in more dimensions}

In more dimensions, corresponding to $n$ random variables,
a PDF can be defined as:
\begin{equation}
  \frac{\mathrm{d}^nP(x_1, \cdots,x_n)}{\mathrm{d}x_1\cdots\mathrm{d}x_n}
  = f(x_1,\cdots,x_n)\,.
\end{equation}
The probability associated to an event which corresponds to a subset
$E\subseteq \mathbbm{R}^n$
is obtained by integrating the PDF over the set $E$, naturally extending Eq.~(\ref{eq:probInt}):
\begin{equation}
  P(E) = \int_Ef(x_1,\cdots,x_n)\,\mathrm{d}^n x\,.
\end{equation}

\subsection{Mean, variance and covariance}
For a PDF that models a random variable $x$, it's useful to define a number of quantities:
\begin{itemize}
\item  The {\it mean} or {\it expected value} of $x$ is defined as:
  \begin{equation}
    \mathbbm{E}[x] = \left<x\right> = \int x\,f(x)\,\mathrm{d}x\,.
  \end{equation}
 More in general, the mean or expected value of $g(x)$ is:
  \begin{equation}
    \mathbbm{E}[g(x)] = \left<g(x)\right> = \int g(x)\,f(x)\,\mathrm{d}x\,.
  \end{equation}
\item The {\it variance} of $x$ is defined as:
    \begin{equation}
      \mathbbm{V}\mathrm{ar}[x] = \left<(x-\left<x\right>)^2\right> = \left<x^2\right>-\left<x\right>^2.
    \end{equation}
    The term $\left<x^2\right>$ is called {\it root mean square}, or {\it r.m.s.}.
  \item The {\it standard deviation} of $x$ is the square root of the variance:
    \begin{equation}
      \sigma_x = \sqrt{\mathbbm{V}\mathrm{ar}[x]} = \sqrt{\left<(x-\left<x\right>)^2\right>}\,.
    \end{equation}
\end{itemize}
Given two random variables $x$ and $y$, the following quantities may be defined:
\begin{itemize}
\item The {\it covariance} of $x$ and $y$ is:
  \begin{equation}
    \mathbbm{C}\mathrm{ov}[x, y] = \left<(x-\left<x\right>)(y-\left<y\right>)\right>\,.
  \end{equation}
\item The {\it correlation coefficient} is:
  \begin{equation}
    \rho_{xy} = \frac{\mathbbm{C}\mathrm{ov}[x, y]}{\sigma_x\sigma_y}\,.
  \end{equation}
\end{itemize}
Two variables with null covariance are said to be {\it uncorrelated}.

\subsection{Commonly used distributions}

Below a few examples of probability distributions are reported 
that are frequently used in physics and more in general in statistical
applications.

\subsubsection{Gaussian distribution}
\begin{figure}[htbp]
\centering\includegraphics[width=.495\linewidth]{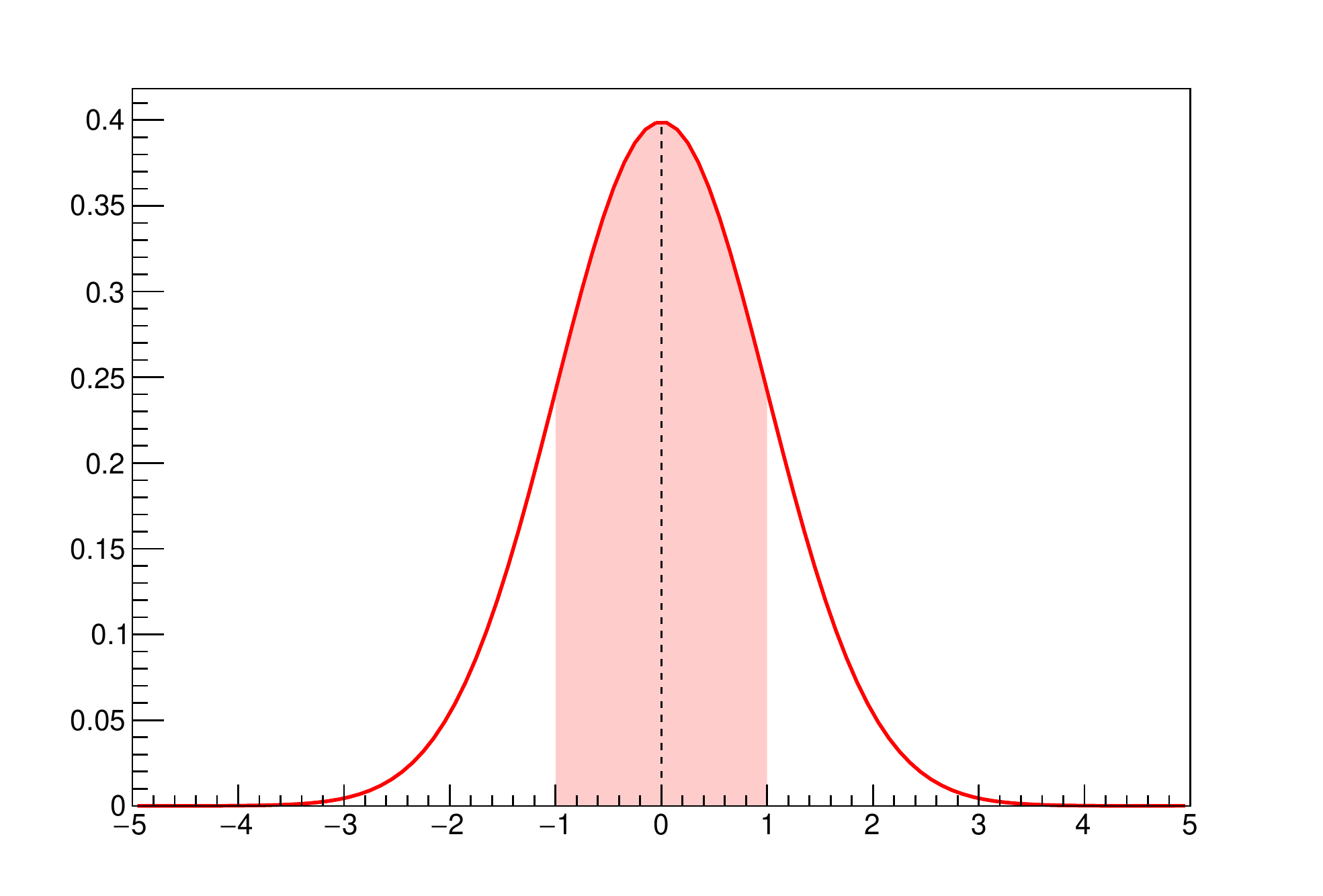}
\caption{Example of Gaussian distribution with $\mu=0$ and $\sigma=1$.
  The shaded interval $[\mu-\sigma, \mu+\sigma]$ correspond to a probability of approximately 0.683.}
\label{fig:gaussianPdf}
\end{figure}

A {\it Gaussian} or {\it normal} distribution is given by:
\begin{equation}
  g(x;\mu,\sigma) = \frac{1}{\sigma\sqrt{2\pi}}e^{-{(x-\mu)^2}/{2\sigma^2}}\,,
  \label{eq:GaussianPDF}
\end{equation}
where $\mu$ and $\sigma$ are parameters equal to the average value and standard deviation of $x$, respectively.
If $\mu=0$ and $\sigma=1$, a Gaussian distribution is also called {\it standard normal distribution}.
An example of
Gaussian PDF is shown in Fig.~\ref{fig:gaussianPdf}.
Probability values corresponding to intervals $[\mu-n\sigma, \mu+n\sigma]$ for a Gaussian
distribution are frequently used as reference, and are reported in Tab.~\ref{tab:GaussianInt}.
\begin{table}
\caption{Probabilities for a Gaussian PDF corresponding to an interval $[\mu-n\sigma, \mu+n\sigma]$.}
\label{tab:GaussianInt}
\centering
\begin{tabular}{cc}\hline\hline
$n$ & Prob.\\\hline
  1 & 0.683 \\
  2 & 0.954 \\
  3 & 0.997 \\
  4 & $1-6.5\times10^{-5}$ \\
  5 & $1-5.7\times10^{-7}$ \\\hline\hline
\end{tabular}
\end{table}
Many random variables in real experiments follow, at least approximately, a Gaussian distribution.
This is mainly due to the {\it central limit theorem} that allows to
approximate the sum of multiple random variables, regardless of their individual distributions,
with a Gaussian distribution.
Gaussian PDFs are frequently used to model detector resolution.

\subsubsection{Poissonian distribution}
\label{sec:Poissonian}
A {\it Poissonian} distribution for an integer non-negative random variable $n$ is:
\begin{equation}
  P(n;\nu) = \frac{\nu^n}{n!}e^{-\nu}\,,
\end{equation}
where $\nu$ is a parameter equal to the average value of $n$.
The variance of $n$ is also equal to $\nu$.

\begin{figure}[htbp]
\centering\includegraphics[width=.65\linewidth]{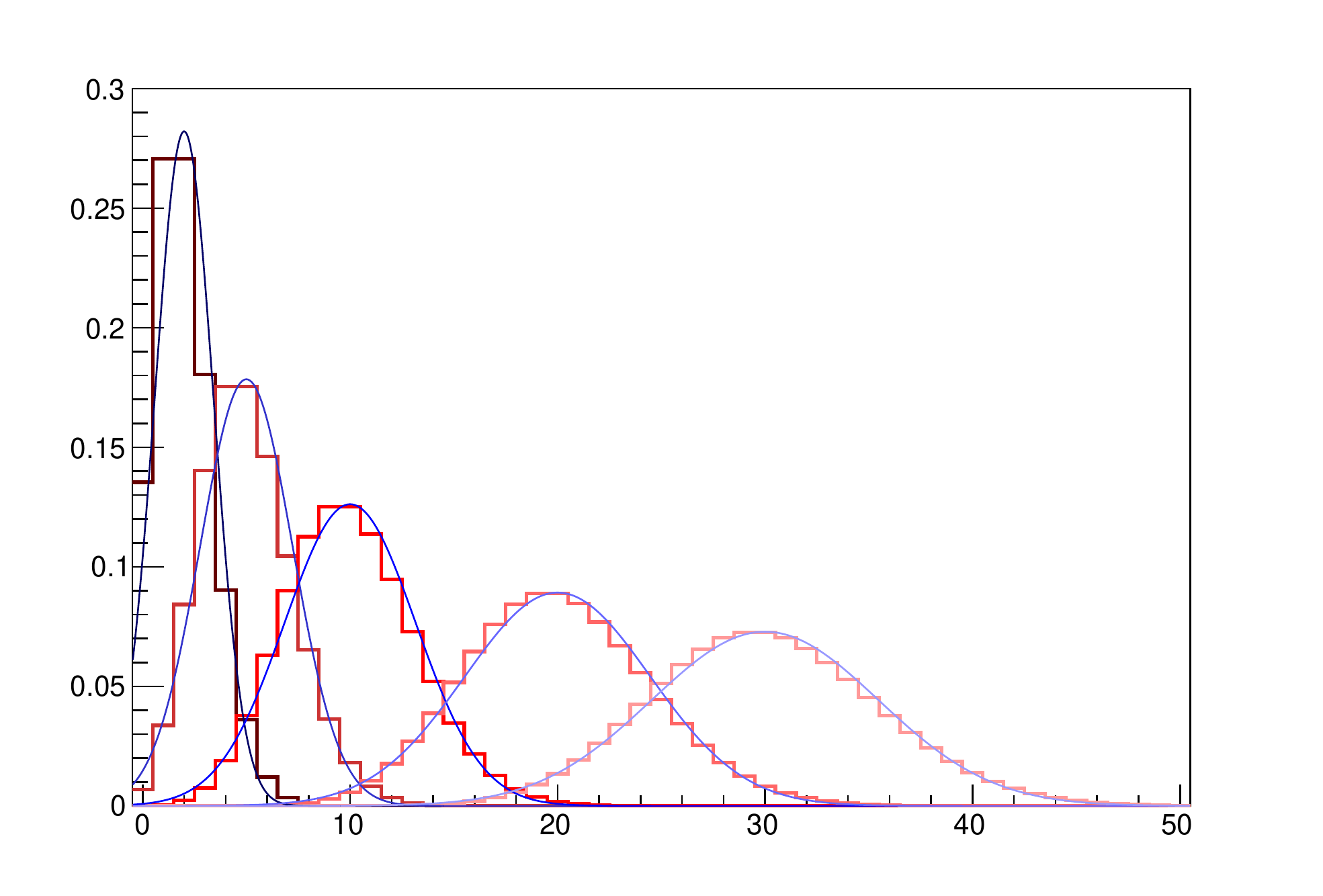}
\caption{Example of Poisson distributions with different values of $\nu$.
  The continuous superimposed curves are Gaussian distributions with
  $\mu=\nu$ and $\sigma=\sqrt{\nu}$.}
\label{fig:poissonianPdf}
\end{figure}

Poissonian distributions model the number of occurrences of random event uniformly distributed
in a measurement range whose rate is known. Examples are the number of rain drops falling in a given area
and in a given time interval or the number of cosmic rays crossing a detector in a given time 
interval. 
Poissonian distributions may be approximated with a Gaussian distribution
having $\mu=\nu$ and $\sigma=\sqrt{\nu}$ for sufficiently large values of $\nu$.
Examples of Poissonian distributions are shown in Fig.~\ref{fig:poissonianPdf} with
superimposed Gaussian distributions as comparison.

\subsubsection{Binomial distribution}

A {\it binomial} distribution gives the probability to achieve $n$ successful outcomes
on a total of $N$ independent trials whose individual probability of success is $p$.
The binomial probability is given by:
\begin{equation}
  P(n;N,p) = \frac{N!}{n!(N-n)!}p^n(1-p)^{N-p}\,.
\end{equation}
The average value of $n$ for a binomial variable is:
\begin{equation}
\left<n\right>=  N\,p
\end{equation}
and the variance is:
\begin{equation}
  \mathbbm{V}\mathrm{ar}[n] =  N\,p\,(1-p)\,.
  \label{eq:binomVar}
\end{equation}
A typical example of binomial process in physics is the case of a detector
with efficiency $p$, where $n$ is the number of {\it detected} particles
over a total number of particles $N$ that {\it crossed} the detector.

\subsection{Conditional probability}

The probability of an event $A$, {\it given} the event $B$
is defined as:
\begin{equation}
  P(A|B) = \frac{P(A\cap B)}{P(B)}\,,
  \label{eq:condProb}
\end{equation}
and represents the probability that an event known to belong to set $B$  also belongs to set $A$.
It's worth noting that, given the sample space $\Omega$ with $P(\Omega)=1$:
\begin{equation}
  P(A|\Omega) = \frac{P(A\cap \Omega)}{P(\Omega)}\,,
\end{equation}
consistently with Eq.~(\ref{eq:condProb}).

An event $A$ is said to be independent on the event $B$ if the 
probability of $A$ given $B$ is equal to the probability of $A$:
\begin{equation}
  P(A|B) = P(A)\,.
\end{equation}
If an event $A$ is independent on the event $B$, then $P(A \cap B) = P(A) P(B)$.
Using Eq.~(\ref{eq:condProb}), it's immediate to demonstrate that
if $A$ is independent on $B$, then $B$ is independent on $A$.

The application of the concept of conditional probability to PDFs in more dimensions
allows to introduce the concept of {\it independent variables}.
Consider a two-variable PDF $f(x,y)$ (but the result can be
easily generalized to more than two variables), two {\it marginal distributions}
can be defined as:
\begin{eqnarray}
  f_x(x) & = & \int f(x,y)\,\mathrm{d}y\,,\\
  f_y(x) & = & \int f(x,y)\,\mathrm{d}x\,.
\end{eqnarray}
If we consider the sets:
\begin{eqnarray}
  A & = & \{ x^\prime: x < x^\prime < x + \delta x\}\,,\\
  B & = & \{ y^\prime: y < y^\prime < y + \delta y\}\,,
\end{eqnarray}
where $\delta x$ and $\delta y$ are very small, if $A$ and $B$ are independent, we have:
\begin{equation}
  P(A\cap B) = P(A)P(B)\,,
\end{equation}
which implies:
\begin{equation}
  f(x,y) = f_x(x)f_y(x)\,.
  \label{eq:indVar}
\end{equation}
From Eq.~(\ref{eq:indVar}), it's possible to define that $x$ and $y$ are independent variables
if and only if their PDF can be factorized into the product of one-dimensional PDFs.
Note that if two variables are uncorrelated they are not necessarily independent.

\subsection{Bayes theorem}
\label{sec:BayesTh}
Considering two events $A$ and $B$, using Eq.~(\ref{eq:condProb}) twice, we can write:
\begin{eqnarray}
  P(A|B) & = & \frac{P(A\cap B)}{P(B)} \,,\\
  P(B|A) & = & \frac{P(A\cap B)}{P(A)} \,,
\end{eqnarray}
from which the following equation derives:
\begin{equation}
  P(A|B)P(B) = P(B|A)P(A)\,.
  \label{eq:bayesInterm}
\end{equation}
Eq.~(\ref{eq:bayesInterm}) can be written in the following form, that
takes the name of {\it Bayes theorem}:
\begin{equation}
  \boxed{
    P(A|B) = \frac{P(B|A)P(A)}{P(B)}\,.
  }
  \label{eq:BayesTheorem}
\end{equation}
In Eq.~(\ref{eq:BayesTheorem}), $P(A)$ has the role of {\it prior} probability and
$P(A|B)$ has the role of {\it posterior} probability.
Bayes theorem, that has its validity in any probability approach, including the frequentist one,
can also be used to assign a posterior probability to a claim $H$ that is necessarily not a random
event, given a corresponding prior probability $P(H)$ and the observation of an event $E$ whose
probability, if $H$ is true, is given by $P(E|H)$:
\begin{equation}
  P(H|E) = \frac{P(E|H)P(H)}{P(E)}\,.
  \label{eq:BayesRule}
\end{equation}
Eq.~(\ref{eq:BayesRule}) is the basis of Bayesian approach to probability. It
defines in a {\it rational way} a role to modify one's prior
belief in a claim $H$ given the observation of $E$.

The following problem is an example of application of
Bayes theorem in a frequentist environment.
Imagine you have a particle identification detector that identifies muons with high efficiency,
say $\varepsilon=95\%$. A small fraction of pions, say $\delta=5\%$, are incorrectly identified as muons ({\it fakes}).
Given a particle in a data sample that is identified as a muon, what is the probability that
it is really a muon? The answer to this question can't be given unless we know more information
about the composition of the sample, i.e.: what is the fraction of muons and pions in the data sample.

Using Bayes theorem, we can write:
\begin{equation}
  P(\mu|+) = \frac{P(+|\mu)P(\mu)}{P(+)}\,,
\label{eq:pMuPlusInt}
\end{equation}
where `$+$' denotes a positive muon identification, $P(\mu|+)=\varepsilon$ is the probability to positively
identify a muon, $P(\mu)$ is the fraction of muons in our sample ({\it purity}) and
$P(+)$ is the probability to positively identify a particle randomly chosen from our sample.

It's possible to decompose $P(+)$ as:
\begin{equation}
  P(+) = P(+|\mu) P(\mu) + P(+|\pi) P(\pi)\,,
  \label{eq:pPlusInt}
\end{equation}
where $P(+|\pi)=\delta$ is the probability to positively identify a pion and $P(\pi)=1-P(\mu)$ is the
fraction of pions in our samples, that we suppose is only made of muons and pions.
Eq.~(\ref{eq:pPlusInt}) is a particular case of the {\it law of total probability} which
allows to decompose the probability of an event $E_0$ as:
\begin{equation}
  P(E_0) = \sum_{i=1}^n P(E_0|A_i) P(A_i)\,,
\end{equation}
where the sets $A_i$ are all pairwise disjoint and constitute a partition of the sample space.

Using the decomposition from Eq.~(\ref{eq:pPlusInt}) in
Eq.~(\ref{eq:pMuPlusInt}), one gets:
\begin{equation}
  P(\mu|+) = \frac{\varepsilon P(\mu)}{\varepsilon P(\mu) + \delta P(\pi)}\,.
\end{equation}

If we assume that our sample contains a fraction $P(\mu)=4\%$ of muons and $P(\pi)=96\%$ of
pions, we have:
\begin{equation}
  P(\mu|+) = \frac{0.95 \cdot 0.04}{0.95 \cdot 0.04 + 0.05\cdot 0.96}\simeq 0.44\,.
\end{equation}
In this case, even if the selection efficiency is very high, given the low sample purity,
a particle positively identified as a muon has a probability less than 50\% to be really a muon.

\subsection{The likelihood function}
\label{sec:likeFun}

The outcome of on experiment can be modeled as a set of random variables $x_1, \cdots, x_n$
whose distribution takes into account both intrinsic physics randomness (theory)
and  detector effects (like resolution, efficiency, etc.).
Theory and detector effects can be described according to some parameters
$\theta_1, \cdots, \theta_m$ whose values are, in most of the cases, unknown.
The overall PDF, evaluated for our observations $x_1, \cdots, x_n$, is called {\it likelihood function}:
\begin{equation}
  L=f(x_1,\cdots,x_n;\theta_1,\cdots,\theta_m)\,.
\end{equation}
In case our sample consists of $N$ {\it independent measurements}, typically each corresponding to a
collision event, the likelihood function can be written as:
\begin{equation}
  L=\prod_{i=1}^Nf(x_1^i,\cdots,x_n^i;\theta_1,\cdots,\theta_m)\,.
  \label{eq:unbinnedLikeFun}
\end{equation}

The likelihood function provides a useful implementation of Bayes rule
(Eq.~(\ref{eq:BayesRule})) in the case of a measurement constituted by the observation of 
continuous random variables $x_1, \cdots, x_n$. The posterior PDF of the unknown parameters
$\theta_1, \cdots, \theta_m$ can be determined as:
\begin{equation}
  P(\theta_1,\cdots,\theta_m|x_1,\cdots,x_n) =
  \frac{
    L(x_1,\cdots,x_n;\theta_,\cdots,\theta_m)\pi(\theta_1,\cdots,\theta_m)
  }{
 \int   L(x_1,\cdots,x_n;\theta_,\cdots,\theta_m)\pi(\theta_1,\cdots,\theta_m)\,\mathrm{d}\theta^m\
    }\,,
\label{eq:likeBayes}
\end{equation}
where $\pi(\theta_1, \cdots, \theta_m)$ is the subjective prior probability and
the denominator is a normalization factor obtained with a decomposition similar to Eq.~(\ref{eq:pPlusInt}).
Equation~(\ref{eq:likeBayes}) can be interpreted as follows:
the observation of $x_1, \cdots, x_n$ modifies the prior knowledge of the unknown parameters $\theta_1,\cdots, \theta_m$.

If $\pi(\theta_1, \cdots, \theta_m)$ is sufficiently smooth and $L$ is sharply peaked around the true values of
the parameters $\theta_1,\cdots, \theta_m$, the resulting posterior will not be strongly dependent on the prior's choice.

Bayes theorem in the form of Eq.~(\ref{eq:likeBayes}) can be 
applied sequentially for repeated independent observations.
In fact, if we start with a prior $P_0(\vec{\theta})$, we can
determine a posterior:
\begin{equation}
  P_1(\vec{\theta})\propto P_0(\vec{\theta})\cdot L_1(\vec{x}_1;\vec{\theta})\,,
\end{equation}
where $L_1(\vec{x_1};\vec{\theta})$ is the likelihood function corresponding to the observation
$\vec{x}_1$. Subsequently, we can use $P_1$ as new prior for a second observation $\vec{x}_2$, and
we can determine a new posterior:
\begin{equation}
  P_2(\vec{\theta})\propto P_1(\vec{\theta})\cdot L_2(\vec{x}_2;\vec{\theta})\,,
\end{equation}
and so on:
\begin{equation}
  P_3(\vec{\theta})\propto P_2(\vec{\theta})\cdot L_3(\vec{x}_3;\vec{\theta})\,.\
  \label{eq:postP3L3Int}
\end{equation}
For independent observations $\vec{x}_1$, $\vec{x}_2$, $\vec{x}_2$, the combined likelihood
function can be written as the product of individual likelihood functions (Eq.~(\ref{eq:indVar})):
\begin{equation}
  P_3(\vec{\theta})\propto P_0(\vec{\theta})\cdot
  L_1(\vec{x}_1;\vec{\theta})\cdot
  L_2(\vec{x}_2;\vec{\theta})\cdot
  L_3(\vec{x}_3;\vec{\theta})\,,
\end{equation}
consistently with Eq.~(\ref{eq:postP3L3Int}).
This allows to use consistently the repeated application of Bayes rule as sequential
improvement of knowledge from subsequent observations.

\section{Inference}
In Sec.~\ref{sec:probability} we presented how probability theory can model the
fluctuation in data due to intrinsic randomness of observable data samples.
Taking into account the distribution of data as a function of the values of
unknown parameters, we can exploit the observed data in order to determine
information about the parameters, in particular to measure their value ({\it central value}) within
some {\it uncertainty}. This process is called {\it inference}.

\subsection{Bayesian inference}
\label{sec:BayesianInference}

\begin{figure}[htbp]
\centering\includegraphics[width=.495\linewidth]{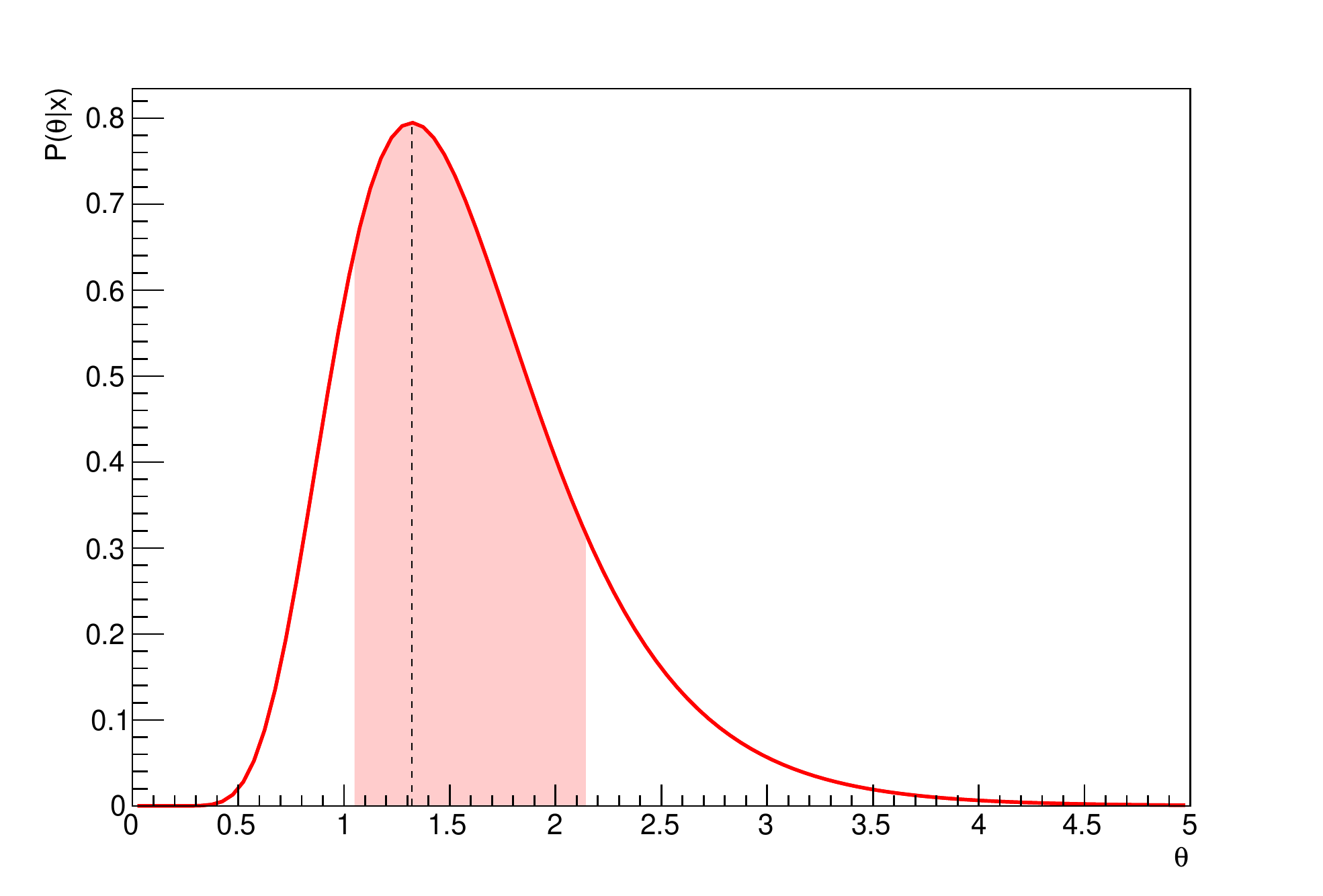}
\centering\includegraphics[width=.495\linewidth]{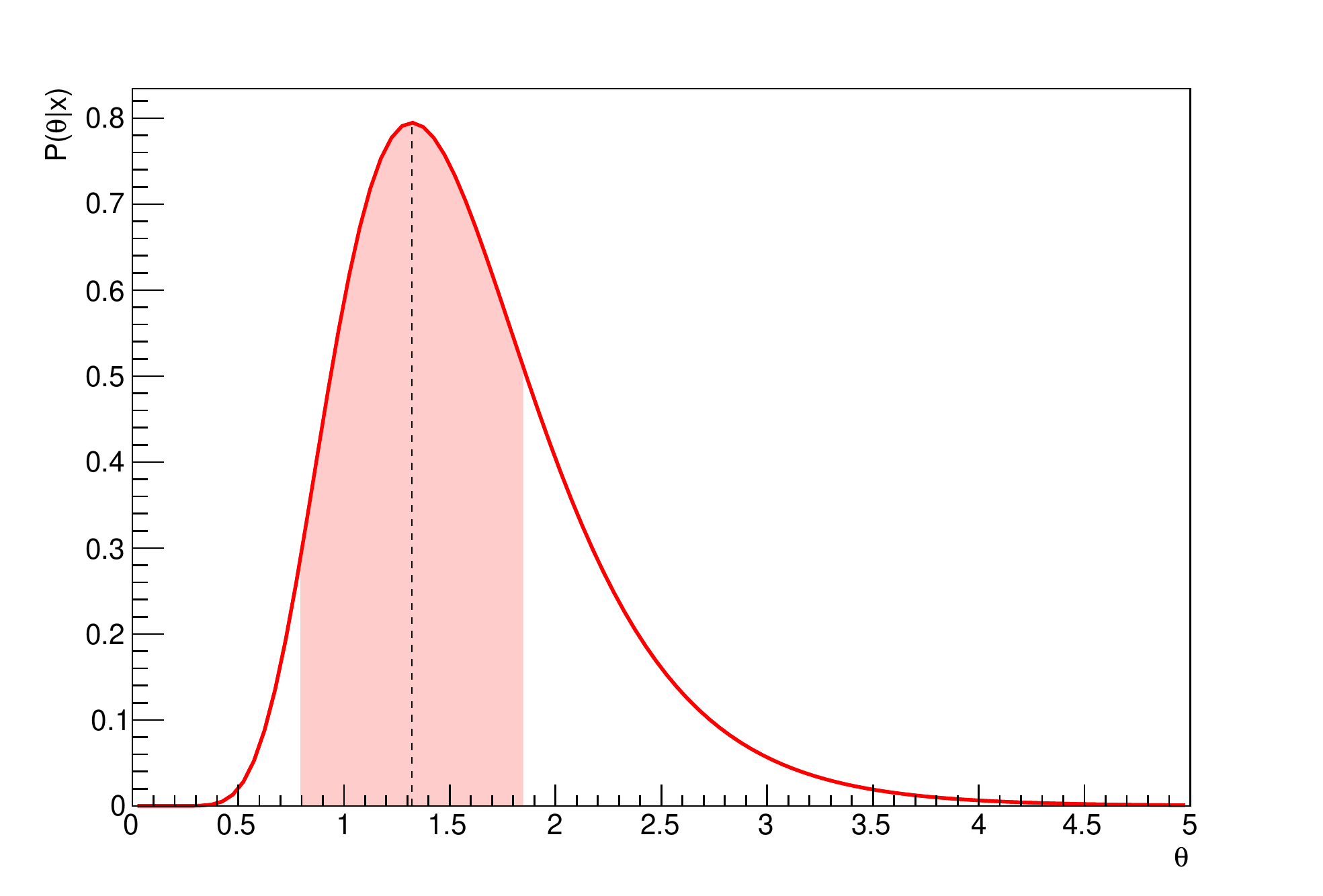}
\caption{Example of a posterior PDF of the parameter $\theta$ with two possible choices of
  a 68.3\% probability interval shown as shaded area: a central interval (left plot)
  and a symmetric interval (right plot).
  The dotted vertical line shows the most probable value (mode).
}
\label{fig:BayesIntCentralSym}
\end{figure}

One example of inference is the use of Bayes theorem to determine the posterior PDF
of an  unknown parameter $\theta$ given an observation $x$:
\begin{equation}
  P(\theta|x) = \frac{L(x;\theta)\pi(\theta)}{
    \int L(x;\theta)\pi(\theta)\,\mathrm{d}\theta}\,,
  \label{eq:BayesianInferenceSimple}
\end{equation}
where $\pi(\theta)$ is the prior PDF.
The posterior $P(\theta|x)$ contains all the information we can obtain from $x$
about $\theta$.
One example of possible outcome for $P(\theta|x)$ is shown in Fig.~\ref{fig:BayesIntCentralSym}
with two possible choices of uncertainty interval  (left and right plots).
The most probable value, $\hat{\theta}$, also called {\it mode}, shown as dashed line
in both plots, can be taken as central value
for the parameter $\theta$.
It's worth noting that if $\pi(\theta)$ is assumed to be a constant,
$\hat{\theta}$ corresponds to the maximum of 
the likelihood function ({\it maximum likelihood estimate}, see Sec.~\ref{sec:maxLik}).
Different choices of 68.3\% probability interval, or uncertainty interval,
can be taken. A central interval $[\theta_1,\theta_2]$,
represented in the left plot in Fig.~\ref{fig:BayesIntCentralSym}
as shaded area, is obtained in order to have equal areas under the two extreme tails:
\begin{eqnarray}
  \int_{-\infty}^{\theta_1} P(\theta|x)\,\mathrm{d}\theta & = & \frac{\alpha}{2}\,,\label{eq:BayesCentralLeft} \\
  \int^{+\infty}_{\theta_2} P(\theta|x)\,\mathrm{d}\theta & = & \frac{\alpha}{2}\,,\label{eq:BayesCentralRight}
\end{eqnarray}
where $\alpha = 1 - 68.3\%$.
Another example of a possible coice of 68.3\% interval is shown in the right plot,
where a symmetric interval is taken, corresponding to:
\begin{eqnarray}
  \int_{\hat{\theta}-\delta}^{\hat{\theta}+\delta} P(\theta|x)\,\mathrm{d}\theta & = & {1-\alpha}\,.\\
\end{eqnarray}

\begin{figure}[htbp]
\centering\includegraphics[width=.495\linewidth]{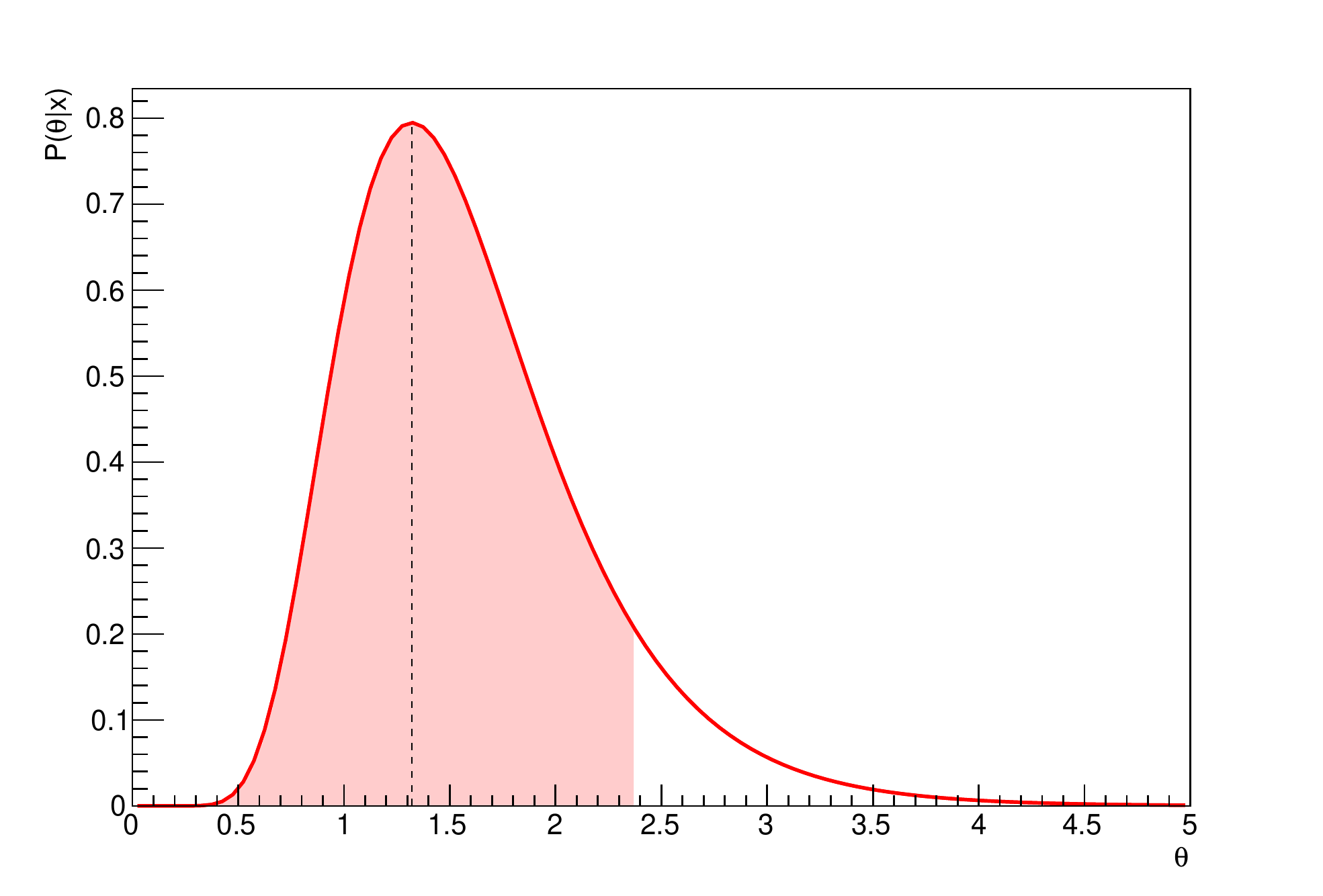}
\centering\includegraphics[width=.495\linewidth]{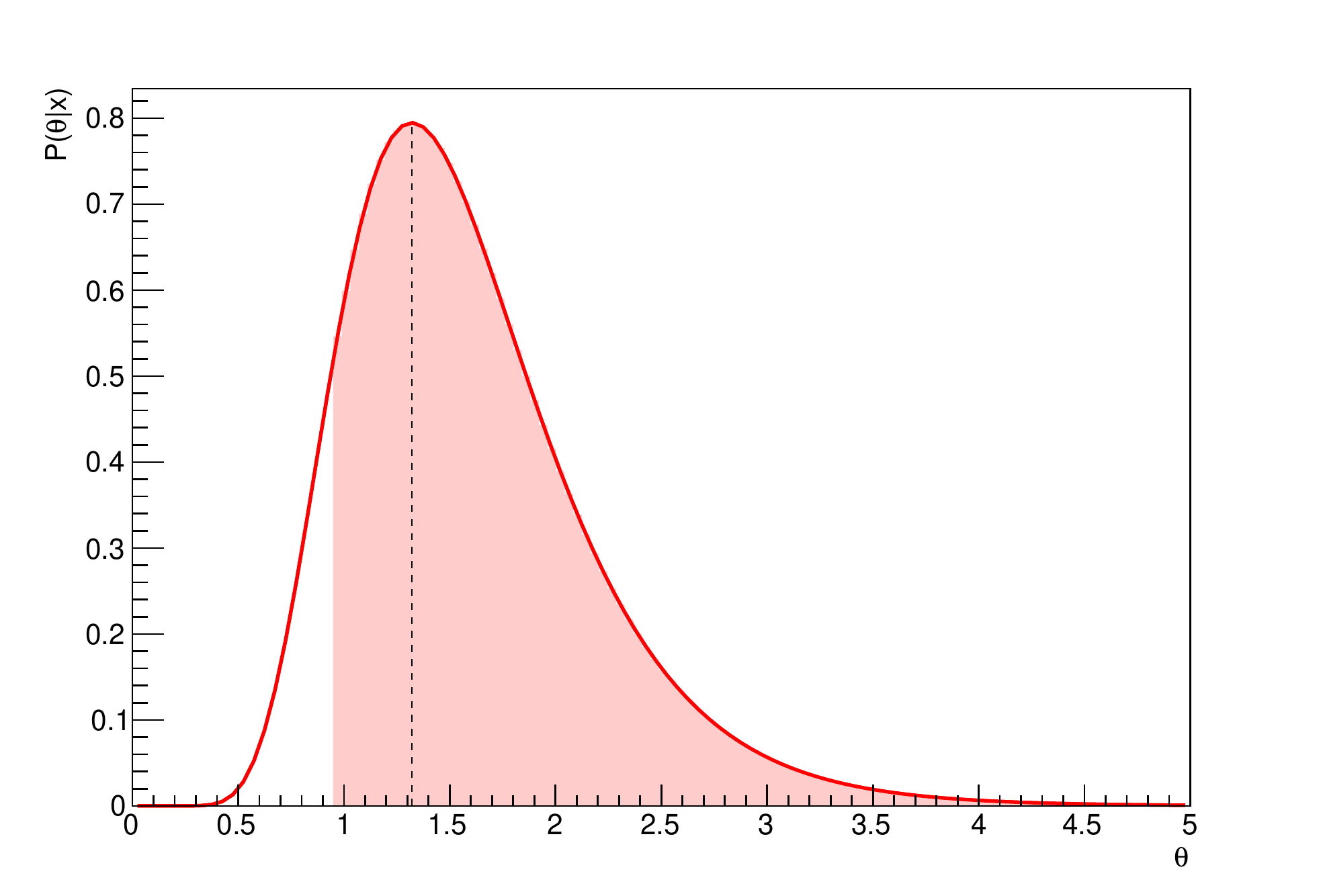}
\caption{Extreme choices of 90\% probability interval leading to an upper limit (left) and
  a lower limit (up) to the parameter $\theta$.
}
\label{fig:BayesIntHiLo}
\end{figure}

Two extreme choices of fully asymmetric probability intervals are shown in Fig.~\ref{fig:BayesIntHiLo},
leading to an upper (left) or lower (right) limit to the parameter $\theta$.
For upper or lower limits, usually a 90\% or 95\% probability interval is chosen instead
of the usual 68.3\% used for central or symmetric intervals.
The intervals in Fig.~\ref{fig:BayesIntHiLo} are chosen such that:
\begin{eqnarray}
  \int_{-\infty}^{\theta^{\mathrm{up}}} P(\theta|x)\,\mathrm{d}\theta & = & {1-\alpha}\quad \text{(left plot)}\,,\\
  \int^{+\infty}_{\theta^{\mathrm{lo}}} P(\theta|x)\,\mathrm{d}\theta & = & {1-\alpha}\quad \text{(right plot)}\,,\\
\end{eqnarray}
where in this case $\alpha = 0.1$.

\subsubsection{Example of Bayiesian inference: Poissonian counting}
\label{sec:BayesianPoissonianCounting}

In a counting experiment, i.e.: the only information relevant to measure the yield of our signal is the
number of events $n$ that pass a given selection, a Poissonian can be used
to model the distribution of $n$ with an expected number of events $s$:
\begin{equation}
  P(n;s) = \frac{s^n e^{-s}}{n!}\,.
  \label{eq:PoissonBayesInference}
\end{equation}
If a particular value of $n$ is measured, the posterior PDF of $s$ is (Eq.~(\ref{eq:BayesianInferenceSimple})):
\begin{equation}
  P(s|n) = \frac{\displaystyle
    \frac{s^n e^{-s}}{n!} \pi(s)
  }{\displaystyle
   \int_0^{\infty} \frac{s^{\prime n} e^{-s^{\prime}}}{n!} \pi(s^\prime)\,\mathrm{d}s^\prime
    }\,,
\end{equation}
where $\pi(s)$ is the assumed prior for $s$. If we take $\pi(s)$ to be uniform, performing the
integration gives a denominator equal to one, hence:
\begin{equation}
  P(s|n) =\frac{s^n e^{-s}}{n!}\,.
  \label{eq:PoissonBayesInferencePost}
\end{equation}
Note that though Eqs.~(\ref{eq:PoissonBayesInference}) and~(\ref{eq:PoissonBayesInferencePost}) lead to the
same expression, the former is a probability for the discrete random variable $n$, the latter is a
posterior PDF of the unknown parameter $s$.
From Eq.~(\ref{eq:PoissonBayesInference}), the mode $\hat{s}$ is equal to $n$, but
$\left<s\right> = n+1$, due to the asymmetric distribution of $s$, and
$\mathbbm{V}\mathrm{ar}[s] = n+1$, while the variance of $n$ for a Poissonian
distribution is $\sqrt{s}$ (Sec.~\ref{sec:Poissonian}).

\begin{figure}[htbp]
\centering\includegraphics[width=.495\linewidth]{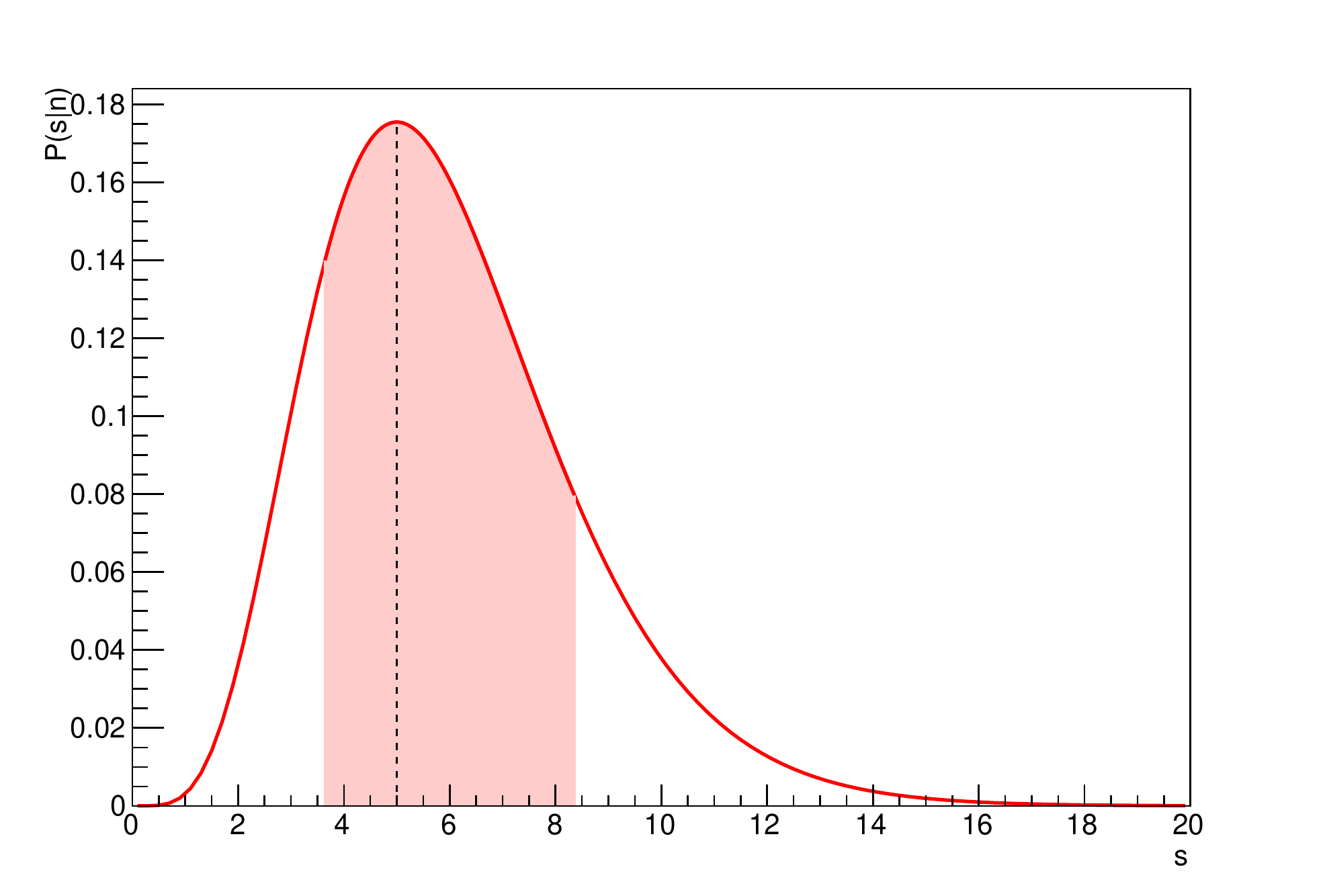}
\centering\includegraphics[width=.495\linewidth]{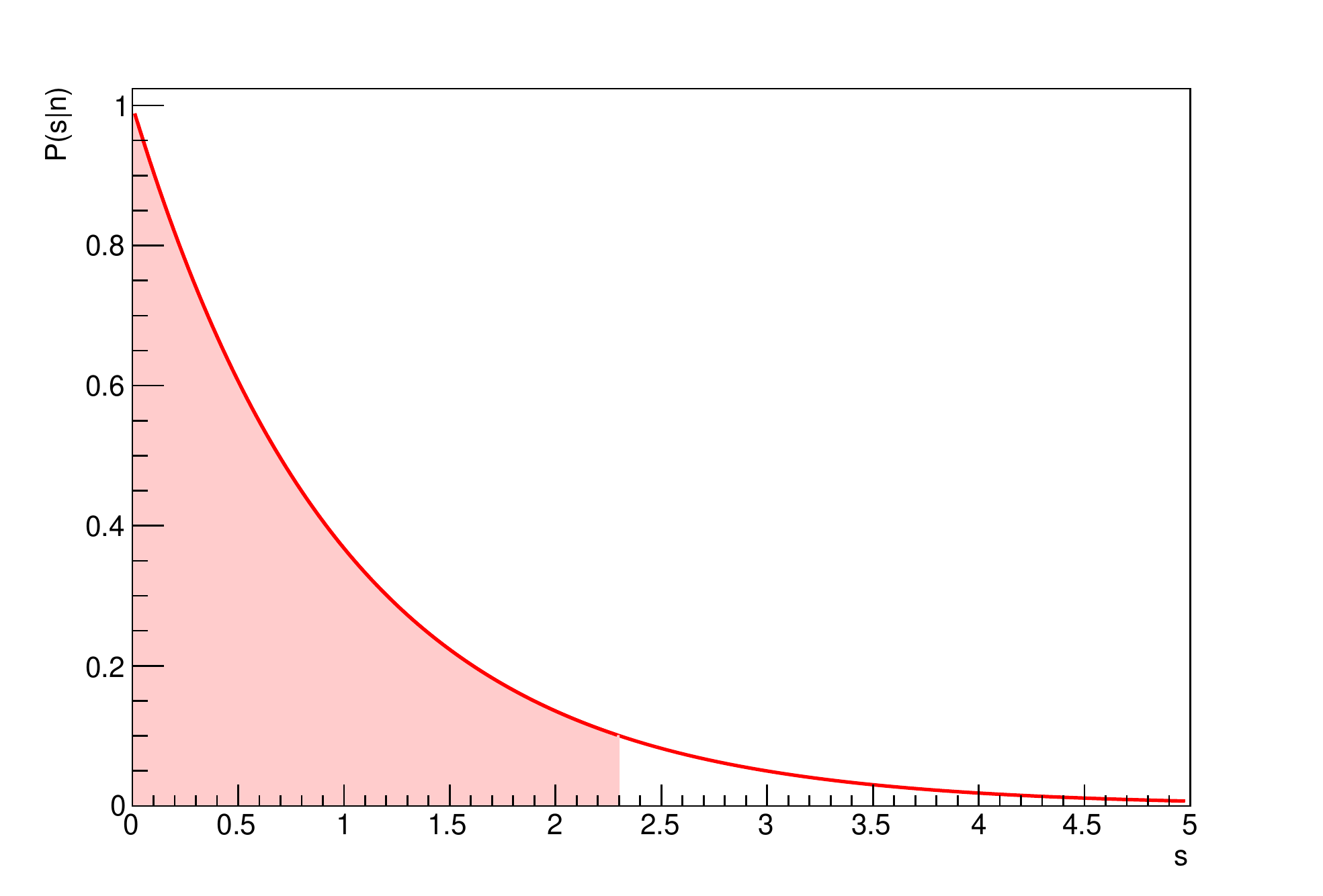}
\caption{Posterior PDF of a Poissonian parameter $s$ for observed number of events $n=5$ (left)
  and for $n=0$ (right). In the left plot, a central 68.3\% probability interval is chosen, while for
  the right plot a fully asymmetric 90\% probability interval, leading to an upper limit, is chosen.}
\label{fig:PoissonBayesPost}
\end{figure}
Figure~\ref{fig:PoissonBayesPost} shows two cases of posterior PDF of $s$, for the cases $n=5$ (left)
and for $n=0$ (right). In the case $n=5$, a central value $\hat{s}=5$ can be taken as most probable value.
In that plot, a central interval was chosen (Eq.~(\ref{eq:BayesCentralLeft},~\ref{eq:BayesCentralRight})).
For the case $n=0$, the most probable value of $s$ is $\hat{s}=0$. A fully asymmetric interval corresponding
to a probability $1-\alpha$ leads to an upper limit:
\begin{equation}
  e^{-s^{\mathrm{up}}} = \alpha\,,
\end{equation}
which then leads to:
\begin{eqnarray}
  s & < & s^{\mathrm{up}} = 2.303\quad\text{for}\quad\alpha=0.1\,\text{(90\% probability)}\,,\label{eq:BayesianPoissonUL90}\\
  s & < & s^{\mathrm{up}} = 2.996\quad\text{for}\quad\alpha=0.05\,\text{(95\% probability)}\,.\label{eq:BayesianPoissonUL95}
\end{eqnarray}

\subsection{Error propagation with Bayiesian inference}
\label{sec:errPropBayes}

Error propagation is needed when applying a parameter transformation, say $\eta = H(\theta)$.
A central value and uncertainty interval need to be determined for the transformed parameter $\eta$.
With Bayesian inference, a posterior PDF of $\theta$, $f(\theta)$ is available, and 
the error propagation can be done transforming the posterior PDF of $\theta$ into a PDF of $\eta$, $f^\prime(\eta)$:
the central value and uncertainty interval for $\eta$ can be computed from $f^\prime$.
In general, the PDF of the transformed variable $\eta$, given the PDF of $\theta$, is given by:
\begin{equation}
  f^\prime(\eta) = \int \delta(\eta-H(\theta))f(\theta)\,\mathrm{d}\theta\,.
  \label{eq:PDFtransform}
\end{equation}
Transformations for cases with more than one variable proceed in a similar way. If we
have two parameters $\theta_1$ and $\theta_2$, and a transformed variable $\eta=H(\theta_1,\theta_2)$,
then the PDF of $\eta$, similarly to Eq.~(\ref{eq:PDFtransform}), is given by:
\begin{equation}
  f^\prime(\eta) =\int \delta(\eta-H(\theta_1,\theta_2))f(\theta_1,\theta_2)\,\mathrm{d}\theta_1\,\mathrm{d}\theta_2\,.
\end{equation}
In case of a transformation from two parameters $\theta_1$ and $\theta_2$ in two other
parameters $\eta_1$ and $\eta_2$: $\eta_1=H_2(\theta_1,\theta_2)$, $\eta_2=H_2(\theta_1,\theta_2)$,
we have:
\begin{equation}
  f^\prime(\eta_1,\eta_2) =\int \delta(\eta_1-H_1(\theta_1,\theta_2))\delta(\eta_2-H_2(\theta_1,\theta_2))f(\theta_1,\theta_2)\,\mathrm{d}\theta_1\,\mathrm{d}\theta_2\,.
\end{equation}

\subsection{Choice of the prior}

One of the most questionable issue related to Bayesian inference is the subjectiveness of
the result, being dependent on the choice of a prior. In particular, there is no unique choice
of a prior that models one's ignorance about an unknown parameter. A choice of a uniform prior,
such as it was done in Sec.~\ref{sec:BayesianPoissonianCounting}, is also questionable:
if the prior PDF is uniform in a chosen variable, it won't necessarily be uniform when applying a coordinate transformation
to that variable. A typical example is  the measurement of a particle's lifetime, which is the inverse of
the particle's width.
Given any choice of a regular prior for a parameter, there is always a transformation that makes the PDF uniform.

Harold Jeffreys provided a method~\cite{Jeffreys} to chose a form of the prior that is invariant under parameter transformation.
The choice uses the so-called Fishers information matrix, which, given a set of parameters
$\vec{\theta}=(\theta_1,\cdots,\theta_m)$, is defined as:
\begin{equation}
  {\cal I}_{ij}(\vec{\theta}) = \left<
  \frac{\partial\ln L(\vec{x};\vec{\theta})}{\partial \theta_i}
  \frac{\partial\ln L(\vec{x};\vec{\theta})}{\partial \theta_j}
  \right>\,.
  \label{eq:FisherInformation}
\end{equation}
Jeffrey's prior is then given by, up to a normalization factor:
\begin{equation}
  \pi(\vec{\theta}) \propto \sqrt{\mathrm{det}\,{\cal I}(\vec{\theta})}\,.
  \label{eq:JeffreysPrior}
\end{equation}
It's possible to demonstrate that Eq.~(\ref{eq:JeffreysPrior}) is invariant under a parameter
transformation $\vec{\eta} = \vec{H}(\vec{\theta})$.

\subsection{Frequentist inference}

Assigning a probability level to an unknown parameter makes no sense in the frequentist approach
since unknown parameters are not random variables.
A frequentist inference procedure should determine a central value and an uncertainty interval that depend
on the observed measurements without introducing any subjective element.
Such central value and interval extremes are random variables themselves.
The function that returns the central value given an observed measurement is called {\it estimator}.
The parameter value provided by an estimator is also called {\it best fit} value.
Different estimator choices are possible, the most frequently adopted is the {\it maximum likelihood
  estimator} because of its statistical properties discussed in Sec.~\ref{sec:estimatorProperties}.

Repeating the experiment will result each time in a different data sample
and, for each data sample, the estimator returns a different central value $\hat{\theta}$.
An uncertainty interval $[\hat{\theta} -\delta, \hat{\theta} +\delta]$ can be associated to
the estimator value $\hat{\theta}$. In some cases, as for the Bayesian inference, an
asymmetric interval choice is also possible with frequentist inference:
$[\hat{\theta} -\delta^-, \hat{\theta} +\delta^+]$.
Some of the intervals obtained with this method contain the fixed and unknown true
value of $\theta$, corresponding to a fraction equal to 68.3\% of the repeated experiments, in the limit of very
large number of experiments. This property is called {\it coverage}.

The simplest example of frequentist inference 
assumes a Gaussian PDF (Eq.~(\ref{eq:GaussianPDF})) with a known $\sigma$ and an unknown $\mu$.
A single experiment provides a measurement $x$, and we can estimate $\mu$ as $\hat{\mu} = x$.
The distribution of $\hat{\mu}$ is the original Gaussian because  $\hat{\mu}$ is just equal to $x$.
A fraction of 68.3\% of the experiments (in the limit of large number of repetitions) will provide an
estimate $\hat{\mu}$ within: $\mu - \sigma < \hat{\mu} < \mu + \sigma$. This means that we can quote:
\begin{equation}
  \boxed{
    \mu = x \pm \sigma\,.
    }
\label{eq:trivialML}
\end{equation}

\subsection{Maximum likelihood estimates}
\label{sec:maxLik}
The maximum likelihood method takes as best-fit values of the unknown parameter
the values that maximize the likelihood function (defined Sec.~\ref{sec:likeFun}).
The maximization of the likelihood function can be performed analytically only in the simplest cases,
while a numerical treatment is needed in most of the realistic cases.
{\sc Minuit}~\cite{minuit} is historically the most widely used minimization software engine in High Energy Physics.

\subsubsection{Extended likelihood function}
\label{sec:extLikFun}
Given a sample of $N$ measurements of the variables $\vec{x}=(x_1, \cdots, x_n)$, the likelihood function expresses the probability
density evaluated for our sample as a function of the unknown parameters $\theta_1,\cdots,\theta_m$:
\begin{equation}
  L(\vec{x}_1,\cdots,\vec{x}_N) =
  \prod_{i=1}^Nf(x_1^i,\cdots,x_n^i;\theta_1,\cdots,\theta_m)\,.
\end{equation}
The size $N$ of the sample is in many cases also a random variable. In those cases,
the {\it extended likelihood function} can be defined as:
\begin{equation}
  L(\vec{x}_1,\cdots,\vec{x}_N) =
  P(N;\theta_1,\cdots,\theta_m) \prod_{i=1}^N f(x_1^i,\cdots,x_n^i;\theta_1,\cdots,\theta_m)\,,
\end{equation}
where $P(N;\theta_1,\cdots,\theta_m)$ is the distribution of $N$, and in practice is always a Poissonian 
whose expected rate parameter is a function of the unknown parameters  $\theta_1,\cdots,\theta_m$:
\begin{equation}
  P(N;\theta_1,\cdots,\theta_m) = \frac{\nu(\theta_1,\cdots,\theta_m)^N e^{-\nu(\theta_1,\cdots,\theta_m)}}{N!}\,.
\end{equation}

In many cases, either with a standard or an extended likelihood function,
it may be convenient to use $-\ln L$ or $-2\ln L$ in the numerical treatment
rather than $L$, because
the product of the various terms is transformed into the sum of the logarithms of
those terms, which may have advantages in the computation.

For a Poissonian process that is given by the sum of a signal plus a background process,
the extended likelihood function may be written as:
\begin{equation}
  L(\vec{x}; s, b, \vec{\theta}) =
  \frac{(s+b)^N e^{-(s+b)}}{N!}
  \prod_{i=1}^N\left(
  f_sP_s(x_i;\vec{\theta}) + f_b P_b(x_i;\vec{\theta})
  \right)\,,
  \label{eq:extLikSB}
\end{equation}
where $s$ and $b$ are the signal and background expected yields, respectively,
$f_s$ and $f_b$ are the fraction of signal and background events, namely:
\begin{eqnarray}
  f_s & = & \frac{s}{s+b} \,,\\
  f_b & = & \frac{b}{s+b} \,,
\end{eqnarray}
and $P_s$ and $P_b$ are the PDF of the variable $x$ for signal and background,
respectively.
Replacing $f_s$ and $f_b$ into Eq.~(\ref{eq:extLikSB}) gives:
\begin{equation}
  L(\vec{x}; s, b, \vec{\theta}) = \frac{e^{-(s+b)}}{N!}
  \prod_{i=1}^N\left(
  sP_s(x_i;\vec{\theta}) + bP_b(x_i;\vec{\theta})
  \right)\,.
  \label{eq:extLikInt}
\end{equation}
It may be more convenient to use the negative logarithm of Eq.~(\ref{eq:extLikInt}),
that should be minimize in order to determine the best-fit values of $s$, $b$ and $\vec{\theta}$:
\begin{equation}
  -\ln  L(\vec{x}; s, b, \vec{\theta}) =
  s + b -\sum_{i=1}^N\ln\left(
  sP_s(x_i;\vec{\theta}) + bP_b(x_i;\vec{\theta})
  \right) +\ln N!\,.
\end{equation}
The last term $\ln N!$ is a constant with respect to the fit parameters,
and can be omitted in the minimization.
In many cases, instead of using $s$ as parameter of interest,
the {\it signal strength} $\mu$ is introduced, defined by the following equation:
\begin{equation}
  s = \mu s_0\,,
\end{equation}
where $s_0$ is the theory prediction for the signal yield $s$.
$\mu=1$ corresponds to the nominal value of the theory prediction for the signal yield.

\begin{figure}[htbp]
\centering\includegraphics[width=.495\linewidth]{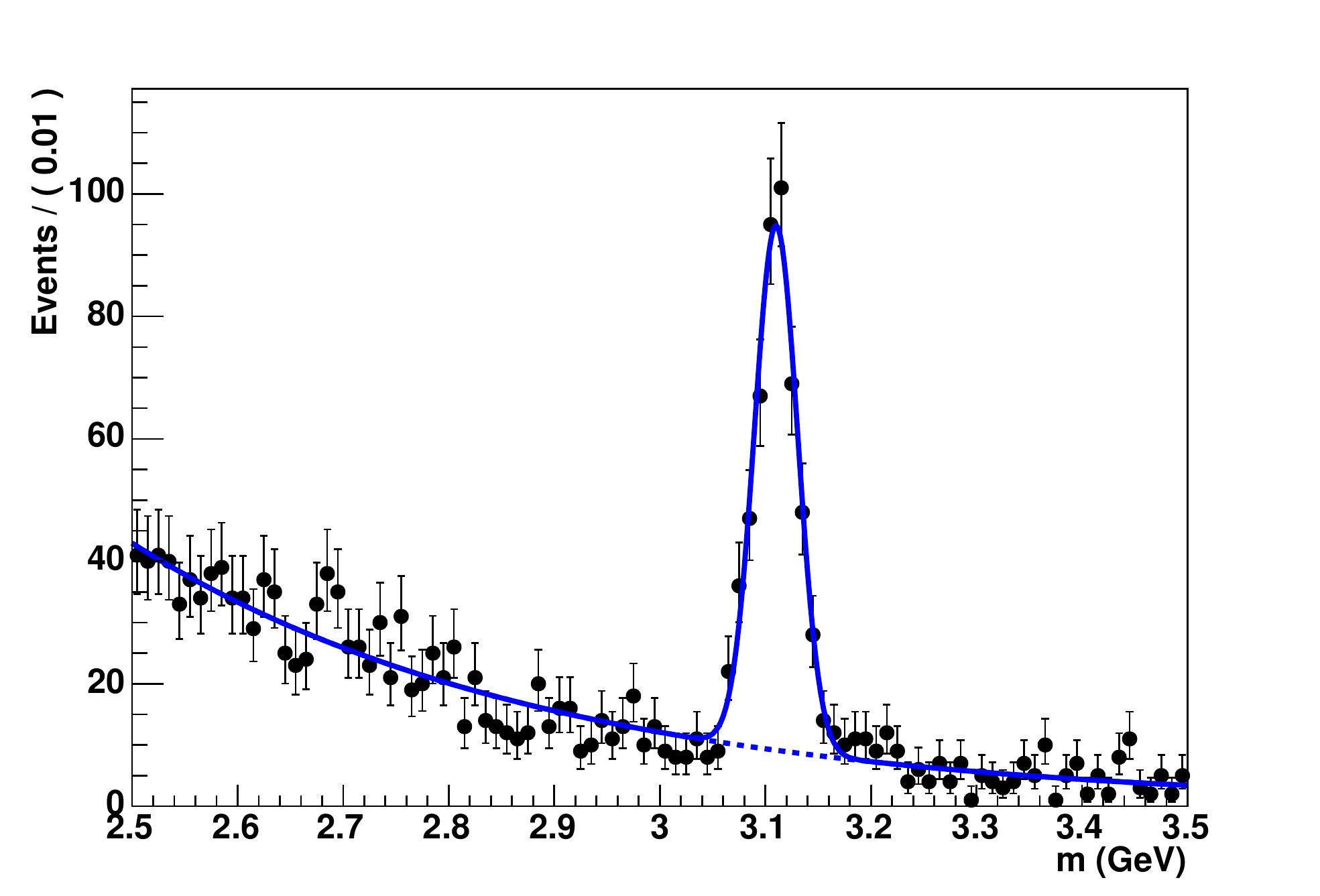}
\caption{Example of an unbinned maximum likelihood fit. Data are fit using
  a Gaussian distribution for the signal and an exponential distribution for the
  background. This figure is taken from Ref.~\cite{LL}.}
\label{fig:sbFit}
\end{figure}
An example of unbinned maximum likelihood fit is given in Fig.~\ref{fig:sbFit},
where the data are fit with a model inspired to Eq.~(\ref{eq:extLikInt}), with
$P_s$ and $P_b$ taken as a Gaussian and an exponential distribution, respectively.
The observed variable has been called $m$ in that case because the spectrum resembles an invariant mass peak,
and the position of the peak at 3.1~GeV reminds a $\mathrm{J}/\psi$ particle.
The two PDFs can be written as:
\begin{eqnarray}
  P_s(m) & = & \frac{1}{\sqrt{2\pi}\sigma}e^{-\frac{(m-\mu)^2}{2\sigma^2}}\,,\\
  P_b(m) & = & \lambda e^{-\lambda m}\,.
\end{eqnarray}
The parameters $\mu$, $\sigma$ and $\lambda$ are fit together with the
signal and background yields $s$ and $b$. While $s$ is our {\it parameter of interest},
because we will eventually determine a production cross section or branching fraction from
its measurement, the other additional parameters, that are not directly
related to our final measurement, are said {\it nuisance parameters}.
In general, nuisance parameters are needed to model background yield,
detector resolution and efficiency, various parameters modeling the
signal and background shapes, etc. Nuisance parameters are also important
to model {\it systematic uncertainties}, as will be discussed more in
details in the following sections.

\subsection{Estimate of Gaussian parameters}

If we have $n$ independent measurements $\vec{x}=(x_1,\cdots,x_n)$ all modeled (exactly or approximatively)
with the same Gaussian PDF,
we can write the negative of twice the logarithm of the likelihood function as follows:
\begin{equation}
  -2\ln L(\vec{x}; \mu) = \sum_{i=1}^n \frac{(x_i-\mu)^2}{\sigma^2} + n(\ln 2\pi + 2\ln\sigma)\,.
  \label{eq:GausChi2}
\end{equation}
The first term, $\sum_{i=1}^n \frac{(x_i-\mu)^2}{\sigma^2}$, is an example of $\chi^2$ variable (see Sec.~\ref{sec:binnedSamples}).

An analytical minimization of $-2\ln L$ with respect to $\mu$,
assuming $\sigma^2$ is known, gives the {\it arithmetic mean} as maximum likelihood estimate of $\mu$:
\begin{equation}
  \hat{\mu} = \frac{1}{n}\sum_{i=1}^n x_i\,.
\end{equation}
If $\sigma^2$ is also unknown, the maximum likelihood estimate of $\sigma^2$ is:
\begin{equation}
  \widehat{{\sigma}^2} = \frac{1}{n} \sum_{i=1}^m(x_i-\hat{\mu})^2\,.
  \label{eq:sigma2MLestimate}
\end{equation}
The estimate in Eq.~(\ref{eq:sigma2MLestimate}) can be demonstrated to have an unpleasant feature, called {\it bias}, that
will be discussed in Sec.~\ref{sec:bias}.

\subsection{Estimator properties}
\label{sec:estimatorProperties}
This section illustrates the main properties of estimators. Maximum likelihood estimators
are most frequently chosen because they have good performances for what concerns those properties.
\subsubsection{Consistency}

For large number of measurements, the estimator $\hat{\theta}$ should converge, in probability, to the true value of $\theta$,
$\theta^{\mathrm{true}}$.
Maximum likelihood estimators are consistent.

\subsubsection{Bias}
\label{sec:bias}
The bias of a parameter is the average value of its deviation from the true value:
\begin{equation}
  \mathbbm{b}[\hat{\theta}] = \left< \hat{\theta} - \theta^{\mathrm{true}}\right> = \left<\hat{\theta}\right> - \theta^{\mathrm{true}}\,.
\end{equation}
An {\it unbiased estimator} has $\mathbbm{b}[\theta]=0$.
Maximum likelihood estimators may have a bias, but the bias decreases with large number of measurements (if the model used in the fit is correct).

In the case of the estimate of a Gaussian's $\sigma^2$,
the maximum likelihood estimate (Eq.~(\ref{eq:sigma2MLestimate})) underestimates the true variance. 
The bias can be corrected for by applying a multiplicative factor:
\begin{equation}
  \widehat{{\sigma}^2}_{\mathrm{unbias.}} = \frac{n}{n-1}\widehat{{\sigma}^2}
  =\frac{1}{n-1}\sum_{i=1}^n (x_i-\hat{\mu})^2\,.
\end{equation}

\subsubsection{Efficiency}

The variance of any consistent estimator is subject to a lower bound
due to Cram\'er~\cite{Cramer} and Rao~\cite{Rao}:
\begin{equation}
  \mathbbm{V}\mathrm{ar}[\hat{\theta}] \ge \frac{\displaystyle
    \left(1 + \frac{\partial \mathbbm{b}[\theta] }{\partial\theta} \right)^2
  }{\displaystyle
    \left<\left(
    \frac{\partial\ln L(\vec{x};\theta)}{\partial\theta}
    \right)\right>
  } = \mathbbm{V}_{\mathrm{CR}}[\hat{\theta}]\,.
  \label{eq:CramerRao}
\end{equation}
For an unbiased estimator, the numerator in Eq.~(\ref{eq:CramerRao}) is equal to one.
The denominator in Eq.~(\ref{eq:CramerRao}) is the Fisher information (Eq.~(\ref{eq:FisherInformation})).

The {\it efficiency} of an estimator $\hat{\theta}$ is the ratio
of the Cram\'er--Rao bound and the estimator's variance:
\begin{equation}
  \varepsilon(\hat{\theta}) = \frac{\mathbbm{V}_{\mathrm{CR}}[\hat{\theta}]}{\mathbbm{V}\mathrm{ar}[\hat{\theta}]}\,.
\end{equation}
The efficiency for maximum likelihood estimators tends to one for large number of measurements.
In other words, maximum likelihood estimates have, asymptotically, the smallest variance
of all possible consistent estimators.

\subsection{Neyman's confidence intervals}
\label{sec:NeymanBelt}

\begin{figure}[htbp]
\centering\includegraphics[width=.495\linewidth]{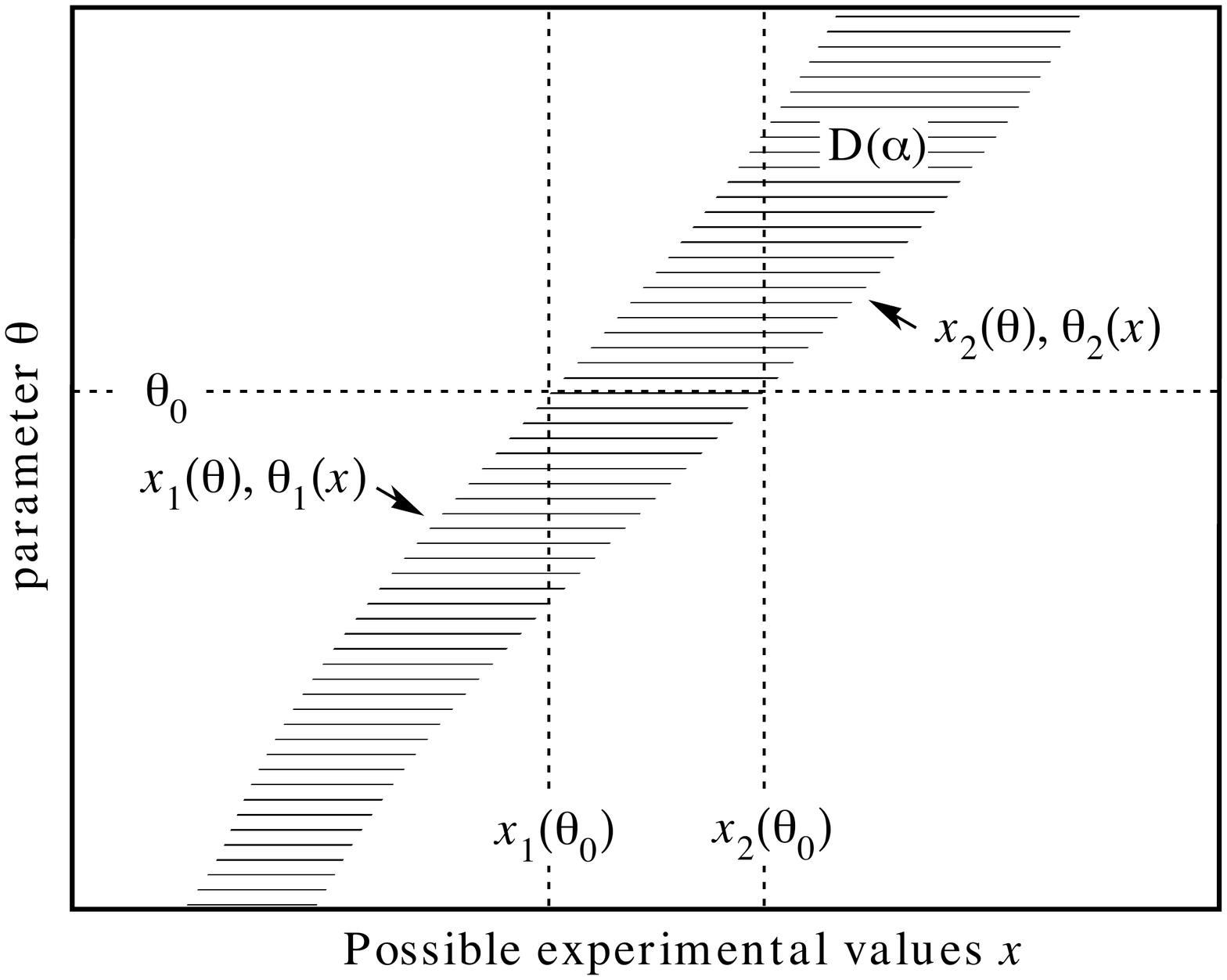}
\caption{Example Neyman's belt construction and inversion.
This figure is taken from Ref.~\cite{PDG}.}
\label{fig:NeymanBelt}
\end{figure}

A procedure to determine frequentist {\it confidence intervals} is due to
Neyman~\cite{neyman_belt}. It proceeds as follows:
\begin{itemize}
  \item Scan the allowed range of the unknown parameter of interest $\theta$.
  \item Given a value $\theta_0$ of $\theta$, compute the interval $[x_1(\theta_0), x_2(\theta_0)]$ that contains $x$ with a probability $1 - \alpha$
    ({\it confidence level}, or CL) equal to 68.3\% (or 90\%, 95\%). 
    For this procedure, a choice of interval ({\it ordering rule}) is needed, as discussed in Sec.~\ref{sec:BayesianInference}.
  \item For the observed value of $x$, invert the confidence belt:  find the corresponding interval $[\theta_1(x), \theta_2(x)]$.
\end{itemize}
By construction, a fraction of the experiments equal to $1 -\alpha$ will measure $x$ such that the corresponding
{\it confidence interval} $[\theta_1(x), \theta_2(x)]$ contains ({\it covers}) the true value of $\theta$.
It should be noted that the random variables are $\theta_1(x)$ and $\theta_2(x)$, not $\theta$.
An example of application of the Neyman's belt construction and inversion is shown in Fig.~\ref{fig:NeymanBelt}.

\begin{figure}[htbp]
\centering\includegraphics[width=.495\linewidth]{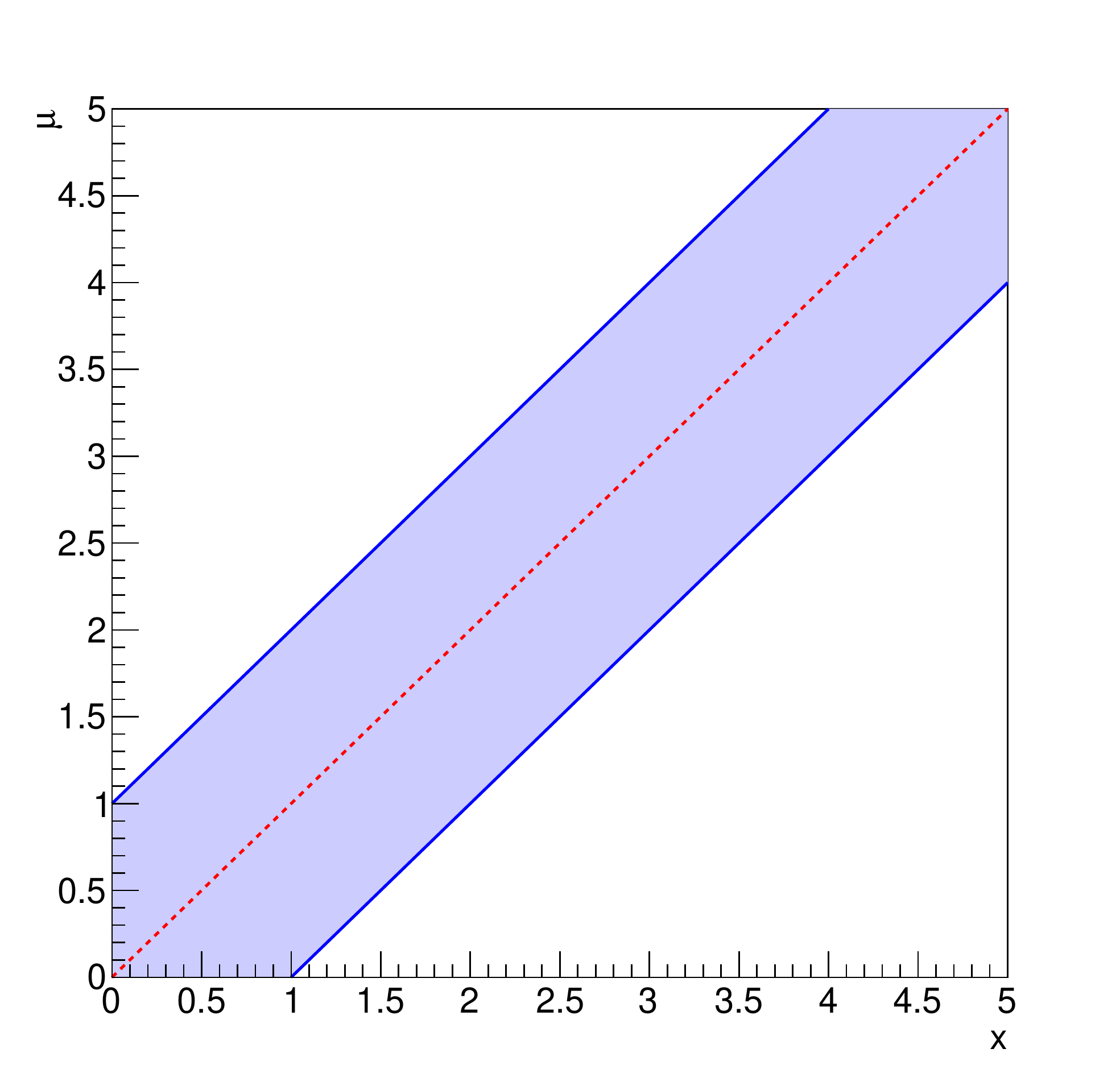}
\caption{Example of Neyman's belt construction for a Gaussian distribution with $\sigma=1$, $1-\alpha =0.683$.}
\label{fig:NeymanGaussianBelt}
\end{figure}

The simplest application of Neyman's belt construction can be done with a Gaussian
distribution with known parameter $\sigma=1$, as shown in Fig.~\ref{fig:NeymanGaussianBelt}.
The belt inversion is trivial and gives the expected result: a central value $\hat{\mu} = x$
and a confidence interval $[\mu_1, \mu_2] = [x - \sigma, x + \sigma]$.
The result can be quoted as $\mu = x\pm\sigma$, similarly to what was determined
with Eq.~(\ref{eq:trivialML}).

\subsection{Binomial intervals}
\label{sec:binInt}

The Neyman's belt construction may only guarantee approximate coverage in case of a discrete
variable $n$. This because the interval for a discrete variable is a set of integer values,
$\{ n_{\mathrm{min}}, \cdots, n_{\mathrm{max}}\}$, and cannot be ``tuned'' like in
a continuous case. The choice of the discrete interval should be such to provide
{\it at least} the desired coverage (i.e.: it may {\it overcover}).
For a binomial distribution, the problem consists of finding the interval such that:
\begin{equation}
  \sum_{n=n_{\mathrm{min}}}^{n_{\mathrm{max}}}
  \frac{N!}{n!(N-n)!} p^n (1-p)^{N-n} \ge 1-\alpha\,.
\end{equation}
Clopper and Pearson~\cite{clopper_pearson} solved the belt inversion problem for
central intervals.
For an observed $n = k$, one has to find the lowest $p^{\mathrm{lo}}$ and highest
$p^{\mathrm{up}}$ such that:
\begin{eqnarray}
  P(n \ge k | N, p^{\mathrm{lo}}) & = & \frac{\alpha}{2}\,,\label{eq:CP1}\\
  P(n \le k | N, p^{\mathrm{up}}) & = & \frac{\alpha}{2}\,.
\end{eqnarray}
\begin{figure}[htbp]
\centering\includegraphics[width=.495\linewidth]{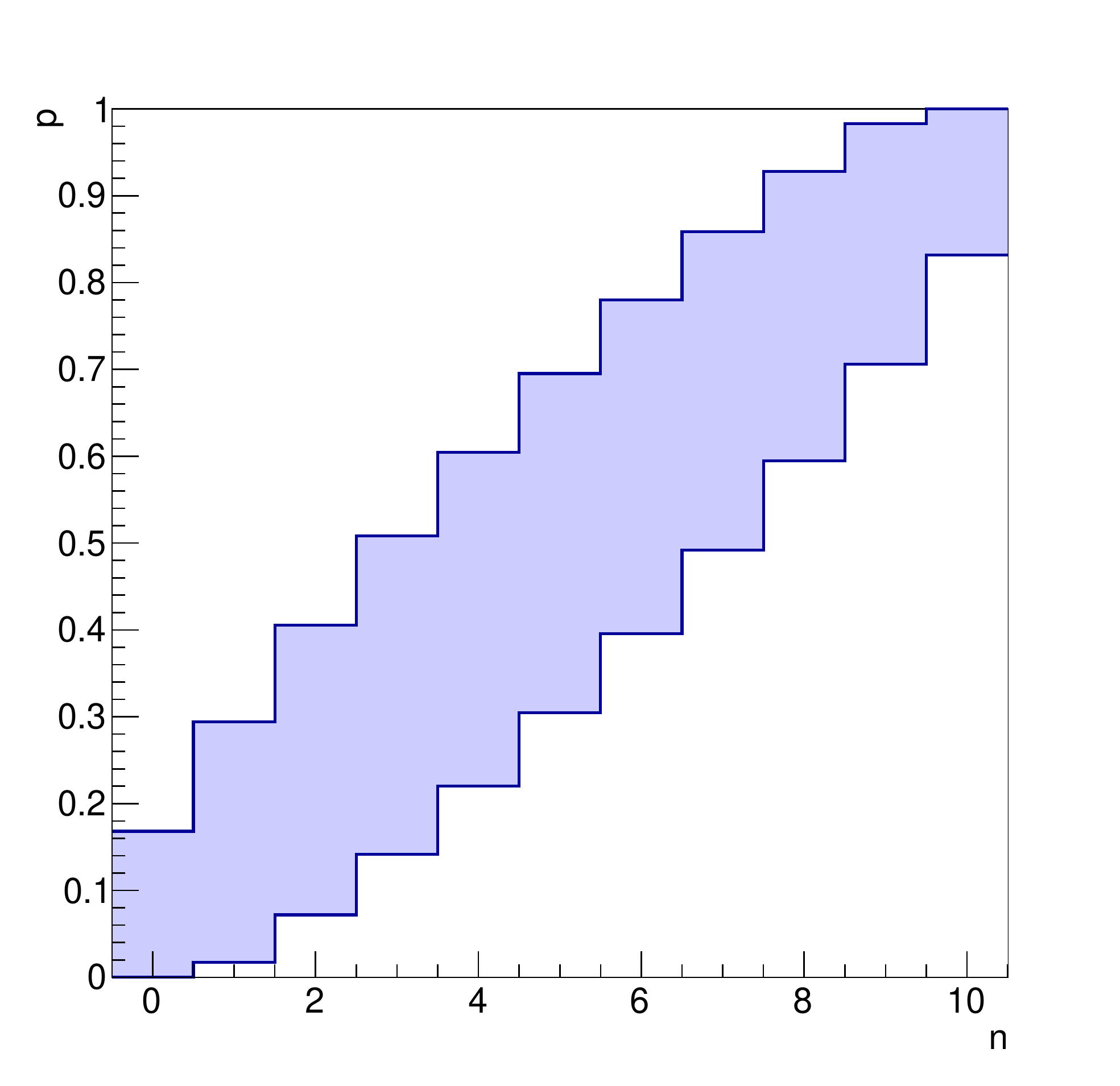}
\caption{Neyman belt construction for binomial intervals, $N=10$, $1-\alpha=0.683$.}
\label{fig:CloPear}
\end{figure}
An example of Neyman belt constructed using the Clopper--Pearson
method is shown in Fig.~\ref{fig:CloPear}.
For instance for $n = N$, Eq.~(\ref{eq:CP1}) becomes:
\begin{equation}
  P(n\ge N|N,p^{\mathrm{lo}}) = P(n=N|N,p^{\mathrm{lo}}) = (p^{\mathrm{lo}})^N = \frac{\alpha}{2}\,,
\end{equation}
hence, for the specific case $N=10$:
\begin{equation}
  p^{\mathrm{lo}} = \sqrt[10]{\frac{\alpha}{2}} = 0.83\,\text(1-\alpha = 0.683), \,0.74\,(1-\alpha = 0.90)\,.
\end{equation}
In fact, in Fig.~\ref{fig:CloPear}, the bottom line of the belt reaches
the value $p=0.83$ for $n=10$.
A frequently used approximation, inspired by Eq.~(\ref{eq:binomVar}) is:
\begin{equation}
  \hat{p} = \frac{n}{N},\,\,\,\sigma_{\hat{p}} \simeq \sqrt{\frac{\hat{p}(1-\hat{p})}{N}}\,.
  \label{eq:varEff}
\end{equation}
Eq.~(\ref{eq:varEff}) gives $\sigma_{\hat{p}}=0$ for $n=0$ or $N=n$, which is
clearly an underestimate of the uncertainty on $\hat{p}$. For this reason,
Clopper--Pearson intervals should be preferred to the approximate
formula in Eq.~(\ref{eq:varEff}).

Clopper--Pearson intervals are often defined as ``exact'' in literature,
though exact coverage is often impossible to achieve for discrete variables.
\begin{figure}[htbp]
\centering\includegraphics[width=.495\linewidth]{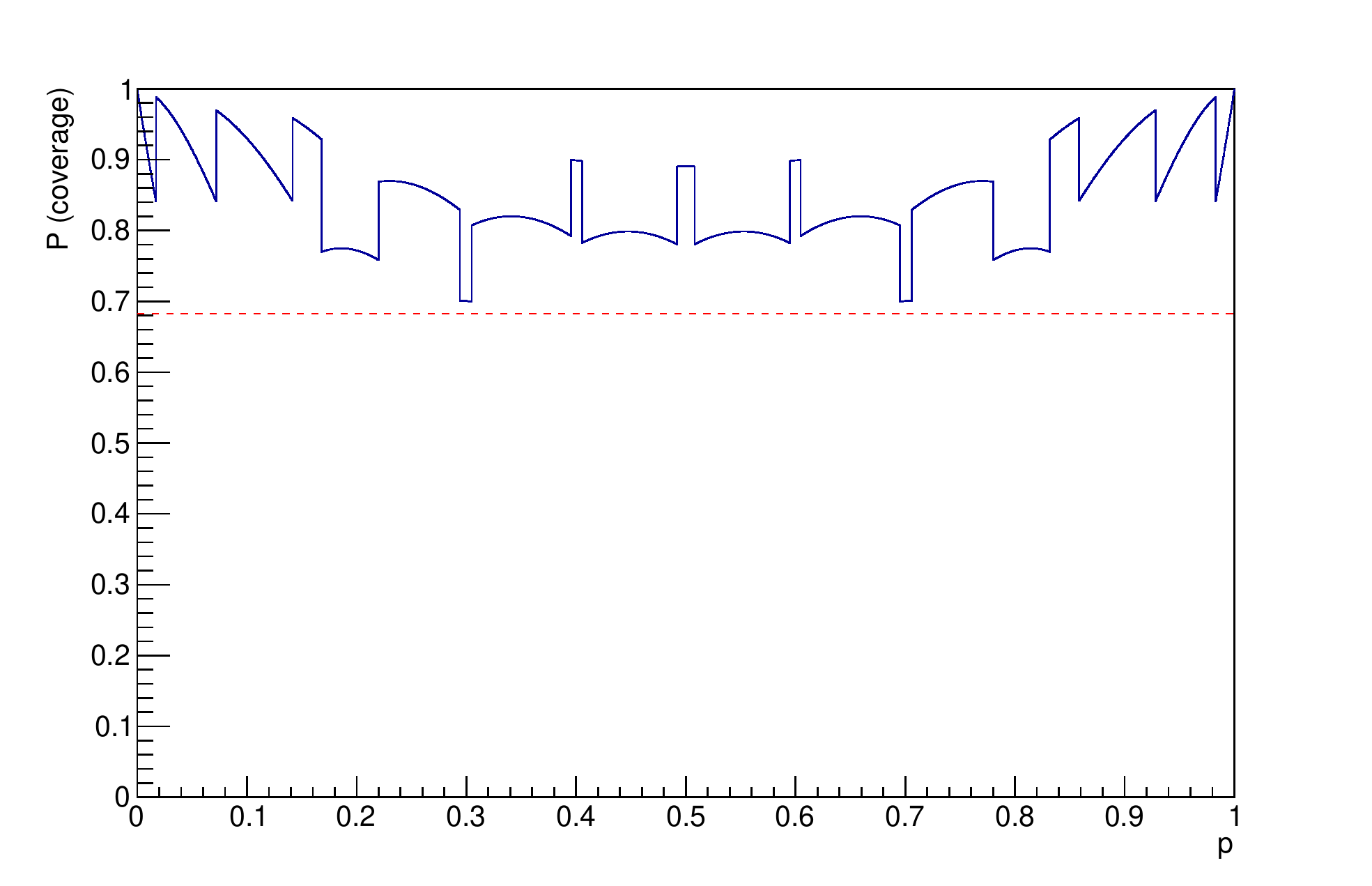}
\centering\includegraphics[width=.495\linewidth]{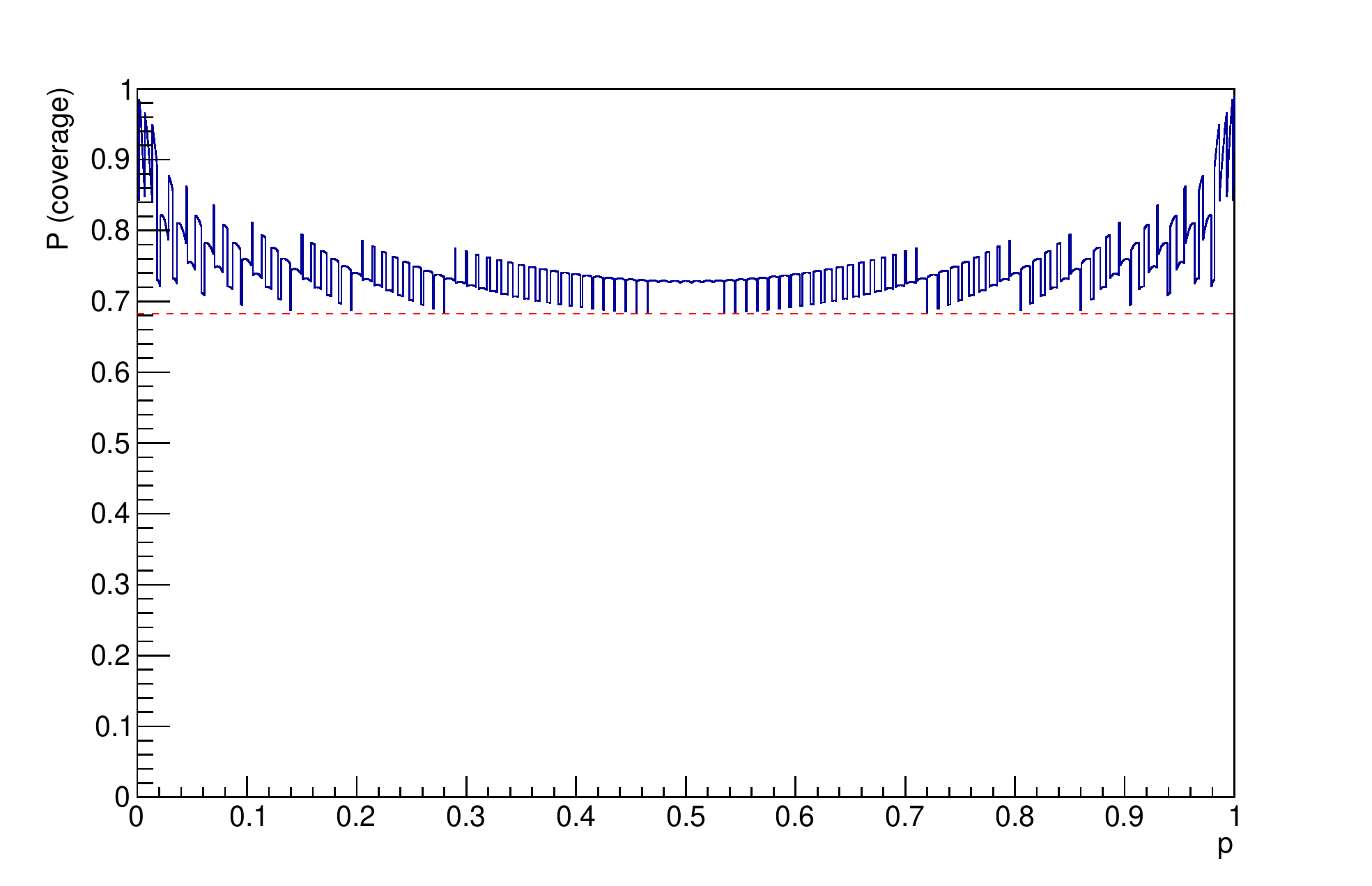}
\caption{Coverage of Clopper--Pearson intervals for $N=10$ (left) and
  for $N=100$ (right).}
\label{fig:CloPearCov}
\end{figure}
Figure~\ref{fig:CloPearCov} shows the coverage of Clopper--Pearson intervals as a
function of $p$ for $N=10$ and $N=100$ for $1-\alpha = 0.683$. A ``ripple'' structure
is present which, for large $N$, tends to gets closer to the nominal 68.3\% coverage.

\subsection{Approximate error evaluation for maximum likelihood estimates}

A parabolic approximation of $-2\ln L$ around the minimum is equivalent to a Gaussian approximation,
which may be sufficiently accurate in many but not all cases. For a Gaussian model, $-2\ln L$
is given by:
\begin{equation}
  -2\ln L(\vec{x};\mu, \sigma) = \sum_{i=1}^n\frac{(x_i-\mu)^2}{\sigma^2}+\text{const.}\,.
\end{equation}
An approximate estimate of the covariance matrix is obtained from the
$2^{\mathrm{nd}}$ order partial derivatives with respect to the fit parameters at the minimum:
\begin{equation}
  V_{ij}^{-1} = -\left.\frac{\partial^2\ln L}{\partial\theta_i\partial\theta_j}\right|_{\theta_k=\hat{\theta}_k,\,\forall k}\,.
\label{eq:errLikVar}
\end{equation}

\begin{figure}[htbp]
\centering\includegraphics[width=.495\linewidth]{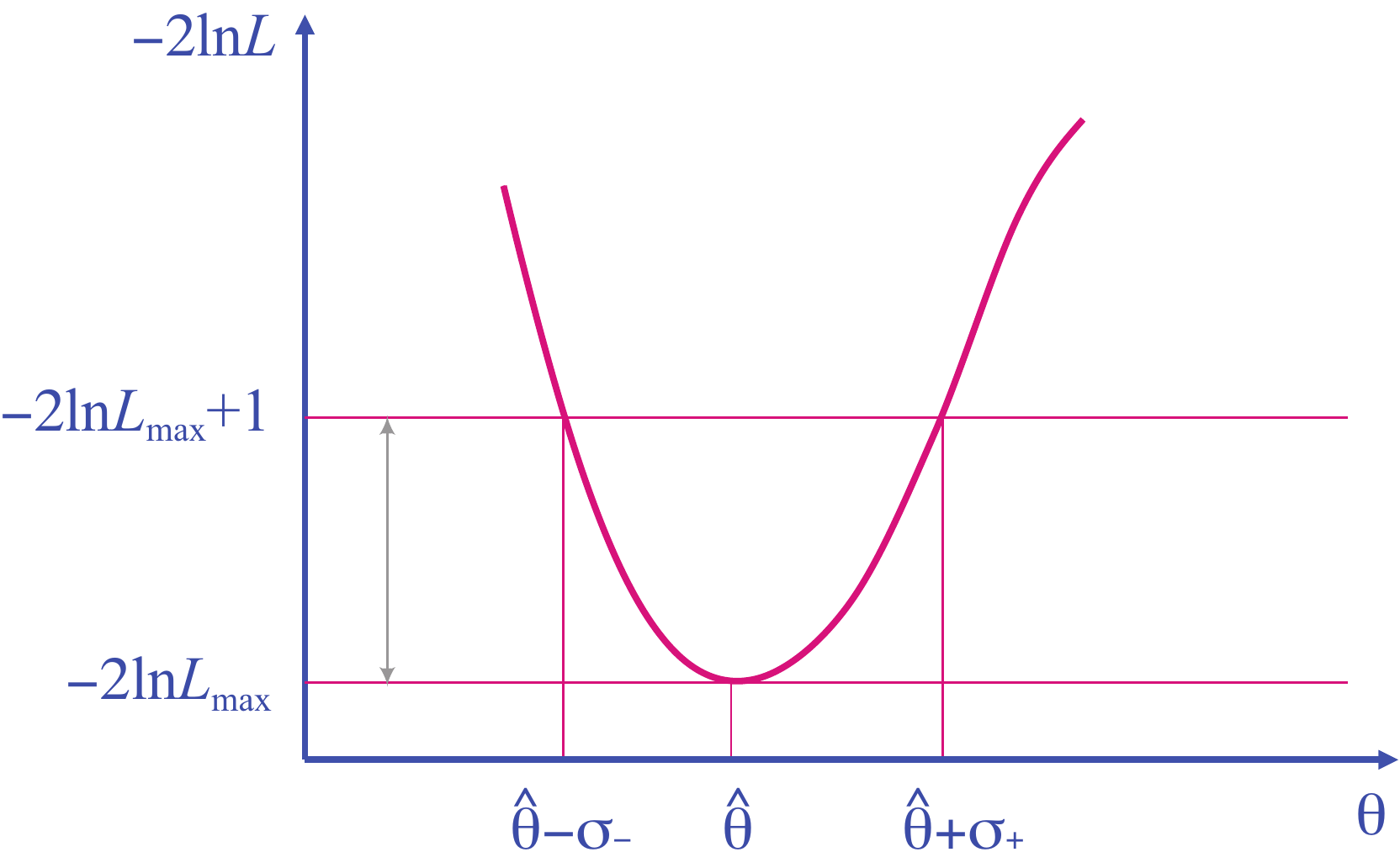}
\caption{Scan of $-2\ln L$ in order to determine asymmetric $1\sigma$ errors. This figure is taken from Ref.~\cite{LL}.}
\label{fig:likeScan}
\end{figure}

Another approximation alternative to the parabolic one from Eq.~(\ref{eq:errLikVar}) is the evaluation of the excursion
range of $-2\ln L$ around the minimum, as visualized in Fig.~\ref{fig:likeScan}.
The uncertainty interval can be determined as the range around the minimum of $-2\ln L$ for which $-2\ln L$ increases by $+1$ (or $+n^2$ for a $n\sigma$ interval).
Errors can be asymmetric with this approach if the curve is asymmetric.
For a Gaussian case the result is identical to the $2^{\mathrm{nd}}$ order derivative matrix (Eq.~(\ref{eq:errLikVar})).

\subsection{Two-dimensional uncertainty contours}

In more dimensions, i.e.: for the simultaneous determination of more unknown parameters from a fit,
it's still possible to determine multi-dimensional contours corresponding to $1\sigma$ or $2\sigma$
probability level. It should be noted that the scan of $-2\ln L$ in the multidimensional
space, looking for an excursion of $+1$ with respect to the value at the minimum, may give
probability levels smaller than the corresponding values in one dimension.
For a Gaussian case in one dimension, the probability associated to an interval $[-n\sigma,+n\sigma]$ is
given, integrating Eq.~(\ref{eq:GaussianPDF}), by:
\begin{equation}
  P_{1\mathrm{D}}(n\sigma)= \sqrt{\frac{2}{\pi}}\int_0^ne^{-\frac{x^2}{2}}\,\mathrm{d}x = \mathrm{erf}\left(\frac{n}{\sqrt{2}}\right)\,.
  \label{eq:GaussInt1D}
\end{equation}
For a two-dimensional Gaussian distribution, i.e.: the product of two independent Gaussian PDF,
the probability associated to the contour with elliptic shape for which $-2\ln L$ increases by $+(n\sigma)^2$ with respect to its
minimum is:
\begin{equation}
  P_{2\mathrm{D}}(n\sigma)= \int_0^ne^{-\frac{r^2}{2}}r\,\mathrm{d}r =  1 - e^{-\frac{n^2}{2}}\,.
  \label{eq:GaussInt2D}
\end{equation}
Table.~\ref{tab:GaussianInt1D2D} reports numerical values for Eq.~(\ref{eq:GaussInt1D}) and
Eq.~(\ref{eq:GaussInt2D}) for various $n\sigma$ levels.
\begin{table}[htbp]
\caption{Probabilities for 1D interval and 2D contours with different $n\sigma$ levels..}
\label{tab:GaussianInt1D2D}
\centering
\begin{tabular}{ccc}\hline\hline
$n\sigma$ & $P_{1\mathrm{D}}$ & $P_{2\mathrm{D}}$ \\\hline
  $1\sigma$ & 0.6827 & 0.3934 \\
  $2\sigma$ & 0.9545 & 0.8647 \\
  $3\sigma$ & 0.9973 & 0.9889 \\
  $1.515\sigma$ & & 0.6827 \\
  $2.486\sigma$ & & 0.9545 \\
  $3.439\sigma$ & 0.9973 \\\hline\hline
\end{tabular}
\end{table}
In two dimensions, for instance, in order to recover a $1\sigma$ probability level in one
dimension (68.3\%), a contour corresponding to an excursion of $-2\ln L$ from its minimum of $+1.515^2$
should be considered, and for a $2\sigma$
probability level in one dimension (95.5\%), the excursion should be $+2.486^2$.
Usualy two-dimensional intervals corresponding to one or two sigma are reported,
whose one-dimensional projection correspond to 68\% or 95\% probability content,
respectively.

\subsection{Error propagation}

In case of frequentist estimates, error propagation can't be performed
with a simple procedure as for the Bayesian case, where the full
posterior PDF is available (Sec.~\ref{sec:errPropBayes}).

Imagine we estimate from a fit the parameter set
$\vec{\theta} = (\theta_1,\cdots,\theta_n) =\hat{\vec{\theta}}$
and we know their covariance matrix $\Theta_{ij}$, for instance
using Eq.~(\ref{eq:errLikVar}).
We want to determine a new set of parameters that are functions of $\vec{\theta}$: 
$\vec{\eta} = (\eta_1,\cdots, \eta_m) = \vec{\eta}(\vec{\theta})$.
The best approach would be to rewrite the original likelihood function
as a function of $\vec{\eta}$ instead of $\vec{\theta}$, and perform the
minimization and error estimate again for $\vec{\eta}$.
In particular, the central value for $\hat{\vec{\eta}}$ will be equal to the transformed
of the central value $\hat{\vec{\theta}}$, but no obvious transformation rule
exists for the uncertainty intervals.

Reparametrizing the likelihood function is not always feasible.
One typical case is when central values and uncertainties for $\vec{\theta}$ are
given in a publication, but the full likelihood function is not available.
For small uncertainties, a linear approximation may be sufficient to
obtain the covariance matrix $H_{ij}$ for $\vec{\eta}$.
A Taylor expansion around the central value $\hat{\vec{\theta}}$ gives, using the error matrix $\Theta_{ij}$, at first order:
\begin{equation}
  H_{ij} = \sum_{k,\,l}\left.\frac{\partial\eta_i}{\partial\theta_k}\frac{\partial\eta_j}{\partial\theta_l}\Theta_{kl}
  \right|_{\vec{\theta}=\hat{\vec{\theta}}}\,.
  \label{eq:ErrPropFreq}
\end{equation}
The application of Eq.~(\ref{eq:ErrPropFreq}) gives well-known error propagation
formulae reported below as examples, valid in case of null (or negligible) correlation:
\begin{equation}
  \sigma_{x+y} =   \sigma_{x-y} = \sqrt{\sigma_x^2+\sigma_y^2}\,,
\end{equation}
\begin{equation}
  \frac{\sigma_{xy}}{xy} =   \frac{\sigma_{{x}/{y}}}{{x}/{y} } = \sqrt{\left(\frac{\sigma_x}{x}\right)^2 + \left(\frac{\sigma_y}{y}\right)^2}\,,
\end{equation}
\begin{equation}
  \sigma_{x^2} = 2x\sigma_x\,,
\end{equation}
\begin{equation}
  \sigma_{\ln x} = \frac{\sigma_x}{\sqrt{x}}\,.
\end{equation}

\subsection{Likelihood function for binned samples}
\label{sec:binnedSamples}

Sometimes data are available in form of a binned histogram. This may be convenient
when a large number of entries is available, and computing an unbinned likelihood function (Eq.~(\ref{eq:unbinnedLikeFun}))
would be too much computationally expansive.
In most of the cases, each bin content is independent on any other bin and all obey Poissonian distributions,
assuming that bins contain event-counting information.
The likelihood function can be written as product of Poissonisn PDFs corresponding to each bin
whose number of entries is given by $n_i$ .
The expected number of entries in each bin depends on some unknown parameters: $\mu_i = \mu_i(\theta_1,\cdots,\theta_m)$.
The function to be minimized, in order to fit $\theta_1, \cdots,\theta_n$, is the following:
\begin{eqnarray}
  -2\ln L(\vec{n};\vec{\theta}) & = &
  -2\ln \prod_{i=1}^{n_{\mathrm{bins}}}\mathrm{Poiss}(n_i;\mu_i(\theta_1,\cdots,\theta_m)) \\
  & = & -2\sum_{i=1}^{n_{\mathrm{bins}}}\ln \frac{
      e^{-\mu_i(\theta_1,\cdots,\theta_m)}\mu_i(\theta_1,\cdots,\theta_m)^{n_i}
    }{n_i!}\\
  & = & 2\sum_{i=1}^{n_{\mathrm{bins}}}\left(\mu_i(\theta_1,\cdots,\theta_m)
  -n_i\ln\mu_i(\theta_1,\cdots,\theta_m) +\ln{n_i!}\right)\,.
  \label{eq:PoisBinLik}
\end{eqnarray}
The expected number of entries in each bin, $\mu_i$, is often approximated by a continuous function $\mu(x)$
evaluated at the center of the bin $x=x_i$.
Alternatively, $\mu_i$  can be given by the superposition of other histograms ({\it templates}),
e.g.: the sum of histograms obtained from different simulated processes.
The overall yields of the considered processes may be left as free parameters in the fit in order to constrain
the normalization of simulated processes from data, rather than relying on simulation prediction,
which may affected by systematic uncertainties.

The distribution of the number of entries in each bin can be approximated,
for sufficiently large number of entries,
by a Gaussian with standard deviation equal to $\sqrt{n_i}$. 
Maximizing $L$ is equivalent to minimize:
\begin{equation}
  \chi^2 = \sum_{i=1}^{n_{\mathrm{bins}}}\frac{\left(n_i-\mu(x_i;\theta_1,\cdots,\theta_m)\right)^2
  }{n_i }
    \label{eq:NeymanChi2}
\end{equation}
Equation~(\ref{eq:NeymanChi2}) defines the so-called Neyman's $\chi^2$ variable.
Sometimes, the denominator $n_i$ is replaced by $\mu_i = \mu (x_i; \theta_1, \cdots, \theta_m)$
(Pearson's $\chi^2$) in order to avoid cases with $n_i$ equal to zero or very small.

Analytic solutions exist in a limited number of simple cases, e.g.: if $\mu$ is a linear function.
In most of the realistic cases, the $\chi^2$ minimization is performed numerically, as for
most of the unbinned maximum likelihood fits.
Binned fits are, in many cases, more convenient with respect to unbinned fits because the number of input
variables decreases from the total number of entries to the number of bins.
This leads usually to simpler and faster numerical implementations,
in particular when unbinned fits become unpractical in cases of very large number of entries.
Anyway, for limited number of entries, a fraction of the information is lost when
moving from an unbinned to a binned sample and a possible loss of precision may occur.

The maximum value of the likelihood function obtained from an umbinned maximum likelihood fit doesn't in general
provide information about the quality ({\it goodness}) of the fit.
Instead, the minimum value of the $\chi^2$ in a fit with a Gaussian underlying model
is distributed according to a known PDF given by:
\begin{equation}
  P(\chi^2;n) =\frac{2^{-{n}/{2}}}{\Gamma\left({n}/{2}\right)}
  \chi^{n-2}e^{-\frac{\chi^2}{2}}\,,
\end{equation}
where $n$ is the {\it number of degrees of freedom}, equal to the number of
bins minus the number of fit parameters.
The cumulative distribution (Eq.~(\ref{eq:cumulative})) of $P(\chi^2; n)$ follows a uniform distribution between from 0 to 1,
and it is an example of {\it p-value} (See Sec.~\ref{sec:HypTest}).
If the true PDF model deviates from the assumed distribution, the distribution of the $p$-value will be more peaked around zero
instead of being uniformly distributed.

It's important to note that $p$-values are not the ``probability of the fit hypothesis'',
because that would be a Bayesian probability, with a completely different meaning, and should be evaluated
in a different way.

In case of a Poissonian distribution of the number of bin entries that may deviate from the Gaussian approximation,
because of small number of entries,
a better alternative to the Gaussian-inspired Neyman's or Pearson's $\chi^2$ has been proposed
by Baker and Cousins~\cite{baker_cousins} using the following likelihood ratio
as alternative to Eq.~(\ref{eq:PoisBinLik}):
\begin{eqnarray}
  \chi^2_{\lambda} & = & -2\ln\prod_i\frac{L(n_i;\mu_i)}{L(n_i;n_i)} = -2\ln\prod_i\frac{e^{-\mu_i}\mu_i^{n_i}}{n_i!}
  \frac{n_i!}{e^{-{n_i}}n_i^{n_i}}  \\
    & = & 2\sum_i\left[
      \mu_i(\theta_1,\cdots,\theta_m)- n_i + n_i\ln\left(
      \frac{n_i}{\mu_i(\theta_1,\cdots,\theta_m)}
      \right) \right]\,.
      \label{eq:BakCous}
\end{eqnarray}
Equation~(\ref{eq:BakCous}) gives the same minimum value as the Poisson likelihood function,
since a constant term has been added to the log-likelihood function in Eq.~(\ref{eq:PoisBinLik}),
but in addition it provides goodness-of-fit information, since it asymptotically obeys a $\chi^2$
distribution with $n - m$ degrees of freedom. This is due to Wilks' theorem, discussed
in Sec.~\ref{sec:profLik}.

\subsection{Combination of measurements}

The simplest combination of two measurements can be performed when no correlation is present between them:
\begin{eqnarray}
  m & = & m_1 \pm \sigma_1 \,,\\
  m & = & m_2 \pm \sigma_2 \,.
\end{eqnarray}
The following $\chi^2$ can be built, assuming a Gaussian PDF model for the two measurements,
similarly to Eq.~(\ref{eq:GausChi2}):
\begin{equation}
  \chi^2 = \frac{(m-m_1)^2}{\sigma_1^2} + \frac{(m-m_2)^2}{\sigma_2^2}\,.
  \label{eq:Chi2Comb}
\end{equation}
The minimization of the $\chi^2$ in Eq.~(\ref{eq:Chi2Comb}) leads to the following equation:
\begin{equation}\
  0 = \frac{\partial\chi^2}{\partial m} =
  2 \frac{(m-m_1)}{\sigma_1^2} + 2 \frac{(m-m_2)}{\sigma_2^2}\,,
\end{equation}
which is solved by:
\begin{equation}
  m = \hat{m} = \frac{
    \frac{m_1}{\sigma_1^2} + \frac{m_2}{\sigma_2^2}
  }{
    \frac{1}{\sigma_1^2} + \frac{1}{\sigma_2^2}
  }\,.
  \label{eq:wAvPart}
\end{equation}
Eq.~(\ref{eq:wAvPart}) can also be written in form of {\it weighted average}:
\begin{equation}
  \hat{m} = \frac{w_1 m_1 + w_2 m_2}{w_1 + w_2}\,,
\end{equation}
where the weights $w_i$ are equal to $\sigma_i^{-2}$.
The uncertainty on $\hat{m}$ is given by:
\begin{equation}
  \sigma_{\hat{m}}^2 = \frac{1}{\frac{1}{\sigma_1^2} + \frac{1}{\sigma_2^2}}\,.
  \label{eq:wAvgErr}
\end{equation}

In case $m_1$ and $m_2$ are correlated measurements, the $\chi^2$ changes from Eq.~(\ref{eq:Chi2Comb}) to the following, including a non-null correlation coefficient $\rho$:
\begin{equation}
  \chi^2 = \left(
  \begin{array}{cc} m - m_1 & m - m_2 \end{array}
  \right)\left(\begin{array}{cc}
    \sigma_1^2 & \rho\sigma_1\sigma_2 \\
    \rho\sigma_1\sigma_2 & \sigma_2^2
    \end{array}
    \right)^{-1}\left(
    \begin{array}{c}
      m - m_1 \\ m - m_2
      \end{array}
    \right)\,.
    \label{eq:chi2BLUE}
\end{equation}
In this case, the minimization of the $\chi^2$ defined by Eq.~(\ref{eq:chi2BLUE}) gives:
\begin{equation}
  \hat{m} = \frac{
    m_1(\sigma_2^2 -\rho\sigma_1\sigma_2) + m_2(\sigma_1^2 -\rho\sigma_1\sigma_2)
  }{
    \sigma_1^2 -2\rho\sigma_1\sigma_2 + \sigma_2^2
  }\,,
  \label{eq:BLUE}
\end{equation}
with uncertainty given by:
\begin{equation}
  \sigma_{\hat{m}}^2 = \frac{
    \sigma_1^2\sigma_2^2(1-\rho)^2
  }{
    \sigma_1^2 -2\rho\sigma_1\sigma_2 + \sigma_2^2
  }\,.
\end{equation}
This solution is also called best linear unbiased estimator (BLUE)~\cite{lyons_gibaut}
and can be generalized to more measurements.
An example of application of the BLUE method is the world combination
of the top-quark mass measurements at LHC and Tevatron~\cite{topMassComb}.

It can be shown that, in case the uncertainties $\sigma_1$ and $\sigma_2$
are estimates that may depend on the assumed central value,
a bias may arise, which can be mitigated by evaluating the uncertainties
$\sigma_1$ and $\sigma_2$ at the central value obtained with the combination,
then applying the BLUE combination, iteratively, until the procedure converges~\cite{LyonsMartinSaxon, ListaBLUE}.

Imagine we can write the two measurements as:
\begin{eqnarray}
  m & = & m_1 \pm \sigma_1^\prime \pm \sigma_C\,, \\
  m & = & m_2 \pm \sigma_2^\prime \pm \sigma_C\,,
\end{eqnarray}
where $\sigma_C^2 = \rho \sigma_1\sigma_2$. This is the case where the two measurements
are affected by a statistical uncertainty, which is uncorrelated between the two measurements,
and a fully correlated systematic uncertainty.
In those case, Eq.~(\ref{eq:BLUE}) becomes:
\begin{equation}
   \hat{m} = \frac{
    \frac{m_1}{\sigma_1^{\prime 2}} + \frac{m_2}{\sigma_2^{\prime 2}}
  }{
    \frac{1}{\sigma_1^{\prime 2}} + \frac{1}{\sigma_2^{\prime 2}}
    }\,,
\end{equation}
i.e.: it assumes again the form of a weighted average with weights $w_i = \sigma^{\prime -2}$
computed on the uncorrelated uncertainty contributions.
The uncertainty on $\hat{m}$ is given by:
\begin{equation}
  \sigma_{\hat{m}}^2 = \frac{1}{\frac{1}{\sigma_1^{\prime 2}} + \frac{1}{\sigma_2^{\prime 2}}} + \sigma_C^2\,,
\end{equation}
which is the sum in quadrature of the uncertainty of the weighted average (Eq.~(\ref{eq:wAvgErr})) and the
common uncertainty $\sigma_C$~\cite{Valassi:2013bga}.

In a more general case, we may have $n$ measurements $m_1,\cdots,m_n$
with a $n\times n$ covariance matrix $C_{ij}$.
The expected values for $m_1, \cdots, m_n$ are $M_1, \cdots, M_n$ and
may depend on some unknown parameters $\vec{\theta}=(\theta_1,\cdots,\theta_m)$.
For this case, the $\chi^2$ to be minimized is:
\begin{eqnarray}
  \chi^2 & = & \sum_{i,j=1}^n (m_i-M_i(\vec{\theta}))\,C_{ij}^{-1}\,(m_j-M_j(\vec{\theta})) \\
  & = & \left(\begin{array}{ccc} m_1 - M_1(\vec{\theta}) & \cdots &  m_n - M_n(\vec{\theta})\end{array} \right)\!\!
    \left(
    \begin{array}{ccc} C_{11} &  \cdots & C_{1n} \\
      \vdots & \ddots & \vdots \\
      C_{n1} & \cdots & C_{nn}
      \end{array}
      \right)^{\!\!\!-1}\!\!\!\!\!\left(
      \begin{array}{c}
        m_1 - M_1(\vec{\theta}) \\ \cdots \\ m_n - M_n(\vec{\theta})
      \end{array}
      \right).
\end{eqnarray}

An example of application of such a combination of measurement
is given by fit of the Standard Model parameters using the
electroweak precision measurements
at colliders~\cite{LEP-2, Baak:2014ora}.

\section{Hypothesis tests}
\label{sec:HypTest}
Hypothesis testing addresses the question whether some observed data sample
is more compatible with one theory model or another alternative one.

The terminology used in statistics may sometimes be not very natural for physics applications,
but it has become popular among physicists as well as long as more statistical methods
are becoming part of common practice. In a test, usually two hypotheses are considered:
\begin{itemize}
\item $H_0$, the {\it null hypothesis}.
  \\Example 1: {\it ``a sample contains only background''}.
  \\Example 2: {\it ``a particle is a pion''}.
\item $H_1$, the {\it alternative hypothesis}.
\\Example 1: {\it ``a sample contains background + signal''}.
\\Example 2: {\it ``a particle is a muon''}.
\end{itemize}
A {\it test statistic} is a variable computed from our data sample that discriminates between the two hypotheses
$H_0$ and $H_1$. Usually it is a  `summary' of the information available in the data sample.

In physics it's common to perform an event selection based on a discriminating variable $x$.
For instance, we can take as signal sample all events whose value of $x$ is above a
threshold, $x > x_{\mathrm{cut}}$. $x$ is an example of {\it test statistic} used to discriminate
between the two hypotheses, $H_1 =$~``signal'' and $H_2 =$~``background''.

The following quantities are useful to give quantitative information about a test:
\begin{itemize}
\item $\alpha$, the {\it significance level}: probability to reject $H_0$ if $H_0$ is assumed to be true (type I error, or false negative).
In physics $\alpha$ is equal to one minus the selection efficiency.
\item $\beta$, the {\it misidentification probability}, i.e.: probability to reject $H_1$ if $H_1$ is assumed to be true
  (type II error, or false negative). $1 - \beta$ is also called  {\it power of the test}.
\item a $p$-value is the  probability, assuming $H_0$ to be true, of getting a value of the test statistic as result
  of our test at least as extreme as the observed test statistic.
\end{itemize}

In case of multiple discriminating variables, a selection of a signal against a background
may be implemented in different ways. E.g.: applying a selection on each individual variable, or on a combination of
those variables, or selecting an area of the multivariate space which is enriched in signal events.

\subsection{The Neyman--Pearson lemma}
The Neyman--Pearson lemma~\cite{Neyman_Pearson} ensures that, for a fixed significance level
($\alpha$) or equivalently a signal efficiency ($1 - \alpha$),
the selection that gives the lowest possible misidentification probability ($\beta$) is based on a likelihood ratio:
\begin{equation}
  \lambda(x) = \frac{L(x|H_1)}{L(x|H_0)} > k_\alpha\,,
  \label{eq:neymanPearsonLemma}
\end{equation}
where $L(x|H_0)$ and $L(x|H_1)$ are the values of the likelihood functions for the two
considered hypotheses. $k_\alpha$ is a constant whose value depends on the fixed significance
level $\alpha$.

The likelihood function can't always be determined exactly.
In cases where it's not possible to determine the exact likelihood function,
other discriminators can be used as test statistics.
Neural Networks, Boosted Decision Trees and other machine-learning algorithms
are examples of discriminators that may closely approximate the performances of the exact likelihood
ratio, approaching the Neyman--Pearson optimal performances~\cite{Roe2005577}.

In general, algorithms that provide a test statistic for samples with multiple variables
are referred to as {\it multivariate discriminators}.
Simple mathematical algorithms exist, as well as complex implementations based on extensive CPU computations.
In general, the algorithms are `trained' using input samples whose nature is known ({\it training samples}),
i.e.: where either $H_0$ or $H_1$ is know to be true.
This is typically done using data samples simulated with computer algorithms (Monte Carlo)
or, when possible, with control samples obtained from data.
Among the most common problems that arise with training of multivariate algorithms,
the size of training samples is necessarily finite, hence the true distributions for the considered hypotheses can't be determined exactly form the training sample distribution. Moreover, the distribution assumed in the simulation of the input samples may not reproduce exactly the
true distribution of real data, for instance because of systematic errors that affect our simulation.

\subsection{Projective likelihood ratio}

In case of independent variables, the likelihood functions appearing in the numerator and
denominator of Eq.~(\ref{eq:neymanPearsonLemma}) can be factorized as product of
one-dimensional PDF (Eq.~(\ref{eq:indVar})). Even in the cases when variables are not
independent, this can be taken as an approximate evaluation of the Neyman--Pearson
likelihood ratio, so we can write:
\begin{equation}
  \lambda(x) = \frac{L(x_1,\cdots,x_n|H_1)}{L(x_1,\cdots,x_n|H_0)}
  \simeq
  \frac{\prod_{i=1}^n f_i(x_i|H_1)}{\prod_{i=1}^n f_i(x_i|H_0)}\,.
\end{equation}
The approximation may be improved if
a proper rotation is first applied to the input  variables in order to
eliminate their correlation. This approach is called {\it principal component analysis}.

\subsection{Fisher discriminant}
\begin{figure}[htbp]
\centering\includegraphics[width=.495\linewidth]{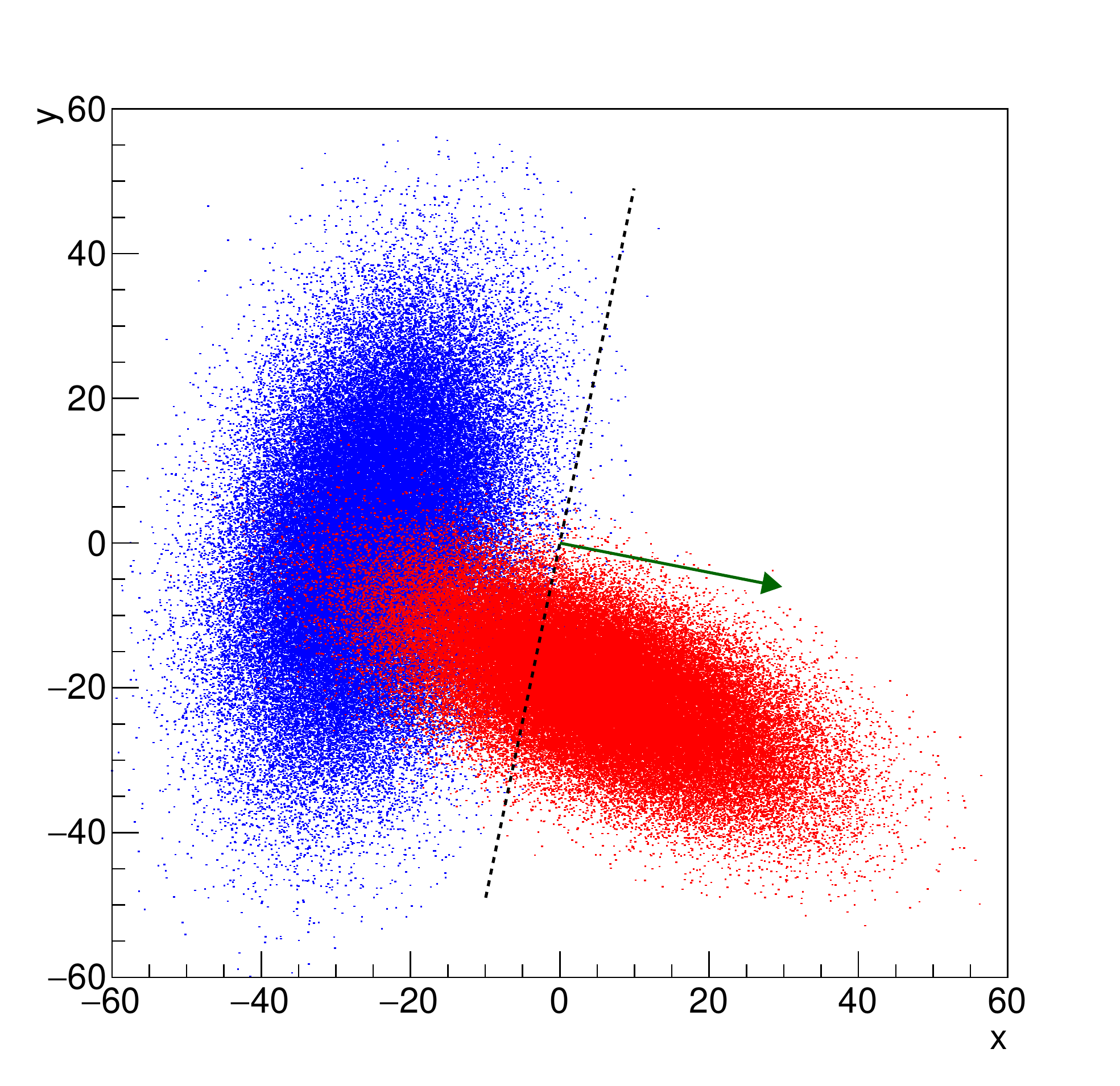}
\centering\includegraphics[width=.495\linewidth]{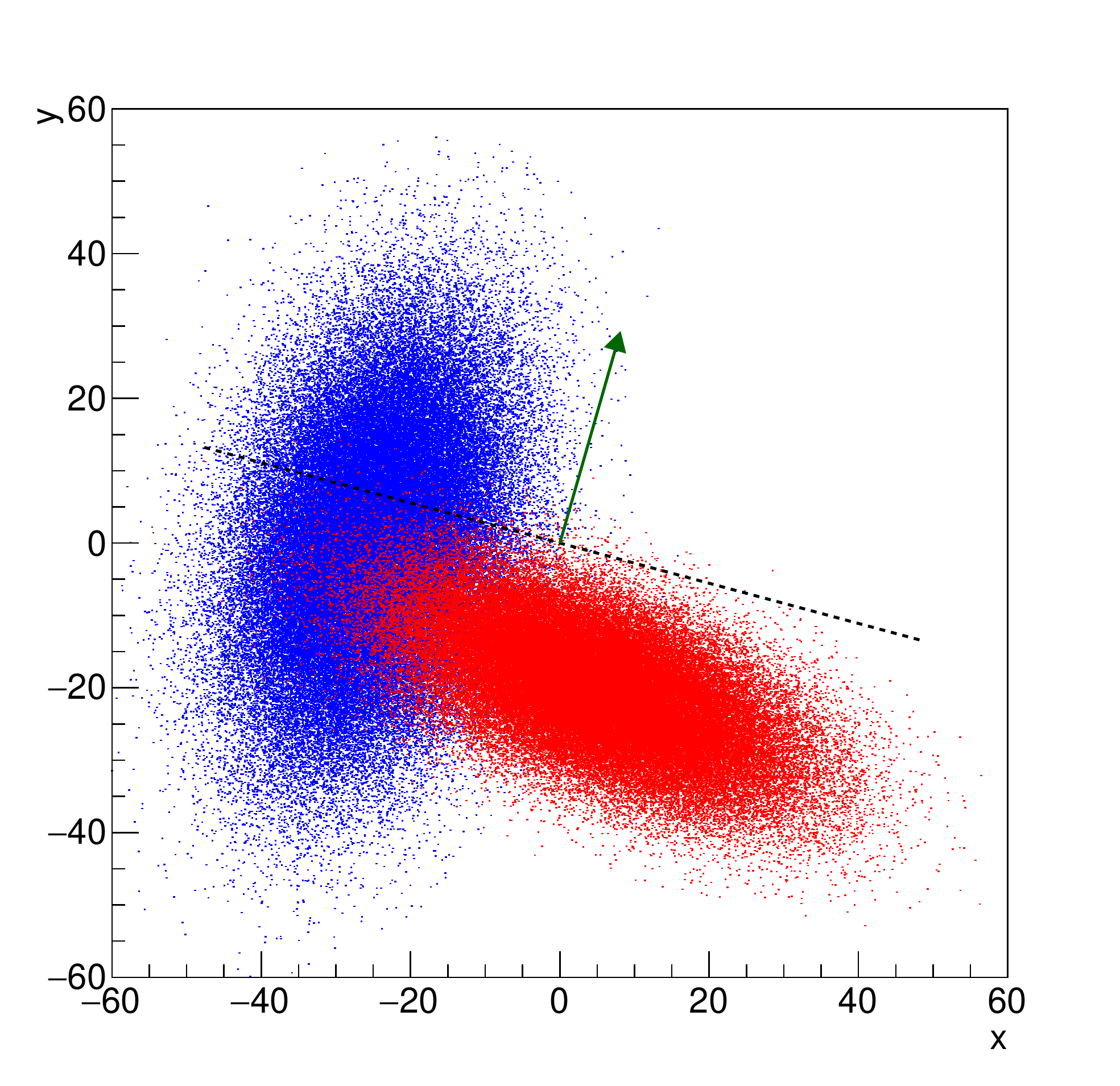}
\centering\includegraphics[width=.495\linewidth]{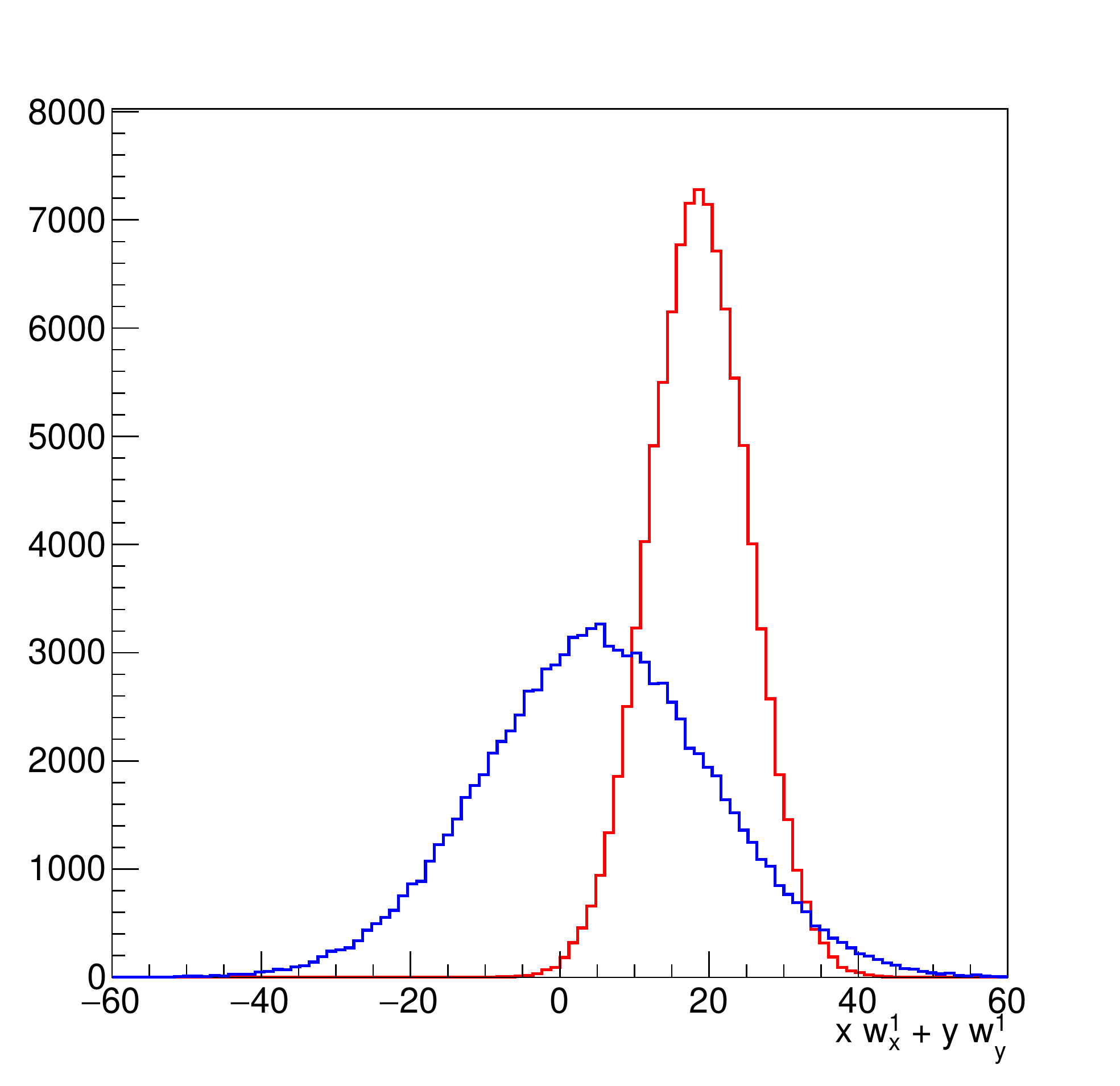}
\centering\includegraphics[width=.495\linewidth]{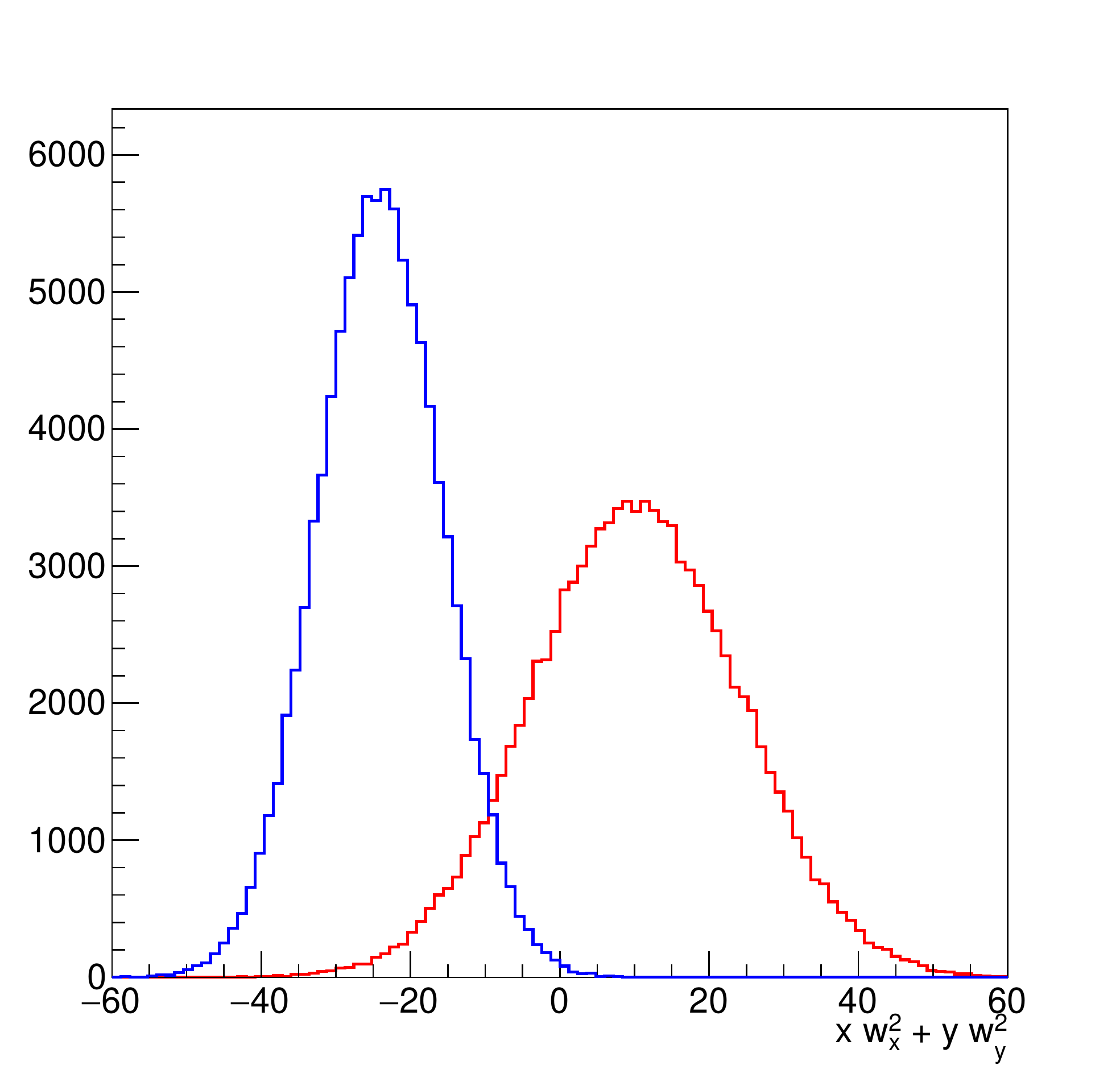}
\caption{Examples of Fisher projections. Two samples are distributed according
  to the red and blue distributions in two dimensions and  two possible projection direction $\mathbf{w}$ are shown
  as dotted line, the green arrows are perpendicular to them (top plots).
  The corresponding one-dimensional projections along the chosen direction show
  different overlap between the red and blue distribution (bottom plots), depending
  on the choice of the projection.
}
\label{fig:Fisher}
\end{figure}

Fisher~\cite{Fisher_discriminant} introduced a discriminator based on a linear combination of input variables
that maximizes the distance of the means of the two classes while minimizing the variance,
projected along a direction $\mathbf{w}$:
\begin{equation}
  J(\mathbf{w}) = \frac{|\mu_0-\mu_1|^2}{\sigma_0^2+\sigma_1^2}
  =\frac{\mathbf{w}^{\mathrm{T}}\cdot(\mathbf{m}_0 - \mathbf{m}_1)}
  {\mathbf{w}^{\mathrm{T}}(\mathbf{\Sigma}_0 + \mathbf{\Sigma}_1)\mathbf{w}}\,.
\end{equation}
The selection is achieved by requiring $J(\mathbf{w}) > J_{\mathrm{cut}}$, which determines an hyperplane
perpendicular to $\mathbf{w}$.
Examples of two different projections for a two-dimensional case is shown in Fig.~\ref{fig:Fisher}.
The problem of maximising $J(\mathbf{w})$ over all possible directions $\mathbf{w}$
can be solved analytically using linear algebra.

\subsection{Artificial Neural Network}
\begin{figure}[htbp]
\centering\includegraphics[width=.7\linewidth]{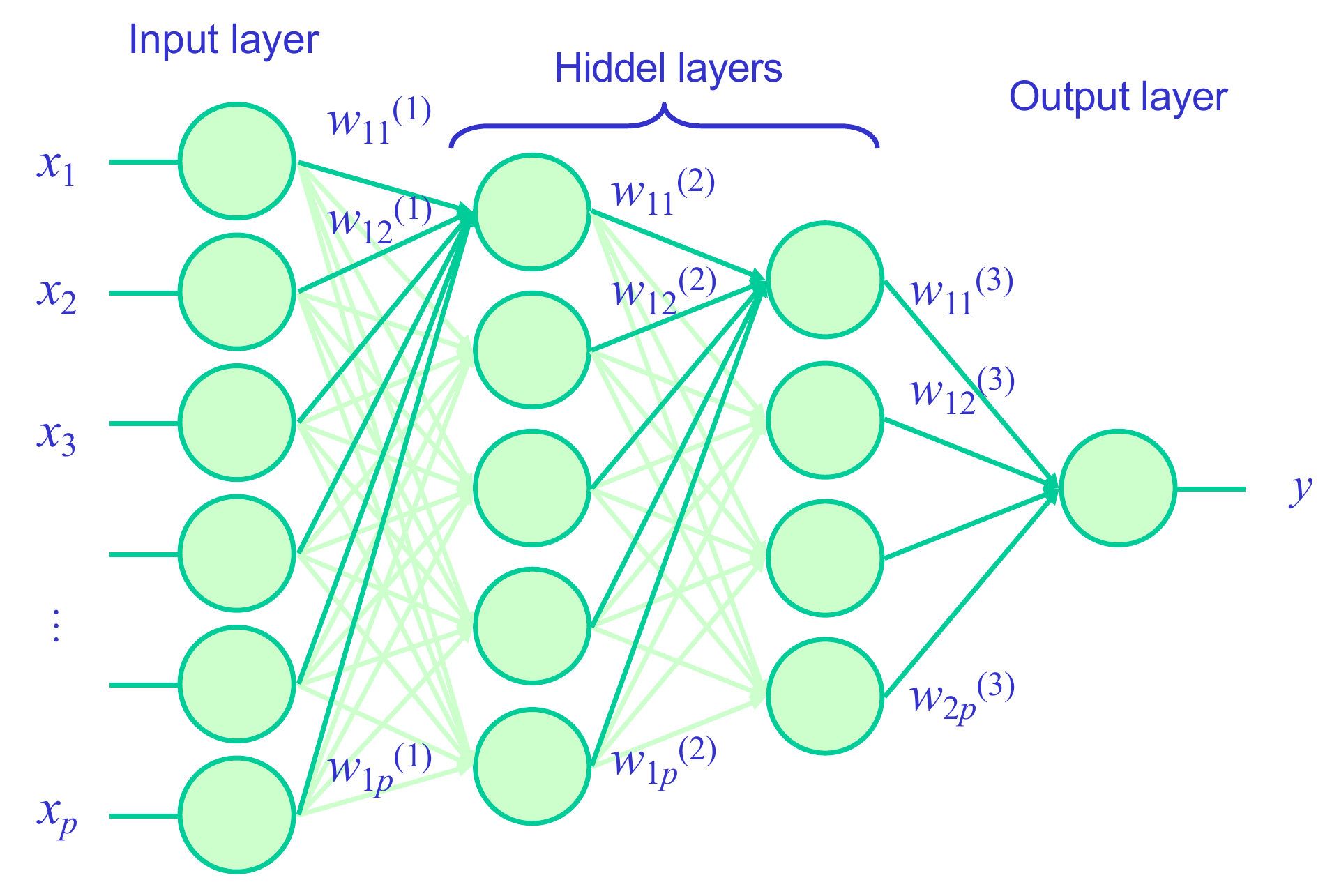}
\caption{Structure of an Artificial Neural Network.
}
\label{fig:ANN}
\end{figure}

Artificial Neural Networks (ANN)
are computer implementations of simplified models of how neuron cells work.
The schematic structure of an ANN is shown in Fig.~\ref{fig:ANN}.
Each node in the network receives inputs from either the input variables
(input layer) or from the previous layer, and provides an output
either of the entire network (output layer) or which is used as input to the next layer.
Within a node, inputs are combined linearly with proper weights
that are different for each of the nodes. Each output is then
transformed using a sigmoid function $\varphi$:
\begin{equation}
  y^{(n)}(\vec{x}) = \varphi\left(
  \sum_{j=1}^p w_{kj}^{(n)} x_j
  \right)\,,
\end{equation}
where $\varphi$ is typically:
\begin{equation}
  \varphi(\nu) = \frac{1}{1+e^{-\lambda\nu}}\,,
\end{equation}
so that the output values are bound between 0 and 1.

In order to find the optimal set of network weights $w_{ij}^{(n)}$, a minimization
is performed on the {\it loss function} defined as the following sum
over a training sample of size $N$:
\begin{equation}
  L(w) = \sum_{i=1}^N(y_i^{\mathrm{true}}-y(\vec{x}_i))^2\,,
\end{equation}
$y_i^{\mathrm{true}}$ being usually equal to 1 for signal ($H_1$) and 0 for background ($H_0$).
Iteratively, weights are modified ({\it back propagation}) for each training event (or each group
of training events) using the {\it stochastic gradient descent} technique:
\begin{equation}
  w_{ij} \rightarrow w_{ij} -\eta\frac{\partial L(w)}{\partial w_{ij}}\,.
\end{equation}
The parameter $\eta$ controls the learning rate of the network.
Variations of the training implementation exist.

Though it can be proven~\cite{ANNproof} that, under some regularity conditions,
neural networks with a single hidden layer can approximate any analytical function
with a sufficiently high number of neurons,
in practice this limit is hard to achieve.
Networks with several hidden layers can better manage complex variables combinations,
e.g.: exploiting invariant mass distributions features using only four-vectors as input~\cite{Baldi:2014kfa}.
Those complex implementation that were almost intractable in the past
can now be better approached  thanks to the availability of improved training algorithms
and more easily available CPU power.

\subsection{Boosted Decision Trees}

A {\it decision tree} is a sequence of simple cuts that are sequentially
applied on events in a data sample. Each cut splits the sample
into nodes that may be further split by the application of subsequent cuts.
Nodes where signal or background is largely dominant are classified as leafs.
Alternatively, the splitting may stop if too few events per node remain, or if the total number of nodes too high.
Each branch on the tree represents one sequence of cuts.
Cuts can be optimized in order to achieve the best split level.
One possible implementation is to maximize for each node the gain of Gini index after a splitting:
\begin{equation}
    G = P(1 - P)\,,
\end{equation}
where $P$ is the purity of the node (i.e.: the fraction of signal events).
$G$ is equal to zero for nodes containing only signal or background events.
Alternative metrics can be used (e.g.: the {\it cross entropy},
equal to: $-(P\ln P+(1-P)\ln(1-P))$ ) in place of the Gini index.

An optimized single decision tree does not usually provide optimal performances
or stability, hence multiple decision trees are usually combined.
Each tree is added iteratively after weights are applied to test events.
{\it Boosting} is achieved by
iteratively reweighting the events in the training sample according to the classifier
output in the previous iteration. The {\it boosted decision tree} (BDT) algorithm usually
proceeds as follows:
\begin{itemize}
\item Events are reweighted using the previous iteration's classifier result.
\item A new tree is build and optimized using the reweighted events as training sample.
\item A score is given to each tree.
\item The final BDT classifier result is a weighted average over all trees:
  \begin{equation}
    y(\vec{x}) = \sum_{k=1}^{N_{\mathrm{trees}}}
      w_iC^{(i)}(\vec{x})\,.
  \end{equation}
\end{itemize}
One of the most popular algorithm is the {\it adaptive boosting}~\cite{AdaBoost}:
misclassified events only are reweighted according to the fraction of classification
error of the previous tree:
\begin{equation}
  \frac{1-f}{f}\,,f=\frac{N_{\mathrm{misclassified}}}{N_{\mathrm{tot}}}\,.
\end{equation}
The weights applied to each tree are also related to the misclassification fraction:
\begin{equation}
  y(\vec{x}) = \sum_{k=1}^{N_{\mathrm{trees}}}\ln\left(\frac{1-f^{(i)}}{f^{(i)}}\right)C^{(i)}(\vec{x})\,.
\end{equation}
This algorithm enhances the weight of events misclassified on the previous iteration
in order to improve the performance on those events.
Further variations and more algorithms are available.

\subsection{Overtraining}

Algorithms may learn too much from the training sample, exploiting features that are
only due to random fluctuations.
It may be important to check for overtraining comparing the discriminator's distributions
for the training sample and for an independent {\it test sample}:
compatible distributions will be an indication that no overtraining occurred.

\section{Discoveries and upper limits}

The process towards a discovery, from the point of view of data analysis,
proceeds starting with a test of our data sample against two hypotheses concerning the theoretical underlying model:
\begin{itemize}
\item $H_0$: the data are described by a model that contains background only;
\item $H_1$: the data are described by a model that contains a new signal plus background.
\end{itemize}
The discrimination between the two hypotheses can be based on a test statistic $\lambda$ whose distribution
is known under the two considered hypotheses.
We may assume that $\lambda$ tends to have (conventionally) large values if $H_1$ is true and small values if $H_0$ is true.
This convention is consistent with using as test statistic the likelihood ratio $\lambda =L(x|H_1)/L(x|H_0)$,
as in the Neyman--Pearson lemma (Eq.~(\ref{eq:neymanPearsonLemma})).
Under the frequentist approach, it's possible to compute a $p$-value equal to the probability that
$\lambda$ is greater or equal to than the value $\lambda^{\mathrm{obs}}$ observed in data.
Such $p$-value is usually converted into an equivalent probability computed as the area
under the rightmost tail of a standard normal distribution:
\begin{equation}
  p = \int_Z^{+\infty} \frac{1}{\sqrt{2\pi}}e^{-{x^2}/{2}}\,\mathrm{d}x = 1 - \Phi(Z)\,,
  \label{eq:significance}
\end{equation}
where $\Phi$ is the cumulative (Eq.~(\ref{eq:cumulative})) of a standard normal distribution.
$Z$ in Eq.~(\ref{eq:significance}) is called {\it significance level}.
In literature conventionally a signal with a significance of at least 3 ($3\sigma$ {\it level})
is claimed as {\it evidence}. It corresponds to a $p$-value of 
$1.35\times 10^{-3}$ or less. If the significance exceeds 5 ($5\sigma$ {\it level}), i.e.: the
$p$-value is below $2.9\times10^{-7}$, one is allowed to claim the {\it observation} of the new signal.
It's worth noting that the probability that background produces a large test statistic is not equal to
the probability of the null hypothesis (background only), which has only a Bayesian sense.

Finding a large significance level, anyway, is only part of the discovery process in the
context of the scientific method. Below a sentence is reported from a recent statement of the
American Statistical Association:
\begin{displayquote}
{\it The p-value was never intended to be a substitute for scientific reasoning. Well-reasoned statistical arguments contain much more than the value of a single number and whether that number exceeds an arbitrary threshold. The ASA statement is intended to steer research into a `post p < 0.05 era'}~\cite{pvalASA}.
\end{displayquote}
This was also remarked by the physicists community, for instance by Cowan {\it et al.}:
\begin{displayquote}
\textit{It should be emphasized that in an actual scientific context, rejecting the background-only hypothesis in a statistical sense is only part of discovering a new phenomenon. One's \textbf{degree of belief} that a new process is present will depend in general on other factors as well, such as the plausibility of the new signal hypothesis and the degree to which it can describe the data}~\cite{asymptotic}.
\end{displayquote}

\subsection{Upper limits}

Upper limits measure the amount of excluded region in the theory's parameter space
resulting from our negative results of a search for a new signal.
Upper limits are obtained by building a fully asymmetric Neyman confidence belt (Sec.~\ref{sec:NeymanBelt})
based on the considered test statistic (Fig.~\ref{fig:BayesIntHiLo}). The belt can be inverted
in order to find the allowed fully asymmetric interval for the signal yield $s$.
The upper limit $s^{\mathrm{up}}$ is the upper extreme of the asymmetric confidence interval $[0, s^{\mathrm{up}}]$.
In case the considered test statistic is a discrete variable (e.g.: the number of selected events $n$ in
a counting experiments), the coverage may not be exact, as already discussed in Sec~\ref{sec:binInt}.

\begin{figure}[htbp]
\centering\includegraphics[width=.9\linewidth]{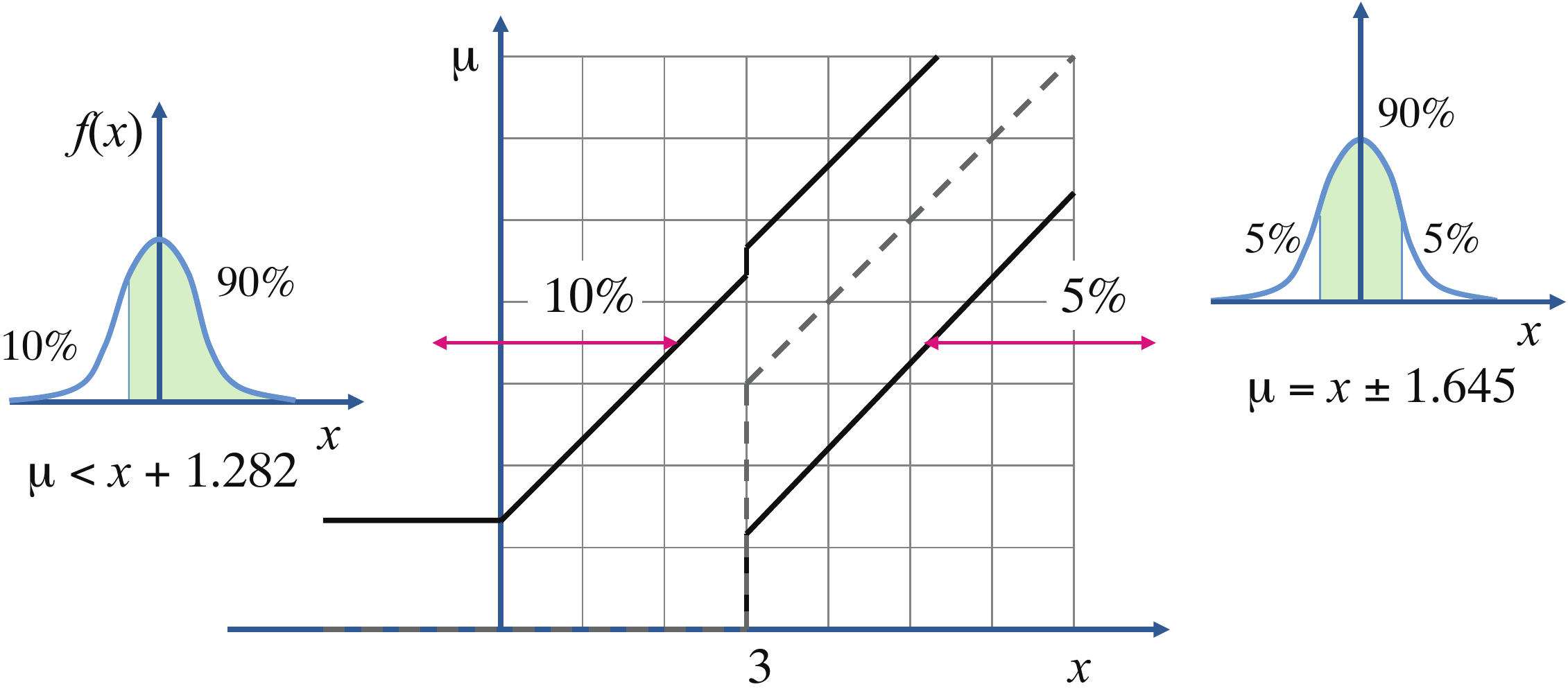}
\caption{Illustration of flip-flopping with a Gaussian confidence belt. This figure is taken from Ref.~\cite{LL}.}
\label{fig:flipFlop}
\end{figure}

The procedure described above to determine upper limits, anyway, may incur the so-called {\it flip-flopping} issue~\cite{feldman_cousins}:
given an observed result of our test statistic, when should we quote a central value or upper limit?
A choice that is sometimes popular in scientific literature is to quote a (90\% or 95\% CL) upper limit
if the significance is below $3\sigma$ or to quote a central value if the significance is
at least $3\sigma$. The choice to ``flip'' from an upper limit to a central value
can be demonstrated to produce an incorrect coverage. This can be visually seen in
Fig.~\ref{fig:flipFlop}, where a Gaussian belt at 90\% CL is constructed, similarly
to Fig.~\ref{fig:NeymanGaussianBelt} (where instead a 68.3\% CL was used).
In Fig.~\ref{fig:flipFlop}, for $x<3$, anyway, the belt is modified because
those values correspond to a significance below the $3\sigma$ level, where an
upper limit (a fully asymmetric confidence interval) was chosen. As shown with the
red arrows in the figures, there are intervals in $x$ that, in this way, have
a coverage reduced to 85\% instead of the required 90\%.

\subsection{Feldman--Cousins intervals}

A solution to the flip-flopping problem was developed by Feldman and Cousins~\cite{feldman_cousins}.
They proposed to select confidence interval based on a likelihood-ratio criterion.
Given a value $\theta=\theta_0$ of the parameter of interest, the
Feldman--Cousins confidence interval is defined as:
\begin{equation}
  R_\mu = \left\{x : \frac{L(x;\theta_0)}{L(x;\hat{\theta})} >k_\alpha\right\}\,,
\end{equation}
where $\hat{\theta}$ is the best-fit value for $\theta$ and the constant $k_\alpha$
should be set in order to ensure the desired confidence level $1-\alpha$.

\begin{figure}[htbp]
\centering\includegraphics[width=.55\linewidth]{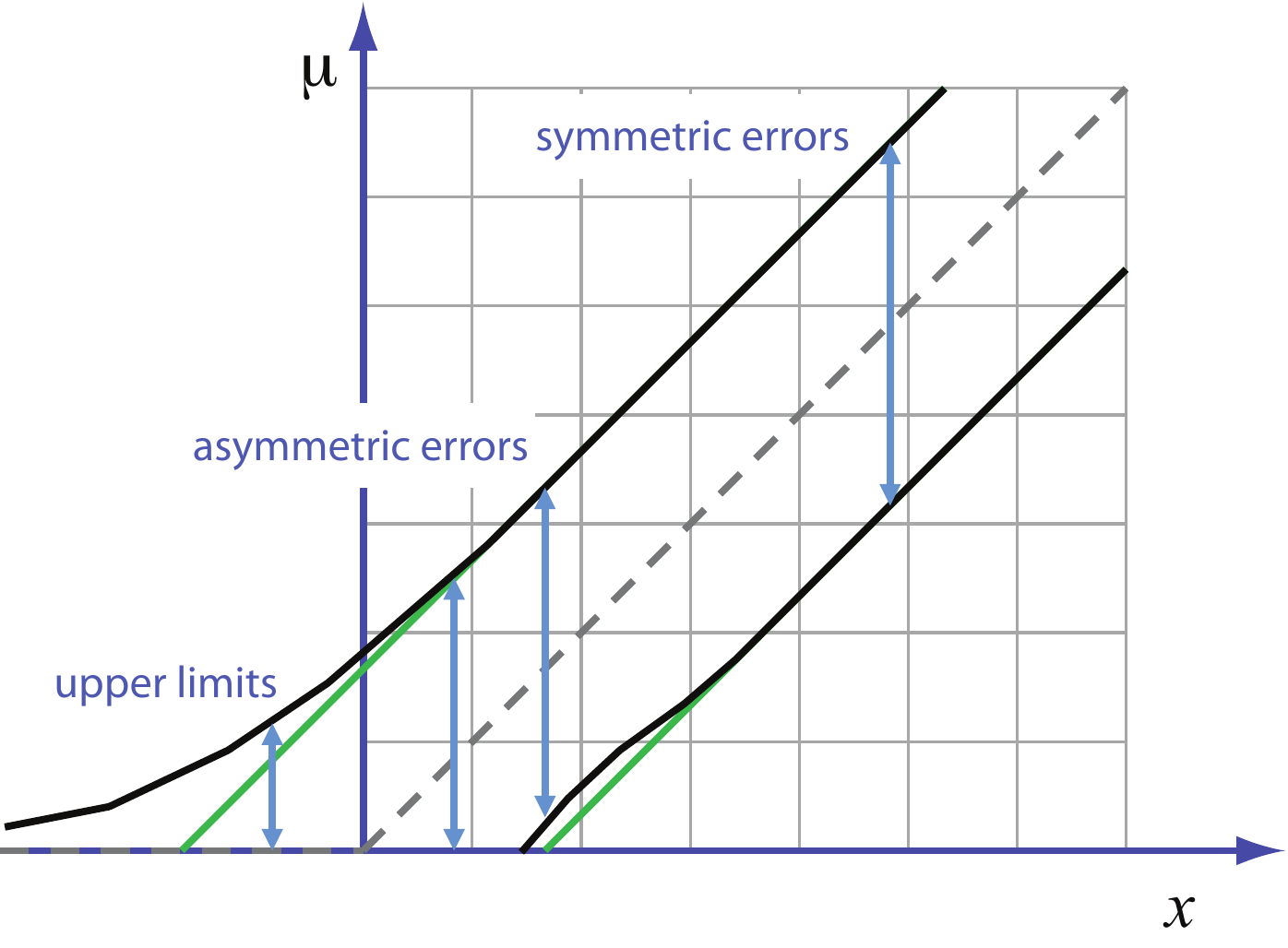}
\caption{Feldman--Cousins belt for a Gaussian case. This figure is taken from Ref.~\cite{LL}.}
\label{fig:fcBelt}
\end{figure}

An example of the confidence belt computed with the Feldman--Cousins approach
is shown in Fig.~\ref{fig:fcBelt}
for the Gaussian case illustrated in Fig.~\ref{fig:flipFlop}.
With the Feldman--Cousins approach, the confidence interval smoothly changes
from a fully asymmetric one, which leads to an upper limit, for low values of $x$, to
an asymmetric interval for higher values of $x$ interval, then finally a symmetric interval
(to a very good approximation) is obtained for large values of $x$, recovering
the usual result as in Fig.~\ref{fig:flipFlop}.

Even for the simplest Gaussian case, the computation of Feldman--Cousins intervals
requires numerical treatment and for complex cases their computation may be very CPU intensive.

\subsection{Upper limits for event counting experiments}

The simplest search for a new signal consists of counting the number of events
passing a specified selection.
The number of selected events $n$ is distributed according to a Poissonian distribution
where the expected value, in case of presence of signal plus background ($H_1$)
is $s + b$, and for background only ($H_0$) is $b$.
Assume we count $n$ events, we then want to compare the two hypotheses $H_1$ and $H_0$.
As simplest case, we can assume that  $b$ is known with negligible uncertainty.
If not, uncertainty on its estimate must be taken into account.

The likelihood function for this case can be written as:
\begin{equation}
  L(n;s) = \frac{(s+b)^n}{n!}e^{-(s+b)}\,.
\end{equation}
$H_0$ corresponds to the case $s=0$.

Using the Bayesian approach, an upper limit
$s^{\mathrm{up}}$ on $s$ can be determined by requiring that the posterior probability
corresponding to the interval $[0,s^{\mathrm{up}}]$ is equal to the confidence
level $1-\alpha$:
\begin{equation}
  1-\alpha = \int_0^{s^{\mathrm{up}}} P(s|n)\,\mathrm{d}s =
  \frac{
    \displaystyle\int_0^{s\mathrm{up}} L(n;a)\pi(s)\,\mathrm{d}s
  }{
    \displaystyle\int_0^{+\infty} L(n;a)\pi(s)\,\mathrm{d}s
  }\,.
\label{eq:poisBayesUL}
\end{equation}
The choice of a uniform prior, $\pi(s)=1$, simplifies the computation and
Eq.~(\ref{eq:poisBayesUL}) reduces to~\cite{Helene}:
\begin{equation}
  \alpha = e^{-s^{\mathrm{up}}}\frac{
    \displaystyle\sum_{m=0}^n\frac{(s^{\mathrm{up}}+b)^m}{m!}
  }{
    \displaystyle\sum_{m=0}^n\frac{b^m}{m!}
  }\,.
  \label{eq:Helene}
\end{equation}
\begin{figure}[htbp]
\centering\includegraphics[width=.495\linewidth]{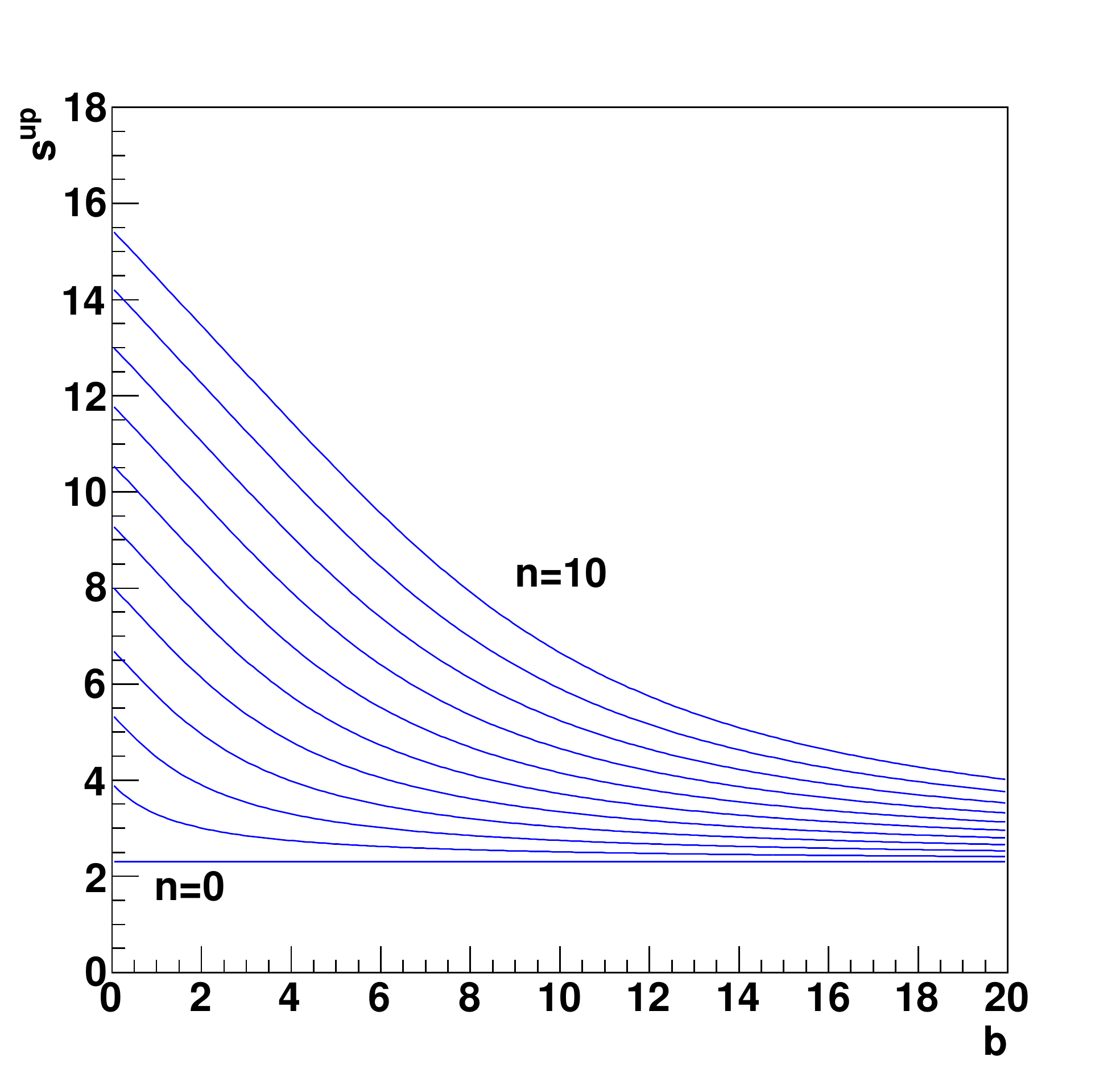}
\centering\includegraphics[width=.495\linewidth]{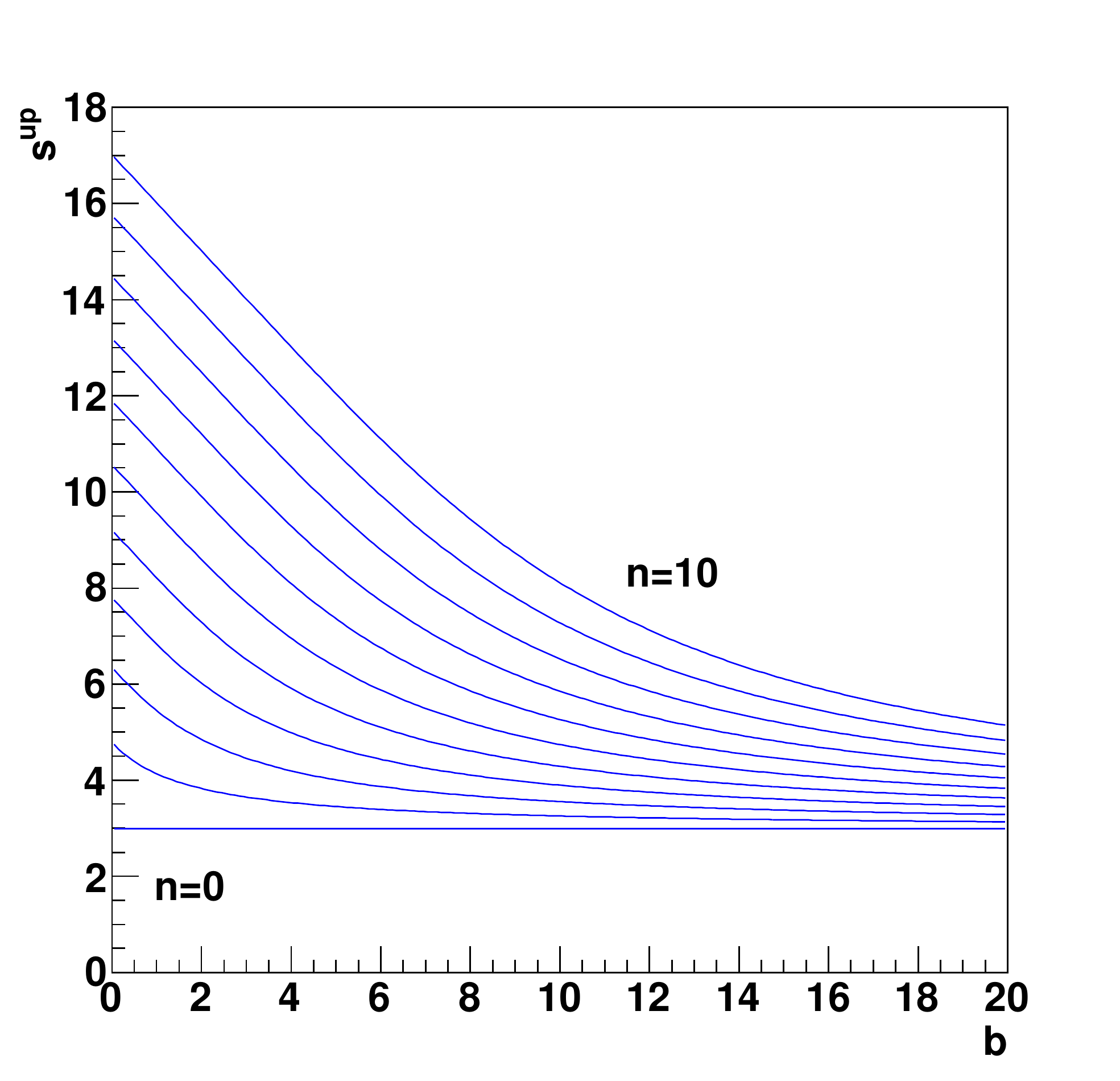}
\caption{Bayesian upper limits at 90\% (left) and 95\% (right) CL for the expected signal
  yield $s$ for a counting experiment with different level of expected background $b$.
  This figure is taken from Ref.~\cite{LL}.
}
\label{fig:Helene}
\end{figure}
Upper limits obtained with Eq.~(\ref{eq:Helene}) are shown in Fig.~\ref{fig:Helene}.
In the case $b=0$, the results obtained in Eq.~(\ref{eq:BayesianPoissonUL90}) and~(\ref{eq:BayesianPoissonUL95}) are
again recovered.

Frequentist upper limits for a counting experiment can be easily computed
in case of negligible background ($b=0$). If zero events are observed ($n=0$),
the likelihood function simplifies to:
\begin{equation}
  L(n=0;s) = \mathrm{Poiss}(0;s) = e^{-s}\,.
\end{equation}
The inversion of the fully asymmetric Neyman belt reduces to:
\begin{equation}
  P(n\le0;s^{\mathrm{up}}) = P(n=0;s^{\mathrm{up}}) = \alpha \implies s^{\mathrm{up}} = -\ln\alpha\,,
\end{equation}
which lead to results that are numerically identical to the Bayesian computation:
\begin{eqnarray}
  s & < & s^{\mathrm{up}} = 2.303\quad\text{for}\quad\alpha=0.1\,\text{(90\% CL)}\,,\label{eq:FreqPoissonUL90}\\
  s & < & s^{\mathrm{up}} = 2.996\quad\text{for}\quad\alpha=0.05\,\text{(95\% CL)}\,.\label{eq:FreqPoissonUL95}
\end{eqnarray}
In spite of the numerical coincidence, the interpretation of frequentist and Bayesian upper limits remain very different.

\begin{figure}[htbp]
\centering\includegraphics[width=.495\linewidth]{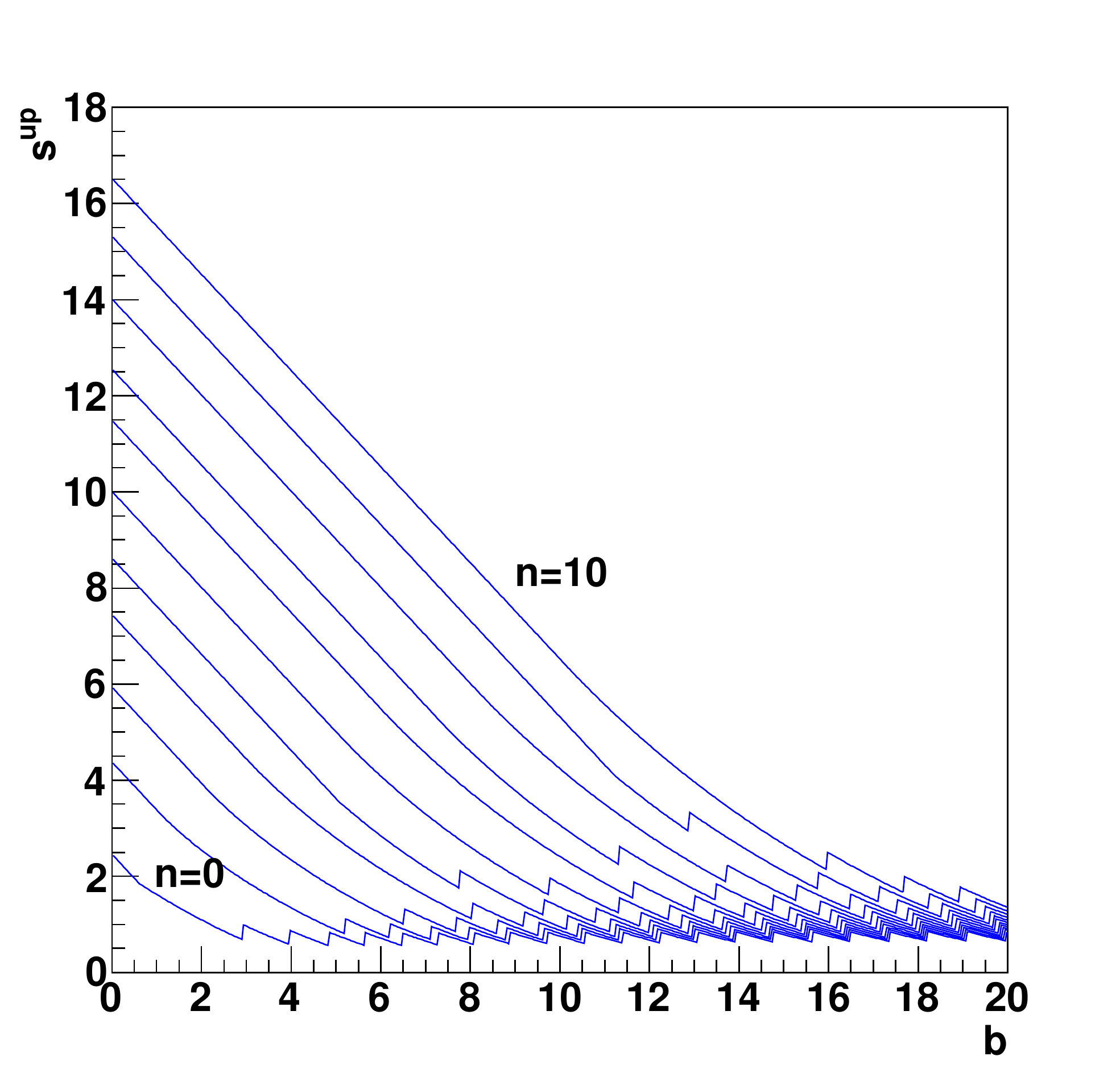}
\caption{Feldman--Cousins upper limits at 90\% CL for a counting experiment.
  This figure is taken from Ref.~\cite{LL}.}
\label{fig:fcUL}
\end{figure}

Upper limits from Eq.~(\ref{eq:FreqPoissonUL90}) and~(\ref{eq:FreqPoissonUL95})
anyway suffer from the flip-flopping problem and the coverage is spoiled when deciding to switch
from an upper limit to a central value depending on the observed significance level.
Feldman--Cousins intervals cure the flip-flopping issue and ensure the correct coverage
(or may overcover for discrete variables).
Upper limits at 90\% computed with the Feldman--Cousins approach for a counting experiment
are reported in Fig.~\ref{fig:fcUL}.
The ``ripple'' structure is due to the discrete nature of Poissonian counting.
It's evident from the figure that, even for $n = 0$, the upper limit decrease as $b$ increases (apart from ripple effects).
This means that if two experiment are designed for an expected background of --say-- 0.1 and 0.01, the
``worse'' experiment (i.e.: the one which expects 0.1 events) achieves the best upper limit in case
no event is observed ($n=0$), which is the most likely outcome if no signal is present.
This feature was noted in the 2001 edition of the PDG~\cite{PDG2001}
\begin{displayquote}
  {\it The intervals constructed according to the unified procedure [Feldman--Cousins] for a Poisson variable $n$ consisting
    of signal and background have the property that for $n = 0$ observed events, the upper limit decreases for increasing
    expected background. This is counter-intuitive, since it is known that if $n = 0$ for the experiment in question,
    then no background was observed, and therefore one may argue that the expected background should not be relevant.
    The extent to which one should regard this feature as a drawback is a subject of some controversy.}
\end{displayquote}
This counter-intuitive feature of frequentist upper limits is one of the reasons that led to the use in High-Energy Physics of
a modified approach, whose main feature is that is also prevents rejecting cases where the experiment has little sensitivity
due to statistical fluctuation, as will be described in next Section.

\subsection{The modified frequentist approach}

A {\it modified frequentist approach}~\cite{CLs} was proposed for the first time for the
combination of the results of searches for the Higgs boson by the four LEP experiments, ALEPH, DELPHI, L3 and OPAL~\cite{Higgs_at_LEP}.
Given a test statistic $\lambda(x)$ that depends on some observation $x$, its distribution should be determined
under the two hypotheses
$H_1$ (signal plus background) and $H_0$ (background only). The following $p$-values can be used,
where we assume that the test statistic $\lambda$ tends to have small values for $H_1$ and
larger values for $H_0$:
\begin{eqnarray}
  p_{s+b}& = & P(\lambda(x | H_1) \ge \lambda^{\mathrm{obs}} )\,,\label{eq:CLSpsb}\\
  p_b & = & P(\lambda(x | H_0) \le \lambda^{\mathrm{obs}} )\,.\label{eq:CLSpb}
\end{eqnarray}
$p_{s+b}$ and $p_b$ can be interpreted as follows:
\begin{itemize}
\item $p_{s+b}$ is the probability to obtain a result which is less compatible with the signal than the observed result, assuming the signal hypothesis;
\item $p_b$ is the probability to obtain a result less compatible with the background-only hypothesis than the observed one, assuming background only.
\end{itemize}
Instead of requiring, as for a frequentist upper limit, $p_{s+b} \le \alpha$,
the modified approach introduces a new quantity, $\mathrm{CL}_s$, defined as:
\begin{equation}
  \boxed{
    \mathrm{CL}_s = \frac{p_{s+b}}{1-p_b}\,,
    }
\end{equation}
and the upper limit is set by requiring $\mathrm{CL}_s \le \alpha$.
For this reason, the modified frequentist approach is also called {\it ``$\mathrm{CL}_s$ method''}.

In practice, in most of the realistic cases, $p_b$  and $p_{s+b}$ are computed from
simulated pseudoexperiments ({\it toy Monte Carlo}) by approximating the probabilities
defined in  Eq.~(\ref{eq:CLSpsb},~\ref{eq:CLSpb}) with the fraction of the total number of pseudoexperiments
satisfying their respective condition:
\begin{equation}
  \mathrm{CL}_s = \frac{p_{s+b}}{1-p_b} = \frac{N(\lambda_{s+b}\ge\lambda^{\mathrm{obs}})}{N(\lambda_{b}\ge\lambda^{\mathrm{obs}})}\,.
\end{equation}

Since $1-p_b \le 1$, then $\mathrm{CL}s \ge p_{s+b}$, hence upper limits computed with the $\mathrm{CL}_s$ method are
always {\it conservative}.

\begin{figure}[htbp]
\centering\includegraphics[width=.495\linewidth]{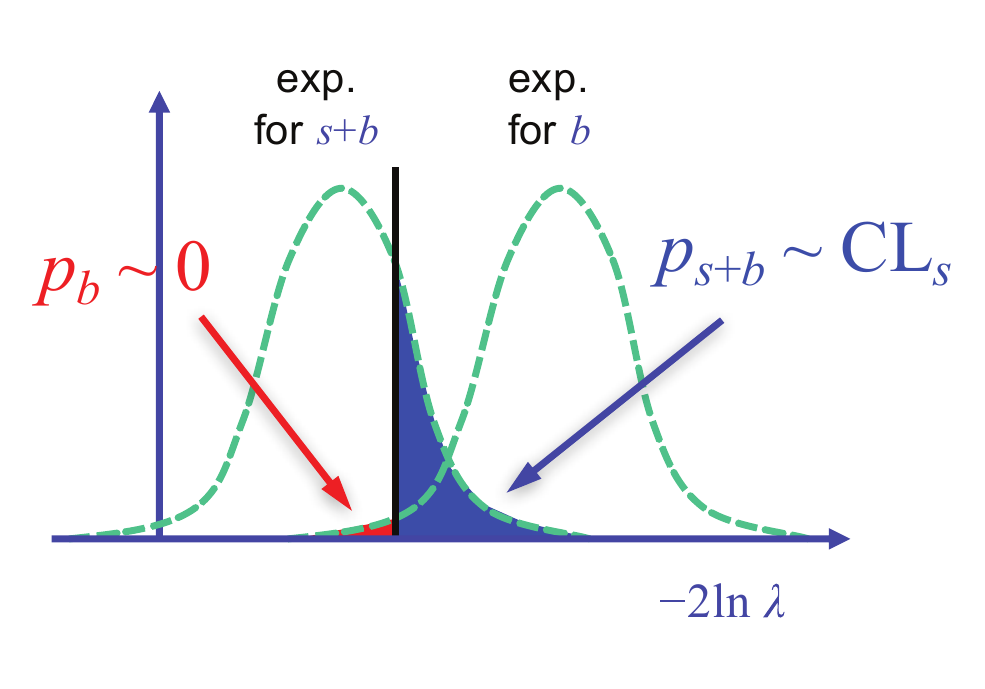}
\centering\includegraphics[width=.495\linewidth]{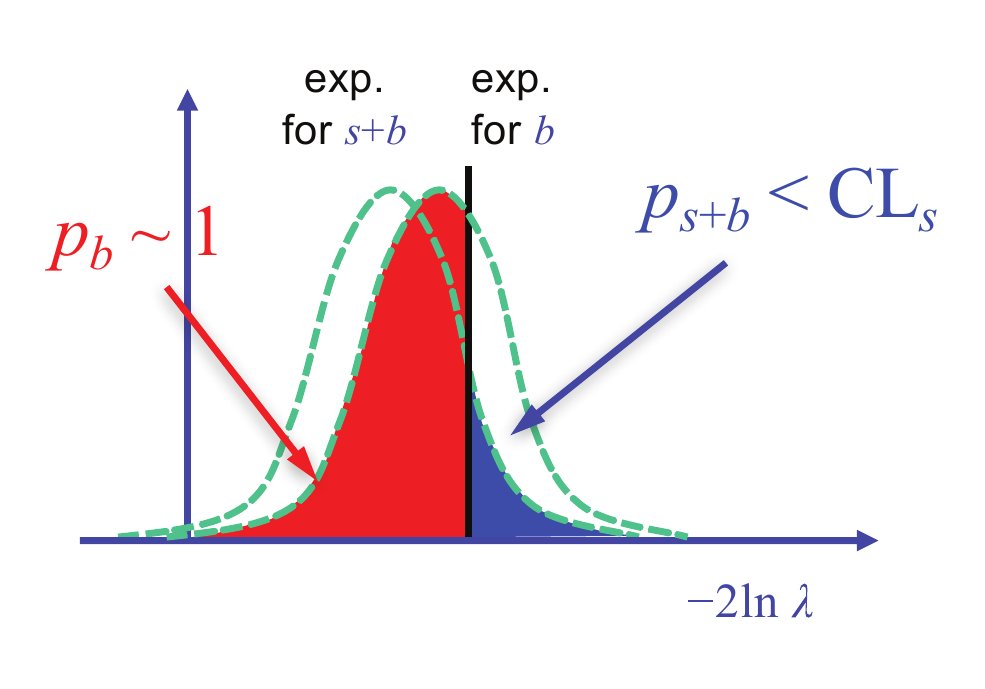}
\caption{Illustration of the application of the $\mathrm{CL}s$ method
  in cases of well separated distributions of the test statistic $-2\ln\lambda$
  for the $s+b$ and $b$ hypotheses (left) and in case of largely
  overlapping distributions (right).
}
\label{fig:CLs12}
\end{figure}

In case the distributions of the test statistic $\lambda$ (or equivalently $-2\ln\lambda$) for the two hypotheses $H_0$ and $H_1$
are well separated (Fig.~\ref{fig:CLs12}, left),
if $H_1$ is true, than $p_b$ will have a very high chance to be very small, hence $1-p_b \simeq 1$ and $\mathrm{CL}_s \simeq p_{s+b}$. In this case
$\mathrm{CL}_s$ and the purely frequentist upper limits coincide.
If the two distributions instead largely overlap (Fig.~\ref{fig:CLs12}, right), indicating that the experiment has poor sensitivity on the
signal, in case $p_b$ is large, because of a statistical fluctuation, then $1 - p_b$ becomes small.
This prevents $\mathrm{CL}_s$ to become too small,
i.e.: it prevents to reject cases where the experiment has little sensitivity.

If we apply the $\mathrm{CL}_s$ method to the previous counting experiment, using 
the observed number of events $n^{\mathrm{obs}}$ as test statistic,
then $\mathrm{CL}s$ can be written, considering that $n$ tends to be large in case of
$H_1$, for this case, as:
\begin{equation}
  \mathrm{CL}_s = \frac{P(n\le n^{\mathrm{obs}} | s+b)}{P(n \le n^{\mathrm{obs}} |b)}\,.
\end{equation}
Explicitating the Poisson distribution, the computation gives the same result as for the Bayesian case with a uniform prior
(Eq.~(\ref{eq:Helene})). In many cases, the $\mathrm{CL}_s$ upper 
limits give results that are very close, numerically, to Bayesian
computations performed assuming a uniform prior.
Of course, this does not allow to interpret $\mathrm{CL}_s$ upper limits
as Bayesian upper limits.
Concerning the interpretation of $\mathrm{CL}_s$, it's worth reporting from Ref~\cite{CLs} the
following statements:
\begin{displayquote}
{\it A specific modification of a purely classical statistical analysis is used to avoid excluding or discovering signals which the search is in fact not sensitive to.}
\end{displayquote}
\begin{displayquote}
  {\it The use of\,  $\mathrm{CL}_s$ is a conscious decision not to insist on the frequentist concept of full coverage (to guarantee that the confidence interval doesn't include the true value of the parameter in a fixed fraction of experiments).}
\end{displayquote}
\begin{displayquote}
{\it Confidence intervals obtained in this manner do not have the same interpretation as traditional frequentist confidence intervals nor as Bayesian credible intervals.}
\end{displayquote}

\subsection{Treatment of nuisance parameters}

Nuisance parameters have been introduced in Sec.~\ref{sec:extLikFun}.
Usually, signal extraction procedures (either parameter fits or upper limits determinations) determine,
together with parameters of interest, also nuisance parameters that model effects
not strictly related to our final measurement, like
background yield and shape, detector resolution, etc.
Nuisance parameters are also used to model sources of systematic
uncertainties.
Often, the true value of a nuisance parameter is not known, but we may have some
estimate from sources that are external to our problem.
In those cases, we can refer to {\it nominal values} of the nuisance parameter and
their uncertainty. Nominal values of nuisance parameters are random variables
distributed according to some PDF that depend on their true value.

A Gaussian distribution is the simplest assumption for nominal values of nuisance parameters.
Anyway, this may give negative values corresponding to the
leftmost tail, which are not suitable for
non-negative quantities like cross sections.
For instance, we may have an estimate of some background yield $b$ given by:
\begin{equation}
b = \beta \sigma_b L_{\mathrm{int}}\,,
\end{equation}
where $L_{\mathrm{int}}$ is the estimate of the integrated luminosity (assumed to be known
with negligible uncertainty), $\sigma_b$ is the background cross section evaluated
from theory, and $\beta$ is a nuisance parameter, whose nominal value is equal to one, representing the
uncertainty on the cross-section evaluation. If the uncertainty on $\beta$ is large, one may
have a negative value of $\beta$ with non-negligible probability, hence an unphysical negative value of
the background yield $b$.
A safer assumption in such cases is to take a log normal distribution for the uncertain non-negative
quantities:
\begin{equation}
b = e^\beta \sigma_b L_{\mathrm{int}}\,,
\end{equation}
where $\beta$ is again distributed according to a normal distribution with nominal value equal to zero,
in this case.

Under the Bayesian approach, nuisance parameters don't require a special treatment.
If we have a parameter of interest $\mu$ and a nuisance parameter $\theta$,
a Bayesian posterior will be obtained as (Eq.~(\ref{eq:BayesianInferenceSimple})):
\begin{equation}
  P(\mu,\theta|\vec{x}) = \frac{
    L(\vec{x};\mu,\theta)\pi(\mu,\theta)
  }{
    \int L(\vec{x};\mu^\prime,\theta^\prime)\pi(\mu^\prime,\theta^\prime)\,\mathrm{d}\mu^\prime\,\mathrm{d}\theta^\prime
  }\,.
\end{equation}
$P(\mu|\vec{x})$ can be obtained as marginal PDF of $\mu$ by integrating $P(\mu,\theta|\vec{x})$ over $\theta$:
\begin{equation}
  P(\mu|\vec{x}) = \int P(\mu,\theta|\vec{x}),\mathrm{d}\theta =
\frac{
   \int L(\vec{x};\mu,\theta)\pi(\mu,\theta)\,\mathrm{d}\theta
  }{
    \int L(\vec{x};\mu^\prime,\theta)\pi(\mu^\prime,\theta)\,\mathrm{d}\mu^\prime\,\mathrm{d}\theta
}\,.
\label{eq:BayesianNuisance}
\end{equation}

In the frequentist approach, one possibility is to introduce in the likelihood
function a model for a data sample that can constrain the nuisance parameter.
Ideally, we may have a control sample $\vec{y}$, complementary to the main
data sample $\vec{x}$, that only depends on the nuisance parameter, and
we can write a global likelihood function as:
\begin{equation}
  L(\vec{x},\vec{y};\mu,\theta) = L_x(\vec{x};\mu, \theta) L_y(\vec{y};\theta)\,.
\end{equation}
Using control regions to constrain nuisance parameters is usually a good method
to reduce systematic uncertainties. Anyway, it may not always be
feasible and in many cases we may just have information abut the nominal
value $\theta^{\mathrm{nom}}$ of $\theta$ and its distribution obtained from a complementary measurement:
\begin{equation}
  L(\vec{x},\vec{y};\mu,\theta) = L_x(\vec{x};\mu, \theta) L_\theta(\theta^{\mathrm{nom}};\theta)\,.
\end{equation}
$L_\theta$ may be a Gaussian or log normal distribution in the easiest cases.

In order to achieve a likelihood function that does not depend on
nuisance parameters, for many measurements at LEP or Tevatron a method
proposed by Cousins and Highland was adopted~\cite{Cousins_Highlands}
which consists of integrating the likelihood function over the nuisance
parameters, similarly to what is done in the Bayesian approach (Eq.~(\ref{eq:BayesianNuisance})).
For this reason, this method was also called hybrid.
Anyway the Cousins--Highland does not guarantee to provide exact coverage,
and was often used as pragmatic solution in the frequentist context.

\subsection{Profile likelihood}
\label{sec:profLik}
Most of the recent searches at LHC use the so-called {\it profile likelihood}
approach for the treatment of nuisance parameters~\cite{asymptotic}.
The approach is based on the test statistic built as the following likelihood ratio:
\begin{equation}
  \lambda(\mu) = \frac{L(\vec{x};\mu,\hat{\hat{\theta}}(\mu))}{L(\vec{x};\hat{\mu},\hat{\theta})}\,,
  \label{eq:profLike}
\end{equation}
where in the denominator both $\mu$ and $\theta$ are fit simultaneously
as $\hat{\mu}$ and $\hat{\theta}$, respectively, and
in the numerator $\mu$ is fixed, and $\hat{\hat{\theta}}(\mu)$ is the best fit of $\theta$
for the fixed value of $\mu$. The motivation for the choice of Eq.~(\ref{eq:profLike})
as the test statistic comes from  Wilks' theorem that allows to approximate asymptotically
$-2\ln\lambda(\mu)$ as a $\chi^2$~\cite{Wilks}.

In general, Wilks' theorem applies if we have two hypotheses $H_0$ and $H_1$ that
are {\it nested}, i.e.: they can be expressed as sets of nuisance parameters
$\vec{\theta}\in\Theta_0$ and $\vec{\theta}\in\Theta_1$, respectively, such that
$\Theta_0\subseteq\Theta_1$. Given the likelihood function:
\begin{equation}
  L = \prod_{i=1}^N L(\vec{x}_i, \vec{\theta})\,,
\end{equation}
if $H_0$ and $H_1$ are nested, then the following quantity,
for $N\rightarrow\infty$, is distributed as a $\chi^2$ with a number of degrees of freedom
equal to the difference of the $\Theta_1$ and $\Theta_0$ dimensionalities: 
\begin{equation}
  \chi_r^2 = -2\ln\frac{\displaystyle
    \sup_{\vec{\theta}\in\Theta_0}\,\prod_{i=1}^NL(\vec{x}_i;\vec{\theta})
  }{\displaystyle
    \sup_{\vec{\theta}\in\Theta_1}\,\prod_{i=1}^NL(\vec{x}_i;\vec{\theta})
  }\,.
  \label{eq:wilks}
\end{equation}
In case of a search for a new signal where the parameter of interest is $\mu$,
$H_0$ corresponds to $\mu = 0$ and $H_1$ to any $\mu\ge0$, Eq.~(\ref{eq:wilks})
gives:
\begin{equation}
  \chi_r^2(\mu) = -2\ln\frac{\displaystyle
    \sup_{\vec{\theta}}\,\prod_{i=1}^NL(\vec{x}_i;\mu,\vec{\theta})
  }{\displaystyle
    \sup_{\mu^\prime,\vec{\theta}}\,\prod_{i=1}^NL(\vec{x}_i;\mu^\prime,\vec{\theta})
  }\,.
\end{equation}
Considering that the supremum is equivalent to the
best fit value, the profile likelihood defined in Eq.~(\ref{eq:profLike}) is obtained.

\begin{figure}[htbp]
\centering\includegraphics[width=.495\linewidth]{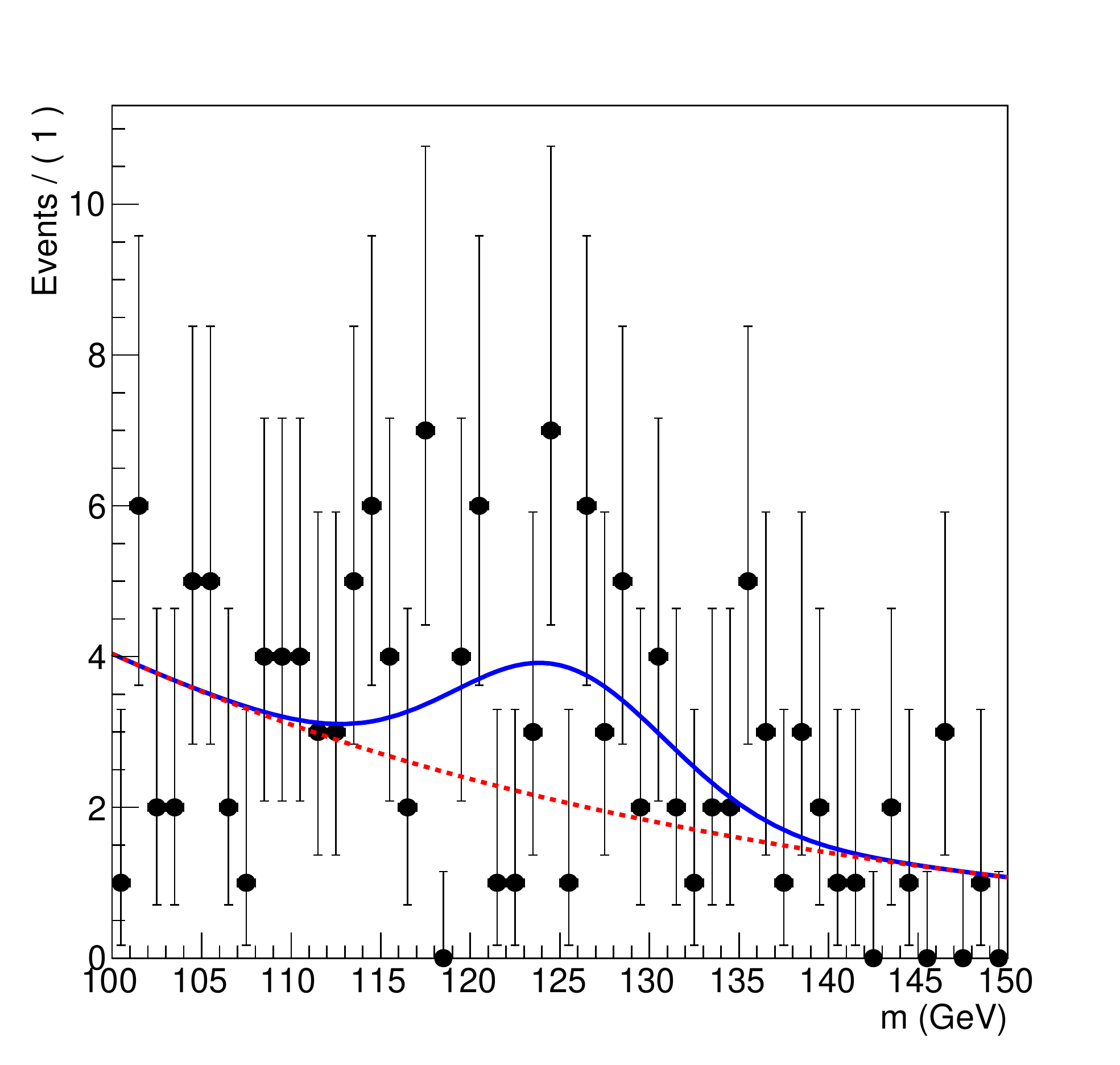}
\caption{Example of pseudoexperiment generated with a Gaussian-distributed signal
  over an exponential background. The assumed distribution for the background
  is the red dashed line, while the distribution for signal plus background is the blue
  solid line.
}
\label{fig:toyGplusB}
\end{figure}

As a concrete example of application of the profile likelihood,
consider a signal with a Gaussian distribution over a background
distributed according to an exponential distribution. A pseudoexperiment that was randomly-extracted
accordint to such a model is shown in Fig.~\ref{fig:toyGplusB}, where a signal yield
$s=40$ was assumed on top of a background yield $b=100$, exponentially
distributed in the range of the random variable $m$ from 100 to 150~GeV.
The signal was assumed centered at 125~GeV with a standard deviation of 6~GeV,
reminding the Higgs boson invariant mass spectrum.
The signal yields $s$ is fit from data.
All parameters in the model are fixed, except the background yield,
which is assumed to be known with some level of uncertainty modeled
with a log normal distribution whose corresponding nuisance parameter is called $\beta$.
The likelihood function for the model, which only depends on two parameters,
$s$ and $\beta$, is, in case of a single measurement $m$:
\begin{equation}
  L(m;s,\beta) = L_0(m;s,b_0 = be^\beta) L_\beta(\beta;\sigma_\beta)\,,
\end{equation}
where:
\begin{eqnarray}
  L_0(m;s,b_0) & = & \frac{e^{-(s+b_0)}}{n!}\left(
    s \frac{1}{\sqrt{2\pi}\sigma} e^{-{(m-\mu)^2}/{2\sigma^2}}+b_0\lambda e^{-\lambda m}
    \right)\,, \\
   L_\beta(\beta;\sigma_\beta) & = & \frac{1}{\sqrt{2\pi}\sigma_\beta}e^{-{\beta^2}/{2\sigma^2}}\,.
\end{eqnarray}
If we measure a set values $\vec{m}=(m_1,\,\cdots\,m_N)$, the likelihood function is:
\begin{equation}
  L(\vec{m};s,\beta) = \prod_{i=1}^N L(m_i;s,\beta)\,.
\end{equation}
\begin{figure}[htbp]
\centering\includegraphics[width=.495\linewidth]{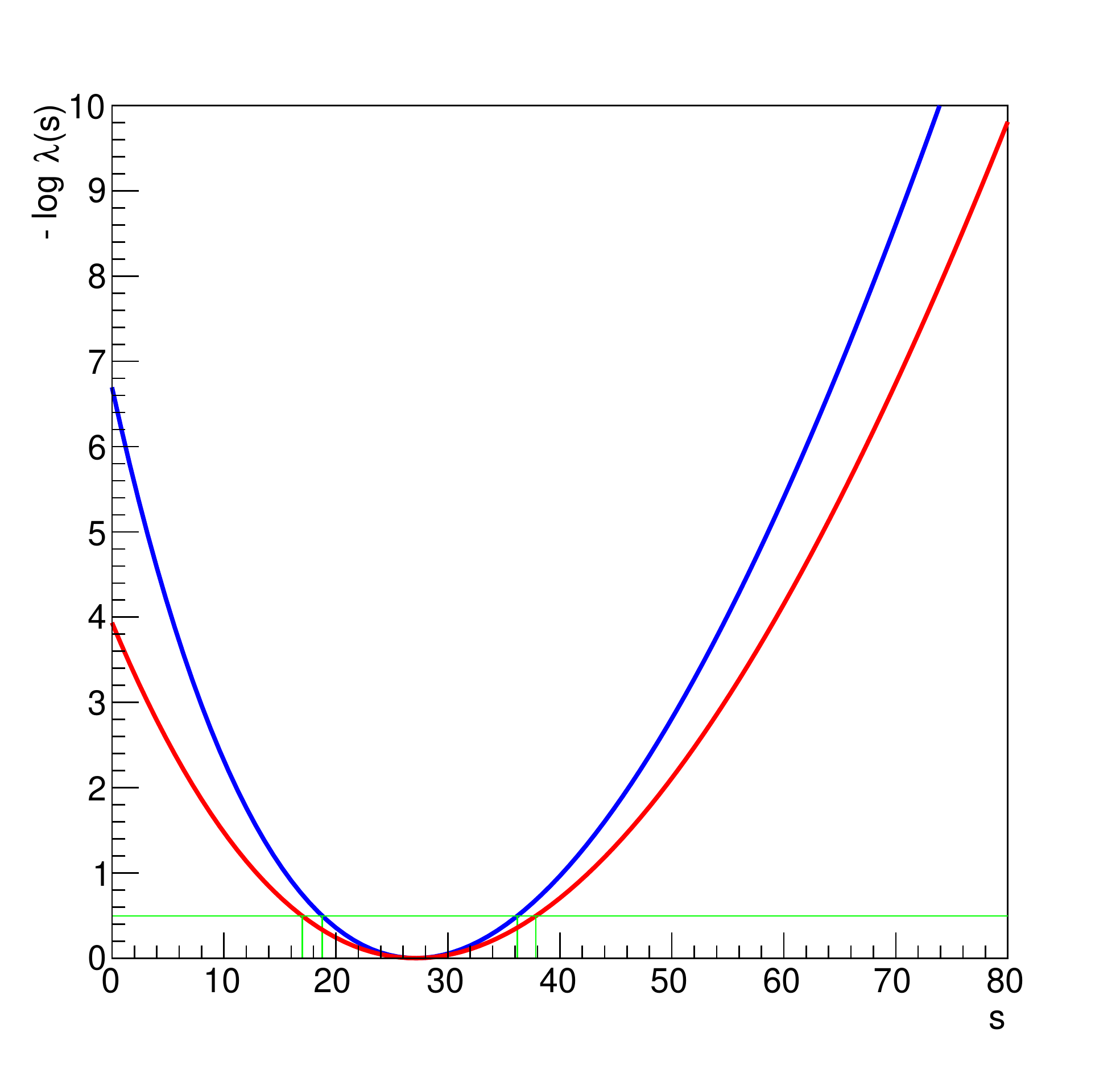}
\caption{Scan of the negative logarithm of the profile likelihood as a function
  of the signal yield $s$. The blue curve shows the profile likelihood
  curve defined assuming the background yield to be a constant (i.e.: $b$ known
  without any uncertainty), the red curve shows the same curve defined with
  $\sigma_\beta=0.3$. The green line at $-\ln\lambda(s)=0.5$ determines
  the uncertainty interval on $s$.
}
\label{fig:plScan}
\end{figure}
The scan of $-\ln\lambda(s)$ is shown in Fig.~\ref{fig:plScan}, where the profile likelihood
was evaluated assuming $\sigma_\beta=0$ (no uncertainty on $b$, blue curve) or $\sigma_\beta=0.3$
(red curve). The minimum value of $-\ln\lambda(s)$ is equal to zero, since
at the minimum numerator and denominator in Eq.~(\ref{eq:profLike}) are identical.
Introducing the uncertainty on $\beta$ (red curve) makes the curve broader.
This causes an increase of the uncertainty on the estimate of $s$, whose uncertainty
interval is obtained by intersecting the curve of the negative logarithm of the profile likelihood
with an horizontal line at $-\ln\lambda(s) = 0.5$ (green line in Fig.~\ref{fig:plScan}\footnote{
  The plot in Fig.~\ref{fig:plScan} was generated with the library {\sc RooStats}
  in {\sc Root}~\cite{Root}, which by default, uses $-\ln\lambda$ instead of $-2\ln\lambda$.
}).

In order to evaluate the significance of the observed signal, Wilks' theorem can be
used. If we assume $\mu=0$ (null hypothesis), the quantity $q_0 = -2\ln\lambda(0)$
can be approximated with a $\chi^2$ having one degree of freedom. Hence, the significance
can be approximately evaluated as:
\begin{equation}
  Z\simeq \sqrt{q_0}\,.
\end{equation}
$q_0$ is twice the intercept of the curve in Fig.~\ref{fig:plScan} with the vertical axis,
and gives an approximate significance of $Z\simeq\sqrt{2\times6.66} = 3.66$,
in case of no uncertainty on $b$, and $Z\simeq\sqrt{2\times3.93} = 2.81$, in case
the uncertainty on $b$ is considered. 
In this example, the effect of background yield uncertainty reduces the
significance bringing it below the evidence level ($3\sigma$).
Those numerical values can be verified
by running many pseudo experiments (toy Monte Carlo) assuming $\mu=0$ and
computing the corresponding $p$-value. In complex cases, the computation
of $p$-values using toy Monte Carlo may become unpractical, and Wilks'
approximation provides a very convenient, and often rather precise,
alternative calculation.

\subsection{Variations on test statistics}

A number of test statistics is proposed in Ref.~\cite{asymptotic} that better
serve various purposes. Below the main ones are reported:
\begin{itemize}
\item {\bf Test statistic for discovery:}
  \begin{equation}
    q_0 = \left\{
      \begin{array}{ll}
        -2\ln\lambda(0), &\hat{\mu}\ge 0\,,\\
        0, & \hat{\mu} < 0\,.
      \end{array}
      \right.
      \label{eq:tsfd}
  \end{equation}
  In case of a negative estimate of $\mu$ ($\hat{\mu}<0$), the test statistic is set to zero in order to
  consider only positive $\hat{\mu}$ as evidence against the background-only hypothesis.
  Within an asymptotic approximation, the significance is given by: $Z\simeq\sqrt{q_0}$.

\item {\bf Test statistic for upper limit:}
  \begin{equation}
    q_\mu = \left\{
    \begin{array}{ll}
        -2\ln\lambda(\mu), &\hat{\mu}\le \mu\,,\\
        0, & \hat{\mu} > \mu\,.
      \end{array}
      \right.
  \end{equation}
  If the $\hat{\mu}$ estimate is larger than the assumed value for $\mu$, an upward fluctuation occurred.
  In those cases, $\mu$ is not excluded by setting the test statistic to zero.
\item {\bf Test statistic for Higgs boson search:}
    \begin{equation}
    \tilde q_\mu = \left\{
    \begin{array}{ll}
      -2\ln\frac{L(\vec{x}|\mu,\hat{\hat{\vec{\theta}}}(\mu))}
      {L(\vec{x}|0,\hat{\hat{\vec{\theta}}}(0))}, & \hat{\mu} < 0\,,\\
      -2\ln\frac{L(\vec{x}|\mu,\hat{\hat{\vec{\theta}}}(\mu))}
      {L(\vec{x}|\mu,\hat{\vec{\theta}}(\mu))}, & 0\le \hat{\mu} < \mu\,,\\
        0, & \hat{\mu} \ge \mu\,.
    \end{array}
      \right.
  \end{equation}
    This test statistics both protects against unphysical cases with $\mu <0$
    and, as the test statistic for upper limits, protects upper limits
    in cases of an upward $\hat{\mu}$ fluctuation.
\end{itemize}

A number of measurements performed at LEP and Tevatron used a
test statistic based on the ratio of the likelihood function evaluated under
the signal plus background hypothesis and under the background only hypothesis,
inspired by the Neyman--Pearson lemma:
\begin{equation}
  q = -2\ln\frac{L(\vec{x}|s+b)}{L(\vec{x}|b)}\,.
\end{equation}
In many LEP and Tevatron analyses, nuisance parameters were treated using the hybrid Cousins--Hyghland approach.
Alternatively, one could use a formalism similar to the profile likelihood, 
setting $\mu=0$ in the denominator and $\mu=1$ in the numerator, and minimizing
the likelihood functions with respect to the nuisance parameters:
\begin{equation}
  q = -2\ln\frac{L(\vec{x}|\mu=1, \hat{\hat{\theta}}(1))}{L(\vec{x}|\mu=0,\hat{\hat{\theta}}(0))}\,.
\end{equation}

For all the mentioned test statistics, asymptotic approximations exist and
are reported in Ref.~\cite{asymptotic}. Those are based either on Wilks' theorem
or on Wald's approximations~\cite{Wald}. If a value $\mu$ is tested, and 
the data are supposed to be distributed according to another value of the signal strength $\mu^\prime$,
the following approximation holds, asymptotically:
\begin{equation}
  -2\ln\lambda(\mu) = \frac{(\mu-\hat{\mu})^2}{\sigma^2} + {\cal O}\left(\frac{1}{\sqrt{N}}\right)\,,
  \label{eq:waldTestStat}
\end{equation}
where $\hat{\mu}$ is distributed according to a Gaussian with average $\mu^\prime$ and
standard deviation $\sigma$. The covariance matrix for the nuisance parameters is
given, in the asymptotic approximation, by:
\begin{equation}
  V_{ij}^{-1} = \left.\left<\frac{\partial^2\ln L}{\partial\theta_i\partial\theta_j}\right>\right|_{\mu=\mu^\prime}\,,
\end{equation}
where $\mu^\prime$ is assumed as value for the signal strength.

In some cases, asymptotic approximations (Eq.~(\ref{eq:waldTestStat})) can be written in terms of an
{\it Asimov dataset}~\cite{Asimov}:
\begin{displayquote}
  {\it We define the Asimov data set such that when one uses it to evaluate the estimators
    for all parameters, one obtains the true parameter values}~\cite{asymptotic}.
\end{displayquote}
In practice, an Asimov dataset is a single ``representative'' dataset obtained by replacing all
observable (random) varibles with their expecteted value. In particular,
all yields in the data sample (e.g.: in a binned case) are replaced with their expected values, that may be non integer values.
The median significance for different cases of test statistics can be computed in this
way without need of producing extensive sets of toy Monte Carlo. The implementation
of those asymptotic formulate is available in the {\sc RooStats} library, released
as part an optional component {\sc Root}~\cite{Root}.

\subsection{The look-elsewhere effect}

When searching for a signal peak  on top of a background that is smoothly distributed over a wide range,
one can either know the position of the peak or not.
One example in which the peak position is known is the
search for a rare decay of a known particle, like $\mathrm{B}_{\mathrm{s}}\rightarrow\mu^+\mu^-$.
A case when the position was not know was the search for the Higgs boson, whose mass is not
prediceted by theory.
In a case like the decay of a particle of known mass, it's easy to compute the peak significance:
from the distribution of the test statistic $f(q)$ computed assuming $\mu=0$ (background only),
given the observed value of the test statistic $q^{\mathrm{obs}}$, a $p$-value can be determined and then
translated into a significance level:
\begin{equation}
  p = \int_{q^{\mathrm{obs}}}^{+\infty} f(q|\mu=0)\,\mathrm{d}q,\quad Z = \Phi^{-1}(1-p)\,.
  \label{eq:localPval}
\end{equation}

In case, instead, the search is performed without knowing the position of the peak,
Eq.~(\ref{eq:localPval}) gives only a {\it local $p$-value}, which means it reflects
the probability that a background fluctuation {\it at a given mass value $m$}  gives a value
of the test statistic greater than the observed one:
\begin{equation}
  p(m) = \int_{q^{\mathrm{obs}}(m)}^{+\infty} f(q|\mu=0)\,\mathrm{d}q\,.
\end{equation}
The {\it global $p$-value}, instead, should quantify the probability that a background
fluctuation {\it at any mass value} gives a value of the test statistic greater than the
observed one.

The chance that an overfluctuation occurs for {\it at least} one mass value increases with
the size of the search range, and the magnitude of the effect depends on the resolution.

One possibility to evaluate a global $p$-value is to let also $m$ fluctuate in the test statistic:
\begin{equation}
  \hat{q} = -2\ln \frac{L(\mu=0)}{L(\hat{\mu};\hat{m})}\,.
\end{equation}
Note that in the numerator $L$ doesn't depend on $m$ for $\mu=0$. This is a case
where Wilks' theorem doesn't apply, and no simple asymptotic approximations exist.
The global $p$-value can be computed, in principle, as follows:
\begin{equation}
  p^{\mathrm{glob}} = \int_{\hat{q}^{\mathrm{obs}}}^{+\infty}f(\hat{q}|\mu=0)\,\mathrm{d}\hat{q}_0\,.
\end{equation}
The effect in practice can be evaluated with brute-force toy Monte Carlo:
\begin{itemize}
\item Produce a large number of pseudoexperiments simulating background-only samples.
\item Find the maximum $\hat{q}$ of the test statistic $q$ in the entire search range.
\item Determine the distribution of $\hat{q}$.
\item Compute the global $p$-value as probability to have a value of $\hat{q}$ greater than the observed one.
\end{itemize}
This procedure usually requires very large toy Monte Carlo samples in order to treat a discovery case:
a $p$-value close to $3\times 10^{−7}$ ($5\sigma$ level) requires
a sample significantly larger than $\sim 10^7$ entries in order to
determine the $p$-value with small uncertainty.

\begin{figure}[htbp]
\centering\includegraphics[width=.495\linewidth]{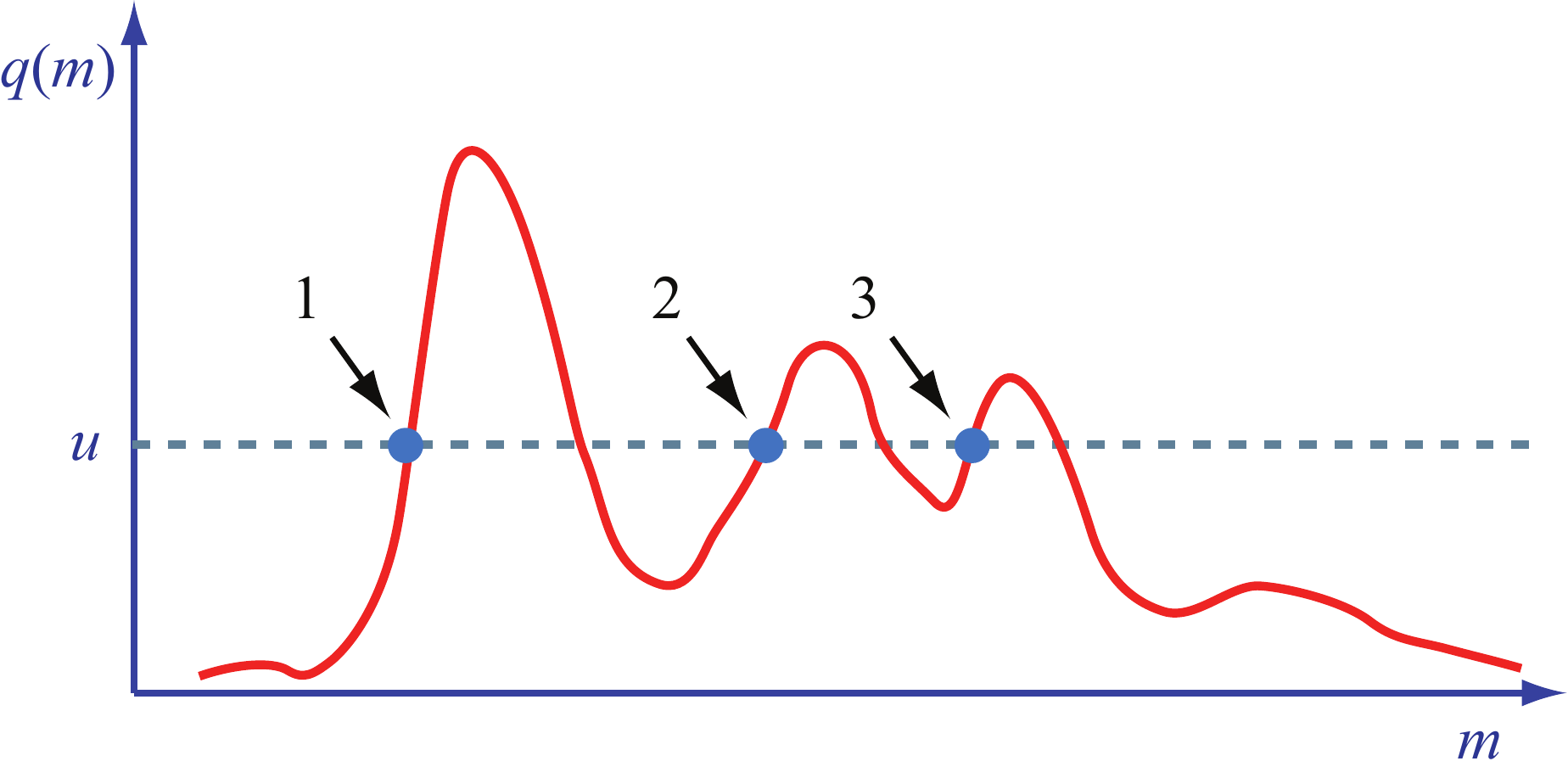}
\caption{Illustration of the evaluation of the number of upcrossing
  of a test statistic curve $q(m)$. The upcrossings are noted as 1, 2, and 3,
  hence the corresponding $N_u$ is equal to 3.
}
\label{fig:lee}
\end{figure}

An asymptotic approximation for the global $p$-value is given by
the following inequation~\cite{lee_trial}~\footnote{
  In case of a test statistic for discovery $q_0$ (Eq.~(\ref{eq:tsfd})), the term $P(\chi^2 > u)$
  in Eq.~(\ref{eq:leeasin}) achieves an extra factor 1/2, which is usually not be present
  for other test statistics.
  }:
\begin{equation}
  p^{\mathrm{glob}} = P(\hat{q}>u ) \le \left<N_u\right> +
  P(\chi^2>u)\,,
  \label{eq:leeasin}
\end{equation}
where $P(\chi^2>u)$ is a standard $\chi^2$ probability and
$\left<N_u\right>$ is the average number of {\it upcrossings} of
the test statistic, i.e.: the average number of times that
the curve $q(m)$ crosses a given horizontal line at a level $u$ with a positive derivative,
as illustrated in Fig.~\ref{fig:lee}.

The number of upcrossings may be very small for some values of $u$,
but an approximate scaling law exists and allows to perform the computation
at a more convenient level $u_0$:
\begin{equation}
  \left<N_u\right> = \left<N_{u_0}\right> e^{-{(u-u_0)}/{2}}\,.
\end{equation}
So, $ \left<N_{u_0}\right>$ can be more conveniently evaluated using
a reasonable number of toy Monte Carlo generations, then it can be extrapolated following
the exponential scaling law.
Numerical comparisons of this approach with the full toy Monte Carlo
show that good agreement is achieved for sufficiently
large number of observations.

In case more parameters are estimated from data, e.g.: when searching for
a new resonance whose mass and width are both unknown, the look-elsewere
effect can be addressed with an extension of the approach described above,
as detailed in Ref.~\cite{leeND}.

\addcontentsline{toc}{chapter}{Bibliography}
\bibliographystyle{ieeetr}
\bibliography{lista}
\end{document}